\documentclass[a4paper,12pt]{article}
\usepackage{jheppub}

\usepackage{amsmath,epsfig}
\usepackage{amssymb,amsfonts}
\usepackage{latexsym}
\usepackage[latin1]{inputenc}
\usepackage{slashed}
\usepackage{empheq}
\numberwithin{equation}{section}
\usepackage{cancel}
\usepackage{overpic}
\usepackage{subcaption}
\usepackage{subeqnarray}
\usepackage{xcolor}

\usepackage{longtable}
\usepackage{color}
\usepackage{multirow}
\usepackage{epstopdf}
\def\nn{\nonumber}

\relax

\def\hri#1#2{\href{http://arxiv.org/abs/#1}{[ArXiv:#1]#2}}
\def\hre#1#2{\href{http://arxiv.org/abs/#1/#2}{[ArXiv:#1/#2]}}
\def\hrj#1#2{\href{https://doi.org/#1}{#2}}

\renewcommand{\theequation}{\arabic{section}.\arabic{equation}}

\def\be{\begin{equation}}
\def\ee{\end{equation}}

\newcommand{\bear}{\begin{eqnarray}}
\newcommand{\bea}{\begin{eqnarray}}
\newcommand{\eear}{\end{eqnarray}}
\newcommand{\eea}{\end{eqnarray}}

\newbox\pippobox

\def\II{\relax{\rm I\kern-.18em I}}

\def\e{\epsilon}
\def\m{\mu}
\def\n{\nu}
\def\r{\rho}
\def\s{\sigma}
\def\pa{\partial}

\def\sp{\;\;\;,\;\;\;}

\def\p{\partial}

\def\f{\varphi}
\def\z{\zeta}
\def\a{\alpha}

\def\l{\lambda}
\def\g{\gamma}
\def\d{\delta}

\def\II{{\cal I}}

\def\bsq{\begin{subequations}}
\def\esq{\end{subequations}}
\def\k{\kappa}
\def\D{\Delta}

\preprint{\\
	\hspace*{12cm} CCTP-2025-4\\ 
	\hspace*{12.1cm} ITCP-2025/4\\ }
\title{On the spectra of holographic QFTs on constant curvature manifolds}

\author[a]{Ahmad Ghodsi}
\author[b,c,d]{, Elias Kiritsis}
\author[a]{, Parisa Mashayekhi}
\author[b]{and Francesco Nitti}
\affiliation[a]{
	Department of Physics, Faculty of Science,	Ferdowsi University of Mashhad,  Mashhad, Iran.}
\affiliation[b]{	\href{http://www.apc.univ-paris7.fr}{Universit\'e Paris Cit\' e, CNRS, Astroparticule et Cosmologie}, F-75013 Paris, France.}
\affiliation[c]{\href{http://hep.physics.uoc.gr}{Crete Center for Theoretical Physics},\\ Institute for Theoretical and Computational Physics,
Department of Physics, \\
University of Crete, Heraklion, Greece.}
\affiliation[d]{\href{https://www.theorie.physik.uni-muenchen.de/}{Arnold Sommerfeld Center for Theoretical Physics}, \\ Ludwig-Maximilians-Universit\"at 
	M\"unchen, 80333
	M\"unchen, Germany.}

\abstract{We analyze linear fluctuations of five-dimensional Einstein-Dilaton theories dual to holographic quantum field theories defined on four-dimensional de Sitter and Anti-de Sitter space-times.  We identify the physical propagating scalar and tensor degrees of freedom. For these,   we write the linearized bulk  field equations as eigenvalue equations. In the dual QFT, the  eigenstates correspond to towers of spin-0 and spin-2 particles propagating on $(A)dS_4$ associated to gauge-invariant composite states.  Using particular care in treating special ``zero-modes,'' we show in general that, for negative curvature, the particle spectra are always discrete, whereas for positive curvature they always have a continuous component starting at $m^2 = (9/4)\alpha^{-2}$, where $\alpha$ is the $(A)dS_4$ radius. We numerically compute the spectra in a concrete model characterized by  a polynomial dilaton  bulk potential admitting holographic  RG-flow solutions with a UV and IR fixed points. In this case, we find no discrete spectrum and no perturbative instabilities.}

\keywords{}
\begin{document}

\maketitle

\section{Introduction and Summary}
When quantum field theories (QFT) are considered on a curved space-time, new phenomena may appear with respect to their Minkowski counterparts. Since curvature is a relevant operator, its effect is more significant in the IR regime of the field theory. Given that many interesting 4-dimensional QFTs are strongly coupled in the IR, one often needs non-perturbative methods to understand how curvature affects these theories.

The holographic gauge/gravity duality \cite{9711200, 9802109, 9802150} provides a tool for such a non-perturbative analysis, at least in the limit of strong coupling and large $N$ for the field theory. In these cases, the existence of a weakly-coupled gravity dual allows one to translate computations of observables in the strong-coupling regime to classical gravity/string theory calculations in a higher dimensional, asymptotically anti-de Sitter (AdS) space-time, whose (conformal) boundary has the same geometry as the space-time on which the QFT is defined.

The extra (holographic) dimension serves as an effective renormalization group (RG) scale in the dual QFT.
This geometrizes the notion of RG flow. In essence, RG flows can be understood as bulk evolution in the holographic dimension \cite{9810126, 9903190, 9904017, 9912012, 0105276, 0404176, 0702088, Papadimitriou:2007sj,iQCD,1010.1264,1010.4036,1106.4826,1112.3356,1205.6205,1310.0858,1401.0888,exotic}, \cite{1006.1263}.

In the context of the gauge/gravity duality, there is no conceptual or technical problem in considering holographic duals of QFTs on a curved space-time $M_d$, as long as one considers gravity duals with the appropriate asymptotic geometry.  Along these lines, a systematic analysis of holographic QFTs defined on curved space-times has been carried out by some of the authors in a series of recent papers  \cite{C,F,Preau,Litos,AdS1,AdS2,AdS3,Jokela:2021knd,Jani,JLR},  which analyze the features of various types of gravity theories admitting   {\em curved holographic RG flow} solutions: these are dual to  QFTs which are defined on a $d$--dimensional constant-curvature space-time, that start at a UV fixed point, and have a non-trivial flow to the IR due to a deformation by a relevant operator. The gravity dual theory may be approximated by a $d+1$--dimensional Einstein gravity coupled to a single scalar field. The latter may be thought of as representing the running relevant coupling of the dual QFT.

On a constant curvature space-time, the physics of such a theory is determined by the interplay between two scales:
\begin{itemize}
	\item The curvature scalar of the UV QFT $R^{(UV)}$;
	\item The mass parameter $\Lambda_{UV}$ which characterizes the relevant coupling.
\end{itemize}
More precisely, the physics depends on their dimensionless ratio,
\be \label{intro1}
{\cal R} \equiv \frac{ R^{(UV)}}{ \Lambda_{UV}^2}\,.
\ee

The details then differ depending on the specific features of the QFT, which, when defined on flat space, may be e.g. a non-perturbative RG flow between two conformal fixed points as in \cite{C,F,AdS1}, or a confining theory as in \cite{AdS3,Jani,JLR}. Also, different signs of the curvature produce {\em vastly} different effects in the IR, giving rise to completely different physics for the QFT defined on $d$--dimensional de Sitter as opposed to Anti-de Sitter space-time.

In all the aforementioned papers on curved holographic RG flows, the focus was on the {\em background} solution, and the main goal was to classify the different types of geometries that solving the gravity dual equations, and establishing the phase diagram as a function of the dimensionless parameter ${\cal R}$ defined in (\ref{intro1}).  This resulted in a rich phase structure and various types of curvature-driven phase transitions both for theories with an IR fixed point \cite{C,AdS1} and for those which in flat space have an IR mass gap \cite{AdS3,Jani,JLR}. Some early examples of similar phase transitions were also found in \cite{1007.3996,Blackman}

In this work, we shall study a first aspect of the dynamics of such theories, namely the spectra of propagating excitations. More precisely, we study linear perturbations around the vacua of curved holographic RG flows.  The motivation is two-fold:
\begin{enumerate}
	
	\item To address the dynamical stability of the background solutions found in previous works;
	
	\item To understand the spectral properties of the dual QFTs when placed on a curved background.
	
\end{enumerate}
The first point is important to establish the consistency of the curved RG flows solutions constituting the phase diagram, i.e., whether these solutions can be taken as {\em credible} ground states of the holographic theory at finite curvature.

Concerning the second point, by the gauge gravity duality, the spectrum of {\em normalizable} (i.e., finite boundary energy) perturbations around a classical gravity solution provides the spectrum of gauge-invariant single-particle-like excitations in the dual field theory above the ground state. In the large-$N$ limit, because the gravity dual is weakly coupled, the spectrum takes the form (schematically) of a   tower of modes $\varphi_\alpha$ satisfying a free $d$--dimensional wave equation on $M_d$,
\be   \label{intro2}
\Box_d  \, \varphi_\alpha(x^\mu) = m^2_\alpha \varphi_\alpha (x^\mu)\,,
\ee
with interactions suppressed by inverse powers of $N$.  The set of mass eigenvalues $m^2_\alpha$ is obtained by solving an appropriate spectral problem in the holographic direction. The spectrum may be continuous or discrete\footnote{This is the case for example in holographic QCD-like theories, where the discrete tower of particles can be thought of as weakly-interacting glueballs \cite{Csaki:1998qr,Kiritsis:2006ua}}, gapped or ungapped, depending crucially on the IR of the geometry (i.e. the interior region far away from the asymptotic AdS$_{d+1}$ boundary.)

Here we take $M_d$ to be a maximally symmetric manifold, i.e., dS$_d$ or AdS$_d$ in the Lorentzian signature, or the corresponding Euclidean manifolds, the $d$--dimensional sphere and Euclidean AdS (or hyperbolic space).

The spectrum of propagating modes of holographic QFTs on a constant curvature manifold $M_d$, can be interesting for several reasons and in different contexts:
\begin{itemize}
	\item When $M_d$ has constant positive curvature, we have two maximally symmetric examples: the sphere $S^d$ when the signature is Euclidean and de Sitter (dS$_d$) space, when the signature is Minkowski-like. The case of dS$_d$ is especially interesting, due to its implications for cosmology, and the fact that massless theories on dS$_d$ have perturbation theories that break down.
	Although some methods were proposed to deal with this issue, \cite{Staro1,Tsamis1,GS}, the problem is still considered open.
	
	Recently, some two-dimensional solvable theories on dS$_2$ were considered, for which this apparent problem is resolved by directly calculating the non perturbative answer that resumes the secular terms of perturbation theory, \cite{Dio}. An interesting feature of spectra on dS$_d$ is that unitary representations of the associated $O(1, d)$ symmetry group can appear with distinct properties: continuous representations, complementary series representations, as well as discrete series representations. It is not yet clear which of these representations play a role in interacting QFTs on dS$_d$.
	
	\item  When a QFT is defined on a constant negative curvature manifold, we have two maximally symmetric cases. Euclidean AdS, EAdS$_{d}$ in Euclidean signature and AdS$_d$ in Minkowski-like signature.
	So far, we know holographic realizations of such theories only when the holographic QFT is confining in flat space, \cite{AdS3}. We have so far no examples of the spectra of strongly-coupled QFTs on AdS$_d$ beyond the somewhat trivial case of near-free theories.
	
	\item In holography, the ansatz that describes QFTs on a constant negative curvature manifold, describes also, generically QFT interfaces in flat space, \cite{Bak,CF,DH2,GW1,GW2,DH,JO,Bobev:2019jbi,Bobev:2020fon,C1,C2}, \cite{AdS1,AdS2,AdS3}, if the slice manifold is non-compact with a boundary (like AdS$_d$).
	When the slice manifold is compact (like a $g>1$ Riemann surface or higher-dimensional Schottky manifolds, then the solution is a wormhole, \cite{maldamaoz,AdS1}.
	In such a case, the spectrum of propagating modes describes both the modes of the different sides of the interface as well as the interface modes.
	
	\item  For generic holographic theories that are confining in flat space, various phases at finite curvature (with different geometrical properties), and first or higher-order phase transitions between them, were recently found in \cite{Jani,JLR}. Calculating the spectrum gives an extra handle to qualitatively distinguish the different phases, for which, as discussed in \cite{Jani}, it is hard to identify a well-defined order parameter. Understanding the spectrum can also help to better understand the corresponding continuous phase transitions (e.g., scaling behavior may be associated with light modes).
\end{itemize}

\subsection{Summary of Results}
The framework we work on is a holographic theory whose gravity side consists of $5$--dimensional Einstein gravity coupled to a scalar field $\Phi$. Our goal will be to study the linear perturbations of this system around curved holographic RG flow solutions of the type described in \cite{C,AdS1,AdS3,Jani}, whose background metric and scalar field take the form
\be \label{intro3}
ds^2 = du^2 + e^{2A(u)}\zeta_{\mu\nu} dx^\mu dx^\nu  \sp \qquad \Phi= \Phi(u) \sp  \qquad \mu, \nu =  1\ldots 4 \,.
\ee
Here $u$ is the holographic coordinate,  the coordinates $x^\mu$ parameterize the constant-$u$ slices, and  the {\em slice metric}  $\zeta_{\mu\nu}$ is a $u$-independent, constant-curvature $4$--dimensional metric on the constant-$u$ slices, whose Ricci tensor is given by:
\be \label{intro3-i}
R^{(\zeta)}_{\mu\nu} = \kappa \zeta_{\mu\nu}\sp \qquad \kappa = \pm \frac{3}{ \alpha^2} \,,
\ee
and we denote by $\alpha$ the radius of the corresponding manifold $S_4$, dS$_4$, AdS$_4$ or EAdS$_4$  (depending on the curvature sign and the metric signature).
\vspace{0.3cm}

The main goals we achieve in this work are:
\begin{enumerate}
	\item To isolate the decoupled physical degrees of freedom describing linearized perturbations around curved holographic RG flows;
	\item   To write the corresponding linearized equation in the form of a spectral problem giving the mass eigenvalues $m^2_\alpha$;
	\item To understand the universal features of the mass spectra as a function of the boundary curvature and the type of IR geometry;
	\item  Finally, we numerically obtain the full spectrum, for positive and negative curvature, for a simple theory admitting (in flat space) holographic RG flows which connect two fixed points.
\end{enumerate}
The analysis follows the same lines which, for flat slice metric $\zeta_{\mu\nu}$, are well established in holography and whose details can be found, e.g., in \cite{Kiritsis:2006ua} for 4--dimensional theories. Here, we generalize that analysis to non-zero curvature slices.

It is worth noting that the problem we study here is related by a Wick-rotation to the problem of cosmological perturbations around a Friedman-Robertson-Walker (FRW) space-time with arbitrary constant spatial curvature. There, the holographic coordinate $u$ is replaced by time.  This problem is conceptually similar; however, an important difference is that, in the cosmological setting, the transverse manifold is Euclidean. This leads to a different (compared to our case) treatment of the zero-modes, which appear in the tensorial decomposition of perturbations, and to which we dedicate a careful treatment. Also, boundary conditions in the $u$-direction are treated differently from the initial data for cosmological perturbations.
The dynamics of cosmological perturbation at non-zero spatial curvature can be found in \cite{Kodama:1984ziu}, and it was recently revisited in \cite{Bianchi:2024mlq}.

In the rest of this introduction, we briefly summarize our results.

\paragraph{Gauge invariant degrees of freedom.}
As in the case of flat slicing, the set of all metric and scalar field fluctuations around the background (\ref{intro3}) is subject to gauge transformations consisting of $5$--dimensional diffeomorphisms. These, together with the constraints from Einstein's equations, remove all but the following propagating degrees of freedom, which are conveniently parameterized in terms of their tensorial properties with respect to the $4$--dimensional slice:

\begin{enumerate}
	
	\item A gauge-invariant {\em scalar mode} $\lambda(u,x^\mu)$, which can be taken as a combination of the metric trace and the perturbation of $\Phi$;
	
	\item A gauge-invariant symmetric {\em  tensor mode} $h_{\mu\nu}^{TT}(u,x^\mu)$, which is transverse and traceless with respect to the slice metric:
	\be \label{intro4}
	\nabla^\mu h_{\mu\nu}^{TT} = \zeta^{\mu\nu} h_{\mu\nu}^{TT} = 0\,,
	\ee
	where $\nabla^\mu$ is the covariant derivative of the slice metric $\zeta_{\mu\nu}$. These modes are the bulk gravitons.
	
\end{enumerate}

This decomposition is universal, i.e., independent of the scalar field potential. It only depends on the two-derivative nature of the Einstein-Dilaton action. In these theories, there are no dynamical vector perturbations.

What is described above is the same gauge-invariant field content of the perturbations around flat-sliced solutions, or around an FRW space-time in a theory with a single scalar field or a single fluid.

\paragraph{Spectral equations.}
By combining various components of Einstein's equation, we obtained decoupled linearized equations for the gauge-invariant scalar and tensor modes.
These take the form of linear second-order partial differential equations, which can be separated in a standard way by setting:
\be \label{intro5}
h_{\mu\nu}^{TT}(u,x^\mu) = h(u)   h_{\mu\nu}^{(M),TT}(x^\mu) \sp \lambda(u,x^\mu) = \lambda(u) Y_m(x^\mu)\,,
\ee
where $h_{\mu\nu}^{(M),TT}(x^\mu)$ and  $Y_m(x^\mu)$ satisfy the spin-2 and spin-0 massive field equation on $M_d$ with metric $\zeta_{\mu\nu}$:
\be \label{intro6}
\left(\nabla^\mu \nabla_\mu - m^2\right) Y_m = 0 \sp \big(\nabla^\mu \nabla_\mu -\frac23 \kappa - M^2\big)  h_{\mu\nu}^{(M),TT} = 0\,,
\ee
where $\kappa$ is the slice curvature, defined in (\ref{intro3-i}). The reader will recognize that the tensor modes in (\ref{intro6}) satisfy the spin-2 Pauli-Fierz equation around a constant curvature background, with $M$ being the Pauli-Fierz mass parameter.

The coefficient functions $h(u)$ and $\lambda(u)$ satisfy  $M$ and $m$-dependent ordinary differential equations, which after a change of variables and coordinates can be put in a Schr\"odinger form:
\be\label{intro7}
\left(-\frac{d^2 }{ dy^2} +V_g(y) \right) \psi_g(y) = M^2 \psi_g(y) \sp \left(-\frac{d^2 }{ dy^2} +V_s(y;m) \right) \psi_s(y) =0\,.
\ee
$\psi_{g,s}$ are related to $h$ and $\lambda$ by an appropriate re-scaling, and $y$ is the bulk conformal coordinate, related to $u$ by the warp factor $A(u)$ of the metric (\ref{intro3}):
\be\label{intro8}
dy = e^{-2A(u)}du.
\ee
The effective {\em Schr\"odinger potentials} $V_g(y)$ and $V_s(y;m)$ depend only on the functions (and their derivatives) specifying the background solution (\ref{intro3}), i.e., $A(y)$ and $\Phi(y)$.

The graviton Schr\"odinger potential takes a particularly simple form, which is in fact the same as for flat slicing \cite{Kiritsis:2006ua}:
\be \label{intro9}
V_g = \left(\frac{dB }{ dy}\right)^2 - \frac{d^2B }{ dy^2} \sp  B(y) \equiv \frac{3}{ 2}A(y)\,.
\ee
i.e., it depends only on the metric scale factor $A(y)$ and not on the dilaton\footnote{Notice, however, that $A(y)$, due to the curvature of the slice, is very different from the flat-slicing scale factor.}. The mass parameter $M^2$ enters as the energy in the graviton  Schr\"odinger equation.

The scalar potential $V_s(y;m)$ is more involved.  Its analytic expression is not particularly illuminating, and we will present some universal features in the next paragraph. It contains the mass parameter $m$ explicitly, and the corresponding equation takes the form of a zero-energy Schr\"odinger equation.

Comparing equations (\ref{intro6}) and (\ref{intro7}), we observe that the set of eigenvalues of the one-dimensional Schr\"odinger spectral problems gives the mass spectra of four-dimensional spin-0 and spin-2 particle excitations of the holographic theory. In maximally symmetric space-times, these are classified by representations of the corresponding (pseudo)-rotation groups: solving the Schr\"odinger equations tells us which representations appear, and also whether the corresponding masses satisfy the appropriate stability bounds (which generalize the flat-slice bounds $m^2\geq 0$, $M^2\geq 0$).

While the details of the actual spectra depend on the specific form of the bulk solution, a great deal of information can be obtained by looking at the asymptotic form of the Schr\"odinger potentials, which, as it turns out, is rather universal. Below we order the discussion according to the sign of the curvature.

\paragraph{Zero Modes.}
Before gauge fixing, the perturbations can be divided into tensors ($h_{\mu\nu}$), vectors ($h_{\mu y} $), and scalars ($h_{yy}, \delta \Phi$) with respect to the transverse manifold coordinate transformations. These modes appeared coupled to each other in Einstein's equation, and the standard way to obtain decoupled equations is to further split them into longitudinal and transverse with respect to the slice covariant derivative. This decomposition is well defined, except when the modes correspond to special eigenvalues ({\em zero-modes}) of the slice-Laplacian $\nabla^2 \equiv \nabla^\mu \nabla_\mu$. These special values are:
\begin{itemize}
	\item {\bf Scalar zero modes:} $\nabla^2_{s} = 0, \,  -\frac43 \kappa$
	\item {\bf Vector zero modes:} $\nabla^2_{v} = - \kappa$
\end{itemize}
where  $\nabla^2_{s}$ and $\nabla^2_{v} $ are the scalar and vector covariant Laplacians of $\zeta_{\mu\nu}$.

In the presence of these zero-modes, the standard transverse/longitudinal decomposition is ill-defined, and one needs to proceed in a different way. To this, we decompose the fluctuations into appropriate bases of eigen-functions of tensor, vector, and scalar Laplacians, and then we derive the radial equations {\em without} having to decompose in longitudinal and transverse modes.  This method actually gives an alternative derivation of the decoupled equations not only for the zero-modes but for {\em any} eigenvalue of the Laplacian, and it is presented in Appendix \ref{SZM}. In the main text, however, we present the derivation using the standard approach.

It is worth noting that the scalar zero-mode corresponding to the eigenvalue
$-\frac43 \kappa$ decouples, since its coefficient in the quadratic action vanishes\footnote{This should extend to all non-linear orders, although we do not investigate this question here.}.

\paragraph{Positive curvature spectra.}
For positive curvature, holographically acceptable background solutions always have one UV boundary (where the metric (\ref{intro3}) is asymptotically AdS$_5$) and a single IR endpoint where the scale factor shrinks to zero size and either $\Phi(y)$ asymptotes a finite value $\Phi_0$, or $\Phi(y) \to +\infty$ \cite{C,Jani}. The latter case can occur in holographic theories which exhibit confinement in flat space \cite{Jani}.

For solutions which extend to $\Phi(y) \to +\infty$, the Schr\"odinger potentials are not very different, qualitatively, from the ones obtained for flat slicing, which have discrete (and gapped) spectra.

More interesting are solutions (called {\em type III} in \cite{Jani}) which are characterized by a regular endpoint $u_0$ near which:
\be\label{intro10}
ds^2 \simeq du^2 + (u-u_0)^2 \zeta_{\mu\nu}dx^\mu dx^\nu \sp \Phi(u) \simeq \Phi_0 + O\left((u-u_0)^2\right)\,,
\ee
and which have no analog flat-sliced solutions. These are the only regular solutions when the flat-space holographic RG flow connects to conformal fixed points \cite{C}, and this is the case we focus on in this paper.  The coordinate value $u_0$ corresponds to a regular endpoint in the Euclidean signature, where the whole space shrinks to zero size. This becomes a horizon in Lorentzian signature, \cite{CdL}.

For these solutions, the Schr\"odinger potential takes the universal form of an infinite barrier in the UV (which is $y\to 0$ in conformal coordinates),
\be \label{intro11}
V_{g}(y) \sim  \frac{15 }{ 4\,y^2} \sp  V_{s}(y) \sim  \frac{K }{ y^2} \sp  y\to 0\,,
\ee
where $K$ is a constant depending on the dimension of the operator dual to $\Phi$, and which can be positive or negative\footnote{If $K<0$, the barrier is actually an infinite well, but it is still repulsive.}.

\begin{figure}[t!]
	\begin{center}
		\includegraphics[width=0.5 \textwidth]{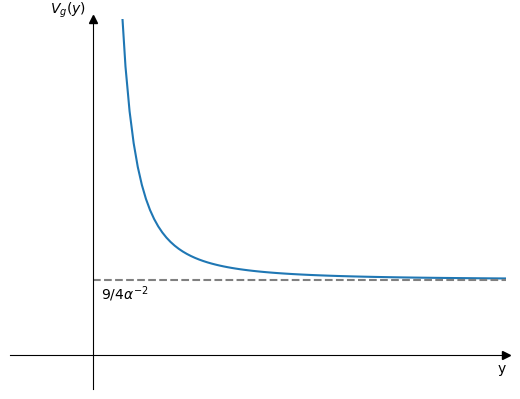}
	\end{center}
	\caption{\footnotesize{Sketch of the Schr\"odinger potential for spin-2 modes on de Sitter slicing with radius $\alpha$}.}\label{fig:intro1}
\end{figure}

\begin{figure}[ht!]
	\begin{center}
		\includegraphics[width=0.49 \textwidth]{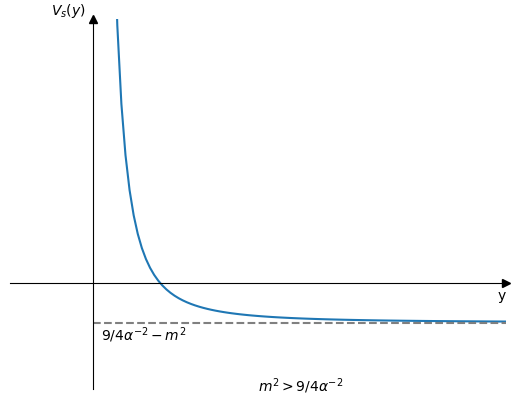}
		\includegraphics[width=0.49\textwidth]{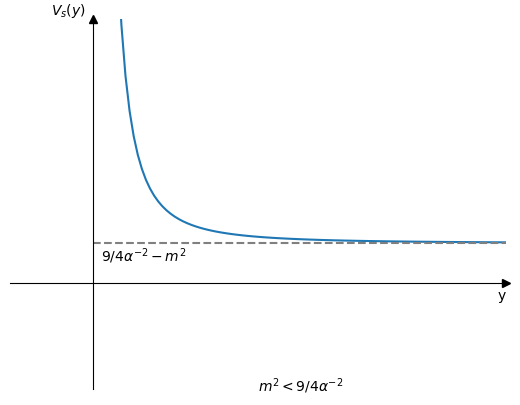}
	\end{center}
	\caption{\footnotesize{Sketch of the Schr\"odinger potential for spin-0 modes on de Sitter slicing with radius $\alpha$, for different masses $m^2$.  The Schr\"odinger equation for scalar modes is to be solved for zero-energy eigenvalue.}} \label{fig:intro2}
\end{figure}

In the IR, the potentials asymptote  to  {\em universal} constants at the  endpoint $u_0$ (which corresponds to $y\to \infty$ in conformal coordinates):
\be \label{intro12}
V_g(y) \to  \frac{9}{ 4\alpha^2}  \sp  V_s(y;m) \to \frac{9}{ 4\alpha^2} - m^2  \sp y\to +\infty\,,
\ee
where $\alpha$ is the curvature radius of the slice metric defined in (\ref{intro3-i}). The qualitative form of the potential for positive curvature slicing is sketched in Figure \ref{fig:intro1} for tensor modes and Figure \ref{fig:intro2} for scalar modes.

Equation (\ref{intro12}) implies that, for both scalars and tensors, the mass spectrum has a continuous component starting\footnote{Recall that for scalars the Schr\"odinger equation is to be solved with zero energy.} at $(9/4)\alpha^{-2}$, above which there exist plane-wave-normalizable solutions.   This value is strictly above the Higuchi bound $2 \alpha^{-2}$ \cite{Higuchi:1986py} for gravitons on de Sitter (below which massive spin-2 modes are unstable).
%  the de Sitter principal series for spin-0 representations.  In addition, there may be extra discrete states below this bound, but this cannot be decided from the asymptotics alone.

\paragraph{Negative curvature spectra.}
For negative curvature, one can again find acceptable solutions with one UV boundary and $\Phi(y) \to +\infty$ \cite{AdS3}. This occurs only for holographic confining theories.  On the other hand, for theories which admit flat holographic RG flows between two conformal fixed points, the situation at negative curvature is very different: now there can be no IR endpoints, but instead there are two UV boundaries as $u \to \pm\infty$, where the scalar field $\Phi$ reaches one of the allowed conformal fixed points $\Phi_{\pm}$ (extrema of the bulk potential):
\be
ds^2 \simeq du^2 + e^{\pm \frac{2u}{ \ell}} \zeta_{\mu\nu} dx^\mu dx^\nu \sp \Phi \to \Phi_{\pm} \sp u\to \pm \infty\,.
\ee
In particular, there are no type-III (regular IR endpoint) solutions for negative curvature.

\begin{figure}[t!]
	\begin{center}
		\includegraphics[width=0.5 \textwidth]{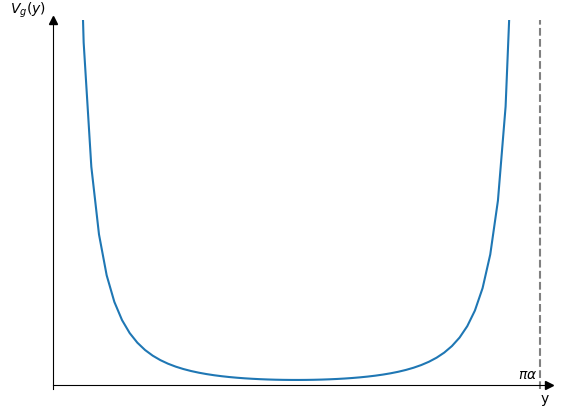}
	\end{center}
	\caption{\footnotesize{Sketch of the Schr\"odinger potential for spin-2 modes on Anti-de Sitter slicing with radius $\alpha$.}}\label{fig:intro3}
\end{figure}

In the conformal coordinate, the graviton and scalar Schr\"odinger potentials both take the form of a ``particle in a box'', with bounded conformal coordinate $y\in (0,y_0)$ and   two universal UV-type asymptotics near the extrema:
\be \label{intro13}
V_{s,g}(y) \sim \frac{1}{ y^2} \sp  y\to 0 \qquad;\qquad V_{s,g}(y) \sim \frac{1}{ (y-y_0)^2} \sp  y\to y_0\,,
\ee
where the coefficients of the leading asymptotics are 15/4 for spin-2 and depend on the dimensions of the operator dual to $\Phi$  for spin-0. The schematic form of the potential is illustrated in Figure \ref{fig:intro3} for spin-2, while for spin-0, there are some features in the middle which depend on the model, see section \ref{mss} for a concrete example. Clearly, in this case, the spectrum is purely discrete. One can check stability in concrete examples by searching for eigenvalues that violate the appropriate bounds ($M^2>0$ for spin-2 and the BF bound on AdS$_4$ for spin-0).

\paragraph{Explicit  results and full spectra.}
We use the general formalism developed so far to obtain the full spectra in two situations: \begin{itemize}
	\item {\bf Exact AdS}\\
	As a warm-up, we recover analytically the spectrum of gravitons on exact AdS$_5$ with constant dilaton.  In this case, there are no propagating scalar modes in the metric, and only the spin-2 survives. In addition there is the standard spin-0 dilaton fluctuation, which is decoupled from the metric  obeys the Klein-Gordon equation on AdS$_5$.
 The spin-2 Schr\"odinger potential $V_g$ has a simple analytic form from which one can obtain the spectra analytically: for dS$_4$ slices, this is a continuum starting at the Higuchi bound; for AdS$_4$ slices, we obtain analytically a tower of discrete eigenvalues which reproduces the known result of \cite{Karch:2000ct}.
	\item {\bf Holographic RG flow}\\
	We perform a full numerical analysis for the case of a polynomial bulk potential with two maxima (thus supporting two possible UV fixed points of the dual QFT). This potential was used in \cite{AdS1} to study holographic RG-flows on negative-curvature space-times. We obtain numerically the background solutions for positive and negative curvature, from which we derive the corresponding spin-0 and spin-2 Schr\"odinger potentials. We solve the Schr\"odinger equations numerically and obtain the spectrum, showing that there are no normalizable eigenstates which violate the stability bound of the curved slice manifold. In the case of positive curvature, we find no states in the discrete spectrum.
\end{itemize}
%\subsection{Paper Structure}

\subsection{Discussion and open problems}

The results in this work give the general equations which can be used to compute the mass spectra of spin-2 and spin-0 fluctuations around any given holographic Einstein-Dilaton background with constant curvature slices. Moreover, we established certain generic, model-independent features:
\begin{itemize}
	\item For positive curvature, the continuous spectrum starts at $m^2 = 9/4 \alpha^{-2}$ for both spin-0 and spin-2.
	\item For negative curvature, the spectrum is discrete.
\end{itemize}
The first point implies that for positive curvature all states in the continuous spectrum are in the spin-0 and spin-2 {\em principal series} representations of the de Sitter group\footnote{See Appendix \ref{reps} for a review of dS$_4$ unitary representations.}. These are the ``heavy'' fields, with $m^2 > (9/4) H^2$, where $H\equiv \alpha^{-1}$ is the dS$_4$ Hubble constant.

 These representations are stable; hence, our results prove in full generality that no instabilities can appear in the continuous part of the spectrum.

Beyond these statements, some interesting questions remain open about what the general features of the spectra are in various situations:
\begin{enumerate}
	\item What are the general features of the spectra for positive curvature, and in particular, can other representations of the de Sitter group appear? Can there be eigenvalues in the {\em complementary series}, i.e. scalars with to $0< m^2 < 9/4 \alpha^{-2}$ and massive gravitons with $2 \alpha^{-2} < m^2 <  9/4 \alpha^{-2}$ ? These would still be stable perturbations, but by our general result, can only occur as discrete eigenvalues.
	
	Finally, are there pathological models which feature discrete representations in the exceptional (discrete) series ( spin-0 with $m^2 = -j(j+3)$ with $j$ integer, and spin-2 with $M^2 = 0, 2$)?  The presence of such modes would lead to a perturbative instability, except for the case of a massless spin-2.

	We note that, for 2-derivative Einstein-dilaton theories on {\em flat} slices, one can prove that the spectrum is non-negative in general, as long as the scalar field satisfies the null-energy condition and under certain mild assumptions on the dilaton potential\footnote{The potential growth as $\Phi\to \infty$   must satisfy a {\em computability bound} \cite{iQCD, exotic}}. Therefore, these putative instabilities, if they arise, would be specific for the curved-space theory.
	\item The questions above are particularly interesting in the case of confining holographic theories in general \cite{Jani} and specifically Improved Holographic QCD \cite{JLR} in the case of positive curvature. These theories may exhibit phase transitions at finite curvature, so one may learn something interesting about the transition by looking at spectra in different phases.
	\item
	Similar questions may be asked for negative curvature, with the additional nuance that here one may have background solutions with two UV boundaries, like the one considered in this paper (holographic interfaces) but also with a single boundary, in the case where the dual theory is confining (on flat space) \cite{AdS3}. What kind of spectra can arise in both cases? May there be instabilities for negative curvature, when none are found for flat slicing?
	\item
	A specific question that arises for the two-boundary case, i.e., holographic interfaces, is how to compute spectra associated to {\em interface operators} in the holographic theories. What is the right holographic setup for the calculation, and what kind of spectra can arise?
	\item Finally, given the fluctuation equations we derived here, one can compute not only spectra but also holographic correlation functions on curved space-times. These require only changing the UV boundary condition in the fluctuation (from normalizable to non-normalizable) and contain a great deal of information beyond the spectra.
\end{enumerate}
%\vspace{0.5cm}
This work is organized as follows. In Section 2, we display our setup and give a short review of curved holographic RG flows. Section 3 presents the linearized Einstein's equations, the decomposition of the fluctuations into scalar, vector, and tensor modes; we identify the propagating degrees of freedom and, for these, we write decoupled second-order equations. We identify special values of the four-dimensional Laplacian eigenvalues (zero-modes) which have to be treated separately (which we do in Appendix D). In   Section 4, we separate variables and decompose the perturbation equations into a one-dimensional Schr\"odinger equation and four-dimensional mode equations. In Section 5, we investigate the normalizability conditions and give the conditions for mode stability. In Section 6, we show, as a first example, how to apply the formalism to recover the exact spectrum of the perturbation of exact five-dimensional Anti-de Sitter in the absence of a bulk dilaton. In Section  7 and 8, we fully solve the spectral problem numerically for a concrete example of a polynomial bulk potential admitting two UV fixed points for negative slice curvature and regular IR to  UV for positive slice curvature.
Several technical details (including a detailed treatment of the zero modes) can be found in the Appendix.

%%%%%%%%%%%%%%%%%%%%%%%%%%%%%%%%%%%%%%%%%%%%%%%%%%%%%
%%%%%%%%%%%%%%%%%%%%%%%%%%%%%%%%%%%%%%%%%%%%%%%%%%%%%
%%%%%%%%%%%%%%%%%%%%%%%%%%%%%%%%%%%%%%%%%%%%%%%%%%%%%
\section{Holographic RG flows on constant curvature manifolds} \label{sec2}

We begin with the 5--dimensional Einstein-dilaton theory with the action\footnote{In this paper we use the convention in \cite{Kiritsis:2006ua} for the scalar field kinetic term.}
\be \label{action}
S=\frac{1}{2\kappa_5^2}\int d^5x\, \sqrt{-G}\Big(R^{(G)}-\partial_A\Phi \partial^A\Phi-V(\Phi)\Big)\,,
\ee
where $G_{AB}$ is the metric and $\Phi$ is a scalar field.
This is a sector of the gravitational description of the holographic theory that contains the metric, dual to the energy-momentum tensor of the holographic QFT, and a scalar dual to a relevant scalar operator of the holographic QFT.

 The equations of motion stemming from this action are
\be  \label{EEin}
R^{(G)}_{AB}-\frac{1}{2}G_{AB} R^{(G)} - T^{(G)}_{AB} =0 \,,\ee
\be\label{EPhi}
-\frac{d}{d\Phi}V(\Phi)+2  G^{AB}\nabla_A^{(G)} \nabla^{(G)}_B \Phi=0 \,.
\ee
Here the energy-momentum tensor is defined as
\be\label{emtensor}
T^{(G)}_{AB}=-\frac{1}{\sqrt{-G}}\frac{\d S_{Matter}}{\d G^{AB}}\,.
\ee
%\subsection{Domain wall coordinates}
We are interested in boundary field theories defined on curved maximally symmetric 4--dimensional space-times, therefore
we consider the following ansatz for the background metric and the scalar field
\be \label{bac2}
ds^2 = G_{AB}dx^A dx^B = du^2 + e^{2A(u)} \zeta_{\m\n} dx^\m dx^\n\,,
\ee
\be \label{Dilaton}
\Phi(u, x)=\Phi_0(u) \,,
\ee
in which $u$ is the holographic dimension and $\zeta_{\mu\nu} $ is a metric describing a 4--dimensional maximally symmetric space-time on constant $u$ slices. The Greek indices raised and lowered with this metric.
As a consequence of the maximal symmetry of the 4--dimensional slices, we have:
\be\label{symspa}
R_{\mu\nu\rho\sigma}=\frac{ R^{(\zeta)}}{12}\left( \zeta_{\mu\rho}\zeta_{\nu\sigma }-\zeta_{\mu\sigma}\zeta_{\nu\rho} \right)\sp
R_{\mu\nu}= R^{(\zeta)}_{\mu\nu} = \kappa \zeta_{\mu\nu}\sp R^{(\zeta)}=4\kappa \,,
\ee
where $\k$ is a constant and relates to the dS/AdS length scale $\a$ as
\be \label{defa}
\k= \pm \frac{3}{\a^2} \,.
\ee
In the rest of this paper, we shall restrict ourselves to $d=4$, although similar results hold for other values.

In most of the paper we use a  conformal holographic coordinate $y$, related to $u$ by:
\be  \label{utoy}
du= e^{A(y)} dy\,,
\ee
so the background metric \eqref{bac2} becomes
\be \label{metr}
ds^2 =  e^{2A(y)} \left( dy^2 + \zeta_{\mu\nu} dx^\mu dx^\nu\right) \,.
\ee
The  equations of motion for the scale factor and the scalar field in these coordinates  are
\be\label{nsEom1}
6 {A'}^2(y) - \frac12 {\Phi'_0}^{\!2}(y) + \frac12 e^{2A(y)}V(\Phi_0) -\frac12 R^{(\zeta)}=0\,,
\ee
\be \label{nsEom2}
3A''(y)+ 3{A'}^2(y)+ \frac12 {\Phi'_0}^{\!2}(y)+ \frac12 e^{2A(y)}V(\Phi_0)-\frac14 R^{(\zeta)}=0\,,
\ee

\be \label{nsEom3}
2  \Phi''_0(y)+ 6A'(y) \Phi'_0(y)-e^{2A(y)}\pa_{\Phi}V(\Phi_0)=0\,,
\ee
where prime denotes derivative with respect to $y$.

\vspace{0.3cm}
%%%%%%%%%%%%%%%%%%%%%%%%%%%%%%%%%%%%%%%%%%%%%%%%%%%%%
%%%%%%%%%%%%%%%%%%%%%%%%%%%%%%%%%%%%%%%%%%%%%%%%%%%%%
%%%%%%%%%%%%%%%%%%%%%%%%%%%%%%%%%%%%%%%%%%%%%%%%%%%%%
\subsection{Near-boundary expansion}

We assume that $V(\Phi_0)$ has at least one maximum, where it can  be expanded as
\be \label{maxV}
V(\Phi_0)= -\frac{12}{\ell^2}-  m_{\Phi}^2 \left(\Phi_0 -\Phi_m \right)^2 + \mathcal{O}\left(\Phi_0^3\right)\,,
\ee
where $\ell$ is a length parameter which determines the size of AdS$_5$ solution at this fixed point  and $m_{\Phi}^2> 0$.
The expansions of $A(u)$ and $\Phi_0(u)$ near the maximum of the potential as ($u\rightarrow \pm\infty$) is, in domain-wall coordinates \cite{C}:
\be \label{yuq2}
A(u)=A_-\pm \frac{u}{\ell} -\frac{1}{48}\mathcal{R}\ell^2\tilde\varphi_-^{2/\Delta_-} e^{\mp 2u/\ell}+\mathcal{O}\left(e^{\mp 2\D_- u/\ell}\right)\,,
\ee
\be \label{yuq3}
\Phi_0(u)= \tilde\varphi_- \ell^{\Delta_-} e^{\mp\Delta_- u/\ell}+\tilde{\mathcal{C}}\ell^{\Delta_+}\tilde\varphi_-^{\Delta_+/\Delta_-}e^{\mp\Delta_+ u/\ell}+\mathcal{O}\left(e^{\mp(2+\D_-) u/\ell}\right)\,,
\ee
where $A_-, \mathcal{R}, \tilde{\varphi}_-$ and $\tilde{\mathcal{C}}$ are integration constants, satisfying the relation
% Note that in deriving the equation \eqref{yuq2}, we have used the following relation
\be \label{yuq4}
 R^{(\z)} e^{-2A_-}=\mathcal{R}\tilde\varphi_-^{2/\Delta_-}\,.
 \ee
% So in total, there are three constants of integration in \eqref{yuq2} and \eqref{yuq3}.
 In the equations above we have defined:
\be \label{Delpm}
\Delta_{\pm} = \frac12 \left(4 \pm \sqrt{16 +4 m_{\Phi}^2 \ell^2} \right) \qquad \text{with} \qquad  0< m_{\Phi}^2 \leq \frac{4}{\ell^2} \,,
\ee
and we assume:
\be \label{dmbounds}
0< \Delta_- \leq2 \, \sp  2\leq \Delta_+ <4 \,.
\ee
The  curvature  $R^{UV}$ of the boundary UV theory is  found by the Fefferman-Graham expansion of the metric
\begin{gather}\label{yuq5}
	ds^{2}=du^2+e^{\pm\frac{2\boldsymbol{u}}\ell}
	\left(ds_{QFT}^2+\cdots\right)
\\
	=du^{2}+e^{\pm\frac{2u}{\ell}}\Big[e^{2A_-}
	\zeta_{\alpha\beta}dx^{\alpha}dx^{\beta}\Big]+\mathrm{subleading}\,,
\end{gather}
which leads to
\be \label{yuq6}
R^{UV}= R^{(\z)} e^{-2A_-}\,.
\ee

We can obtain a similar expansion in conformal coordinates (\ref{metr}).   Using \eqref{utoy} and the expansion of $A(u)$ near the AdS boundary \eqref{yuq2}, one finds
\be \label{chyu}
y\simeq \ell e^{\frac{u+c}{\ell}}+\frac{1}{144}\ell^3 R^{(\z)} e^{\frac{3(u+c)}{\ell}}+\mathcal{O}\left(e^{\frac{2\D_-(u+c)}{\ell}}\right)\,.
\ee
The AdS boundary is reached as $u\rightarrow - \infty$, corresponding to $y \rightarrow 0^+$.  The expansion of the scale factor and the scalar field is obtained by solving equations (\ref{nsEom1}--\ref{nsEom3}) asymptotically near $y=0$:
\begin{gather}
	A(y)= -\log\left(\frac{y}{\ell}\right) - \frac{\Delta_-}{6(1+ 2\Delta_-)}\, \varphi_-^2 y^{2 \Delta_- }-\frac{1}{72}  R^{(\z)} y^2
\nn\\
	+ \frac{1}{432}\frac{\Delta_- ^2 \left(2 \Delta_- ^2-11 \Delta_- -18\right)}{(\Delta_--1)(2\Delta_-+1)(2\Delta_-+3)} R^{(\z)}\varphi_-^{2} y^{2 \Delta_- +2}
 \nn\\
	 -\frac{1}{30} \mathcal{C}\Delta_+ \Delta_- \varphi_-^{4/\Delta_-} y^4 + \mathcal{O}\left(y^{2 \Delta_+}, y^6\right)  \,,\label{Ayexp}
\end{gather}
\begin{gather}
	\Phi_0(y)= \varphi_-\, y^{\Delta_-}+\mathcal{C} \varphi_-^{\Delta_+/\Delta_-}\,y^{\Delta_+}
	-\frac{1}{144} \frac{(\Delta_- -7) \Delta_-}{(\Delta_- -1) }  R^{(\z)}\varphi_- y^{2+\Delta_-}
\nn\\
	+ \frac{\Delta_-(\Delta_- + 2) }{12(\Delta_- -1)(2\Delta_- +1)}\,\varphi_-^{3} y^{3\Delta_-} +\mathcal{O}\left(y^{2+ \Delta_+}, y^{4+\Delta_-}\right) \,, \label{phiyexp}
\end{gather}
where $\f_-$ and $\mathcal{C}$ are constants of integration (for simplicity we have considered $1<\Delta_-<2$).  These expansions match those in  \eqref{yuq2} and  \eqref{yuq3} if we relate the respective integration constants by:
\be\label{yuq1}
c=-A_- \ell \sp \varphi_-=\tilde\varphi_- e^{\Delta_- A_-}\sp   \mathcal{C}=\tilde{\mathcal{C}}\,.
\ee

\vspace{0.3cm}
%%%%%%%%%%%%%%%%%%%%%%%%%%%%%%%%%%%%%%%%%%%%%%%%%%%%%
%%%%%%%%%%%%%%%%%%%%%%%%%%%%%%%%%%%%%%%%%%%%%%%%%%%%%
%%%%%%%%%%%%%%%%%%%%%%%%%%%%%%%%%%%%%%%%%%%%%%%%%%%%%
\subsection{IR  expansion}

Away from the UV fixed point, the holographic RG flows solutions can have different form, depending on the behavior of the potential and the sign of the slice curvature $\kappa$.

For $\kappa >0$, we can either have a regular endpoint at finite $\varphi_0$ , \cite{C}, or (if the potential has exponential large-$\varphi$ asymptotics) solutions that run to $\varphi \to \pm \infty$ in a ``controlled'' way \cite{Jani,JLR}.

For $\kappa<0$ we may still have asymptotic solutions with    $\varphi \to \pm \infty$ \cite{AdS3} but no regular end-points. In contrast, the flow may have a turning point (corresponding to a field value $\varphi_0$) and reach a second asymptotic boundary \cite{C,AdS1,AdS3}.

In this work we will mostly discuss the perturbations around solutions with either regular IR end-points ($\kappa >0$)  or turning points ($\kappa<0$).

For $\kappa>0$, an IR end-point   is a point where at a finite value of the $u$ coordinate, say $u=u_0$, the scalar field $\Phi_0(u)$ reaches a finite value, and the scalar factor $e^{A(u)}$ shrinks to zero size  \cite{C}. This is a regular end-of-space in the Euclidean theory and a coordinate horizon in the Lorentzian signature.

Suppose the IR end-point is located at $\Phi_0(u_0)=\f_0$ and the potential at this point is expanded as ($\Phi_0(u)\rightarrow \f_0^-$)
\be \label{VIR}
V=V_0+V_1\left(\f_0-\Phi_0(u)\right)+V_2 \left(\f_0-\Phi_0(u)\right)^2+\cdots\,,
\ee
where $V_0, V_1$ and $V_2$ are constants. The expansions of the scalar field and scale factor near the IR end-point, are given by ($u\rightarrow u_0^+$)
\be \label{fuIR}
\Phi_0(u)=\f_0-\frac{V_1}{20}\left(u-u_0\right)^2+\mathcal{O}\left(u-u_0\right)^4\,,
\ee
\be \label{AuIR}
A(u)=\log\left(\frac{u-u_0}{\a}\right)-\frac{V_0}{72} \left(u-u_0\right)^2+\mathcal{O}\left(u-u_0\right)^4\,,
\ee
where we have used
\be \label{Ralfa}
R^{(\z)}= \frac{12}{\a^2}\geq 0\,.
\ee
Using the relation \eqref{utoy} and the expansion of $A(u)$ in \eqref{AuIR} we find
\be \label{yurel1}
y=\a \left(\log\frac{u-u_0}{\bar{c}}+\frac{V_0}{144}\left(u-u_0\right)^2+\mathcal{O}\left((u-u_0)^3\right)\right)\,,
\ee
where $\bar{c}$ is a constant of integration with dimension of length. Inversely
\be\label{yurel2}
u-u_0= \bar{c} e^{\frac{y}{\a}}+\mathcal{O}\big( e^{3\frac{y}{\a}}\big)\,.
\ee
From the above relations between the two coordinates, we observe that in the $y$-coordinate, the IR end-point is reached as $y\rightarrow-\infty$.
If we solve the equations of motion \eqref{nsEom1}-\eqref{nsEom3} near the end of space, we  obtain the following expansions
\be \label{fidirect}
\Phi_0(y)=\f_0-\frac{V_1}{20 a_1^2}e^{2(a_0+ a_1 y)}+\mathcal{O}\left(e^{4a_1 y}\right)\,,
\ee
\be \label{Adirect}
A(y)= a_0 +a_1 y  -\frac{V_0}{48 a_1^2}\,e^{2(a_0+ a_1 y)} +\mathcal{O}\left(e^{4a_1 y}\right)\,,
\ee
where $a_0$ and $a_1$ are two new constants of integration.
Knowing the dependence of $u$ in terms of $y$ near the IR end-point, given in \eqref{yurel2},  on one hand and the expansions of $A(u)$ and $\Phi_0 (u)$ in \eqref{fuIR} and \eqref{AuIR}, we can read the IR end-point expansions in the $y$ coordinate
\be \label{fyIR}
\Phi_0(y)=\f_0-\frac{V_1 \bar{c}^2}{20}e^{\frac{2y}{\a}}+\mathcal{O}\left(e^{\frac{4y}{\a}}\right)\,,
\ee
\be \label{AyIR}
A(y)=\frac12 \log \frac{\bar{c}^2}{\a^2}+\frac{1}{\a} y - \frac{\bar{c}^2}{48}V_0 e^{\frac{2y}{\a}} +\mathcal{O}\left(e^{\frac{4y}{\a}}\right)\,,
\ee
which coincides with the equations \eqref{fidirect} and \eqref{Adirect} if we set
\be \label{cbar}
a_1 = \frac{1}{\a} \sp \bar{c}=\a e^{a_0}\,.
\ee
From \eqref{AyIR} we observe that near the IR end-point, that is when $y\rightarrow -\infty$, the geometry is given by
\be \label{IRmetr}
ds^2\approx \frac{\bar{c}^2}{\a^2}\,e^{\frac{2y}{\a}}\left(dy^2+ds^2_{dS_4}\right)\,.
\ee

\vspace{0.5cm}
%%%%%%%%%%%%%%%%%%%%%%%%%%%%%%%%%%%%%%%%%%%%%%%%%%%%%
%%%%%%%%%%%%%%%%%%%%%%%%%%%%%%%%%%%%%%%%%%%%%%%%%%%%%
%%%%%%%%%%%%%%%%%%%%%%%%%%%%%%%%%%%%%%%%%%%%%%%%%%%%%
\section{The fluctuation equations\label{flu}}

We now go to the main topic of the paper and  consider  perturbations around to the  solution \eqref{bac2} and \eqref{Dilaton}.

First, in this section, we derive the equations for the linear fluctuations around the background solution % \eqref{metr} and \eqref{scalar}

The perturbed metric $g_{AB}$,  in terms of the background metric $g_{AB}^{(0)}$ and its fluctuations $h_{AB}$ is:
\be\label{bac+fl}
ds^2   = a^2 (y) \left[g^{(0)}_{AB} + h_{AB}(y, x^\m) \right]dx^A dx^B\,,
\ee
where $g^{(0)}_{AB}$ is given in \eqref{metr} and $a(y)$  related to $A(y)$ in \eqref{metr} through
\be \label{defA}
a(y) = e^{A(y)}\,.
\ee
By decomposing the coordinates as $x^A=(y, x^\m)$, a generic perturbation of the metric around the background can be parameterized as
\be \label{change1}
h_{yy} = 2\phi \sp  h_{\m y}=
 A_\mu \sp  h_{\m\n}\,.
\ee
We parametrize the scalar  field fluctuation as:
\be \label{fixi}
\Phi=\Phi_0(y) + \chi (y, x^\m)\,.
\ee
There are two ways to obtain linearized field equations. We can take the equations of motion \eqref{EEin} and \eqref{EPhi}, and then linearize these equations up to the first order in fluctuations. Alternatively, it is possible to expand the Lagrangian up to the second order of the field fluctuations and then obtain the field equations by variation of this Lagrangian. These methods are equivalent if one is interested in the equations of motion, but having the full quadratic action offers some advantages if one is interested in boundary term, normalizability, evaluating solutions on-shell, etc, which is frequently the case in holography. For this reason, below we present both approaches (leaving the details in appropriate sections in the appendix).

We start by the first method. The  components of the linearized Einstein equations are (see appendix \ref{linz} for the details of calculations)
\begin{gather}
	\hspace{-1.8cm}(\mu \nu) \quad\quad\quad h_{\mu\nu}'' + 3\frac{a'}{ a} h_{\mu\nu}' + \nabla_\r \nabla^\r h_{\mu\nu} \!-\!2 \nabla^{\rho} \nabla_{(\mu}h_{\nu)\rho} +\nabla_\mu\nabla_\nu h + 2\nabla_\nu\nabla_\mu \phi
\nn\\
	-2 a^{-3}\left(a^3 \nabla_{(\mu} A_{\nu)}\right)'\, + \zeta_{\mu\nu}\bigg[-h'' - 3\frac{a'}{a}h' - \nabla_\r \nabla^\r h +
	\nabla^\rho \nabla^\sigma h_{\rho\sigma}
\nn\\
	 -2 \nabla_\r\nabla^\r \phi + 6\frac{a'}{a}\phi' + 6\bigg(\frac{a''}{a}+2 \Big(\frac{a'}{a}\Big)^2\bigg)\phi  -2 a^{-3}\left(a^3 \Phi_0' \chi\right)'
 \nn\\
	+ 2 a^{-3}\left(a^3 \nabla_{\rho} A^{\rho}\right)'\bigg]- \kappa \zeta_{\mu\nu}( 2\phi + h \big) + 2 \kappa h_{\mu\nu} =0\,,\label{munu1}
\end{gather}

\be\label{muy1}
(\mu y) \quad\quad  \nabla_\n \nabla^\n A_\mu - \nabla^{\nu} h_{\nu\mu}'  +\nabla_{\mu}\left[ - 6\frac{a'}{a}\phi + 2\Phi_0' \chi + h' - \nabla^{\nu} A_\nu\right] + \kappa A_{\mu} =0 \,,
\ee

\begin{gather}
	\hspace{-1.5cm}(y y) \quad\quad   -\nabla_\m \nabla^\m h + \nabla^{\mu}\nabla^\nu h_{\mu\nu} - 3\frac{a'}{a}h' + 6\bigg(\frac{a''}{a}+2 \Big(\frac{a'}{a}\Big)^2\bigg)\phi + 6 \frac{a'}{a} \nabla^\nu A_\nu
\nn\\
	+ 4 \Phi_0' \chi'- 2 a^{-3}\left(a^3 \Phi_0' \chi\right)'  +\kappa\left(2 \phi -h\right)= 0 \,.\label{yy1}
\end{gather}
The linearized field equation for the scalar field $ \Phi $ originates from the equation \eqref{EPhi}:
\be \label{Dil1}
\chi''  + \nabla_\m \nabla^\m \chi+ 3 \frac{a'}{a}\chi' - \frac{1}{2}a^2 \partial^2_{\Phi}V \chi
  -\frac{2}{a^3} \left(a^3 \Phi_0' \phi\right)' +\Phi_0'\phi' + \frac{1}{2}\Phi_0'h'   -
 \Phi_0' \nabla^\mu A_\mu=0\,.
\ee

We now present the quadratic action for the perturbations.  Under suitable assumptions  for boundary terms (see details in Appendix \ref{ApB}), we find that it is given by:
\begin{gather}
	S^{(2)} =  \frac{1}{2k_5^2}\int d^4x dy\,\sqrt{-g^{(0)}}\Bigg\{ a^3 \bigg[ L_{EH}^{(2)}  - \frac{1}{4} h'_{\m\n}h'^{\m\n} + \frac{1}{4} (h')^2
\nn\\
	-\frac{1}{4}F_{\mu\nu}F^{\mu\nu}
	-\partial_{\mu}\chi \partial^{\mu}\chi -\chi'^2- \frac{1}{2}a^2 \partial^2_{\Phi}V \, \chi^2+ 2 \Phi_0' \phi' \chi
\nn\\
	+ \Phi_0' h' \chi
	+ 4 \Phi_0'\phi \chi'+ 2 \Phi_0' A^{\mu}\partial_{\mu}\chi - \partial^{\mu}\phi\left(\nabla^{\nu}h_{\mu\nu} - \partial_\mu h\right) \bigg]
\nn\\
	-\left(a^3 A^{\mu}\right)' \left(\nabla_{\mu}h - \nabla^{\nu}h_{\mu\nu}\right) - \left(a^3\right)' \left(2 A_{\mu}\partial^\mu\phi + 2 \phi\phi' + \phi h'\right)
\nn\\
	+ a^3 \kappa \big( \phi^2 + A_{\mu} A^{\mu} - \phi h -\frac{1}{6} h_{\mu \nu} h^{\mu \nu}-\frac{1}{12} h^2 \big)\Bigg\}, \label{squad2}
\end{gather}
where $F_{\mu\nu} = \nabla_{\mu} A_\nu - \nabla_\nu A_\mu$,  and
\be \label{Lagein}
L_{EH}^{(2)} = -\frac{1}{4}\nabla^\mu h_{\rho\sigma}\nabla_\mu h^{\rho\sigma} + \frac{1}{2} \nabla^\mu h_{\rho\mu}\nabla_\nu h^{\rho\nu}- \frac{1}{2} \nabla_\mu h \nabla_\rho h^{\rho\mu} +
\frac{1}{4} \nabla^\mu h \nabla_\mu h\,,
\ee
is the quadratic part of the four dimensional Einstein-Hilbert Lagrangian. One can check that the equations of motion \eqref{munu1}--\eqref{Dil1}  can be obtained by variation of \eqref{squad2} with respect to $h_{\m\n}, A_\mu, \phi$ and $\chi$.

\vspace{0.3cm}
%%%%%%%%%%%%%%%%%%%%%%%%%%%%%%%%%%%%%%%%%%%%%%%%%%%%%
%%%%%%%%%%%%%%%%%%%%%%%%%%%%%%%%%%%%%%%%%%%%%%%%%%%%%
%%%%%%%%%%%%%%%%%%%%%%%%%%%%%%%%%%%%%%%%%%%%%%%%%%%%%
\subsection{Decomposition of fluctuations}\label{Dec}

It is convenient to  introduce a decomposition of the vector and tensor fields in transverse and longitudinal components, which will end up satisfying decoupled equations:
\be
 A_{\mu} = \nabla_\mu W + A^{T}_\mu \,, \label{split1}
\ee
\be
h_{\mu\nu} = 2\zeta_{\mu\nu}\psi + 2 \nabla_{(\mu} \nabla_{\nu)} E + 2 \nabla_{(\mu} V^{T}_{\nu)} + h^{TT}_{\mu\nu},\label{split2}
\ee
with
\be \label{dAT}
\nabla^\mu A^{T}_\mu=0\,,
\ee
\be \label{dVT}
\nabla^\mu V^{T}_\mu=0\,,
\ee
\be \label{dhTT}
\nabla^\mu h^{TT}_{\mu\nu} = h^{TT\mu}_{\mu}=0\,,
\ee
where $W, E$ and $\psi$ are new scalar fluctuations together with the old scalars $\phi$ and $\chi$.

As we will show below, the decomposition \eqref{split1} and \eqref{split2} are well defined, except in the presence of certain special eigenvalues of the slice Laplacian on scalars and vectors. These special values   must  be carefully treated separately, which we do in Appendix \ref{SZM}.

\begin{enumerate}
\item Decomposition of $A_\m$:

Taking the divergence of \eqref{split1} and use \eqref{dAT} we obtain the relation:
\be\label{RW}
\nabla_\m \nabla^\m W =\nabla^\m A_\m\,.
\ee
The solution of this equation exists, and is unique, up to the presence of zero eigenmodes of the scalar Laplacian operator\footnote{If the right hand side of the equation contains zero eigenmodes, then a solution for $W$ does not exist.}. Barring such zero eigenmodes, then $W$ can be found by inverting the slice Laplacian and  one can read $ A^T_\m$ in terms of $A_\m$ from \eqref{split1}.

\item Decomposition of the tensor $h_{\m\n}$:

Similarly, there are obstructions to decomposing the tensor as in  \eqref{split2}.  Taking the   trace of that equation we find:
\be \label{Trh}
{h^\m}_\m  \equiv h=8\psi + 2 \nabla_\m \nabla^\m E\,.
\ee
From the above equation, it can be seen that just like the $W$ field, to obtain $E$ we should consider the case in which the field $E$ is not the zero mode of the Laplacian operator.

By taking a divergence from \eqref{split2} we also find that
\be \label{Divh}
\nabla^\m h_{\mu\nu} = 2\nabla_\n \psi + 2 \nabla_\m \nabla^\m \nabla_\n E +\left( \nabla_\m \nabla^\m +\kappa\right) V^{T}_\nu\,,
\ee
where we have used the following identity
\be \label{Divh1}
2 \nabla^\m \nabla_{(\n} V^T_{\m)}= \left(\nabla_\m \nabla^\m + \k\right)V^T_\n\,.
\ee
Substituting $\psi$ from \eqref{Trh} into \eqref{Divh} we have
\be\label{ZV}
\nabla^\m h_{\mu\nu}-\frac14\nabla_\n h = \frac32  \nabla_\n \!\big(\nabla_\m \nabla^\m +{\frac43}\k \big)\!E+\left( \nabla_\m \nabla^\m +\kappa\right) V^{T}_\nu\,.
\ee

Again in the equation \eqref{Divh}, we should consider the zero and non-zero modes of $V^T_\m$ when the operator $(\nabla_\m \nabla^\m +\k)$ acting on it. By taking another divergence we find that
\be \label{2Divh}
\nabla^\n \nabla^\m h_{\mu\nu} = 2\nabla_\m \nabla^\m \psi + 2 (\nabla_\m \nabla^\m)^2 E + 2\kappa \nabla_\m \nabla^\m E \,.
\ee
By substituting $\nabla_\m \nabla^\m E$ from \eqref{Trh} into the \eqref{2Divh} we find
\be \label{asub}
\big(\nabla_\m \nabla^\m + \frac43 \k\big)\psi = \frac16 \left(\left(\nabla_\m \nabla^\m+\k\right)h -\nabla^\m\nabla^\n h_{\m\n}\right)\,.
\ee
Solving for $\psi$ provided that there is no zero mode for $\nabla_\m \nabla^\m + \frac43 \k$ operator, we can find $E$ from \eqref{Trh} if $E$ does not have a zero mode of the Laplace operator.

On the other hand by substituting $ \psi$ from \eqref{Trh} into \eqref{2Divh} we find $E$ in terms of $h_{\m\n}$
\be\label{Eeq}
\big(\nabla_\m \nabla^\m + \frac43 \kappa\big)\nabla_\n \nabla^\n E= \frac23 \nabla^\n \nabla^\m h_{\mu\nu} -\frac16 \nabla_\m \nabla^\m h\,.
\ee

Knowing $E$, we can obtain $\psi$ from \eqref{Trh}. Knowing  $E$ and $\psi$,  the field $V^T_\m$ is determined from \eqref{Divh}. Finally after determining all these fields, $h^{TT}_{\m\n}$ can be obtained from \eqref{split2}.
\end{enumerate}

According to the discussion above,  to determine all the fields in the decomposition of $A_\m$ and $h_{\m\n}$, we should consider the cases in which $W$ does not contain zero modes of $\nabla_\m \nabla^\m$, the field $E$ does not contain the zero modes of the $\nabla_\m \nabla^\m$ or $(\nabla_\m \nabla^\m+\frac43\kappa)$ and $V^T_\m$ does not contain the zero modes of the operator $(\nabla_\m \nabla^\m+\k)$.

In the rest of this section we assume the decomposition \eqref{split1} and \eqref{split2}  are well defined. At the linear level, different eigenvalues of the slice Laplacian do not mix, so we can treat separately the  cases of the special scalar Laplacian eigenvalues $\nabla^2 = 0, -4\kappa/3$ and vector Laplacian eigenvalue  $\nabla^2 =-\kappa$.  This will be done in appendix \ref{SZM}.

The equations of motion \eqref{munu1}--\eqref{Dil1} after substituting the decomposition \eqref{split1} and \eqref{split2} become:
\begin{itemize}
\item $\m\n$ component (equation \eqref{munu1}):
\begin{gather}
	\!\!\!\!	- \!\left[\big(\nabla_\r \nabla^\r+\k\big)\big(4\psi+2\phi\big)+\frac{6}{a^3}\bigg( \Big[ \psi'\!-\!\frac{a'}{a}\phi-\frac13 \nabla_\r \nabla^\r(W\!-E')+\frac13 \Phi'_0 \chi \Big]a^3 \bigg)' \right] \zeta_{\m\n}
\nn\\
	\!\!\!\!	+\nabla_\m \nabla_\n \left(4\psi +2 \phi - \frac{2}{a^3}\left(a^3\left(W-E'\right)\right)'\right) -\frac{2}{a^3}\left(a^3 \left[\nabla_{(\m}{A^T_{\n)}}- \nabla_{(\m}{V^T_{\n)}}'\right]\right)'
\nn\\
	\!\!\!\!	+{h^{TT}_{\m\n}}''+ 3\frac{a'}{a}{h^{TT}_{\m\n}}'+ \left(\nabla_\r \nabla^\r - \frac23 \k\right) h^{TT}_{\m\n}=0\,.
	\label{munu}
\end{gather}
\item $\mu y$ component (equation \eqref{muy1}):
\be
\left(\nabla_\n \nabla^\n+ \k\right)\left( A^T_\m - {V^T_\m}'\right)+ \nabla_\m \left(6 \psi'- 6\frac{a'}{a}\phi +2 \k \left(W-E'\right)+ 2\Phi'_0 \chi \right)=0\,. \label{muy}
\ee
\item{$yy$ component} (equation\eqref{yy1}):
\begin{gather}
	0=\left(\nabla_\m \nabla^\m +\frac{4}{3} \kappa\right)\psi +4 \frac{a'}{a}\psi'- \frac{a'}{a} \nabla_\m \nabla^\m\left(W-E'\right)- \left(\frac{a''}{a}+2 \left(\frac{a'}{a}\right)^2\right)\phi
\nn\\
	- \frac{1}{3}\kappa \phi + \frac13 a^{-3}\left(a^3 \Phi_0' \chi\right)' - \frac{2}{3} \Phi_0' \chi'\,. \label{yy}
\end{gather}
\item{Dilaton } (equation \eqref{Dil1}):
\begin{gather}
	0=\chi'' + 3 \frac{a'}{a}\chi' + \nabla_\m \nabla^\m \chi - \frac{1}{2}a^2 \pa^2_{\Phi}V
	\chi -2 a^{-3} \left(a^3 \Phi_0' \phi\right)'
\nn\\
	+\Phi_0'\phi' + 4\Phi_0'\psi'   - \Phi_0' \nabla_\m \nabla^\m \left(W-E'\right)\,. \label{dilaton}
\end{gather}
\end{itemize}

Substituting the decompositions \eqref{split1} and \eqref{split2} in the action \eqref{squad2} and after some integrating by part it will become
\be
S^{(2)}= S^{(2)TT}+ S^{(2)T}+S^{(2)scalar}\,,
\ee
where
\begin{gather}\label{acTT}
	S^{(2)TT} =  \frac{1}{ 2k_5^2}\int d^4x dy\,\sqrt{-g^{(0)}}\Bigg\{ a^3(y)  \bigg[-\frac{1}{4}\nabla^\mu h^{(TT)}_{\rho\sigma}\nabla_\mu h^{(TT)\rho\sigma}
	\\ \nn
	- \frac14 {h^{(TT)}_{\rho\sigma}}' {h^{(TT)\rho\sigma}}'- \frac{1}{6}\kappa h_{\mu \nu}^{(TT)} h^{(TT)\mu \nu}\bigg]\Bigg\}\,,
\end{gather}
\begin{gather}
	S^{(2)T} =  \frac{1}{ 2k_5^2}\int d^4x dy\,\sqrt{-g^{(0)}}\Bigg\{ a^3(y)  \bigg[ -\frac{1}{2} \nabla_\mu ({V_{\nu}^T})'\nabla^\mu ({V^{T\nu}})'+\frac{1}{2} \kappa ({V_{\mu}^T})'(V^{T\mu })'
\nn\\
	- \frac{1}{4}\left(\nabla_\mu A^{T}_\nu -\nabla_\nu A^{T}_\mu\right)\left(\nabla^\mu A^{T\nu} -\nabla^\nu A^{T\mu}\right)+\kappa A^{T}_\mu A^{T\mu}\bigg]
\nn \\
	+\left(a^3(y)A^{T\mu}\right)'\left( \nabla_\n \nabla^\n +\kappa\right){V_\m^T}\Bigg\}\,,\label{acT}
\end{gather}
\begin{gather}
	S^{(2)scalar} =  \frac{1}{ 2k_5^2}\int d^4x dy\,\sqrt{-g^{(0)}}\Bigg\{ a^3(y)  \bigg[6 \pa_\mu \psi \pa^\mu \psi +12 \psi'^2 +6\pa_\mu \psi \pa^\mu \phi
\nn \\
	-6\psi' \nabla_\m \nabla^\m \left(W-E'\right)-\pa_\mu \chi \pa^\mu \chi - \chi'^2 -\frac{1}{2}a^2(y)\chi^2 \pa^2_{\Phi} V(\Phi) + 4\Phi'_0 \phi \chi'
\nn \\
	+2\Phi'_0 \chi \left( \phi' + 4  \psi'  -   \nabla_\m \nabla^\m \left(W-E'\right)\right)+\kappa\nabla_\mu \left(W-E'\right)\nabla^\mu \left(W-E'\right)
\nn \\
	-8\kappa \psi^2+\kappa \phi^2 -8\kappa \phi \psi\bigg] -6a^2(y)a'(y)\phi \Big( \phi' + 4  \psi'  -   \nabla_\m \nabla^\m \left(W-E'\right)\Big)\Bigg\}\,.\nn \\ \label{acSca}
\end{gather}

\vspace{0.3cm}
%%%%%%%%%%%%%%%%%%%%%%%%%%%%%%%%%%%%%%%%%%%%%%%%%%%%%
%%%%%%%%%%%%%%%%%%%%%%%%%%%%%%%%%%%%%%%%%%%%%%%%%%%%%
%%%%%%%%%%%%%%%%%%%%%%%%%%%%%%%%%%%%%%%%%%%%%%%%%%%%%
\subsection{The action of diffeomorphisms}

Under a 5--dimensional diffeomorphisms, $(\delta y \equiv \xi^5, \delta x^\mu \equiv \xi^\mu)$, the fluctuations defined by \eqref{change1} and \eqref{fixi} transform as
\be
\delta h_{\mu\nu} = -\nabla_\nu \xi_\mu - \nabla_\mu \xi_\nu - 2 \frac{a'}{a} \xi^5 \zeta _{\mu\nu}\,,\label{dif h}
\ee
\be
\label{dif A}
\delta A_\mu = -\partial_\mu \xi^5 - \xi'_\m \,,
\ee
\be \label{difphi}
\delta \phi = -{\xi^5}' - \frac{a'}{a}\xi^5\,,\ee
\be
\label{difchi}
\delta \chi =-\Phi'_0\, \xi^5 \,.
\ee
Because of diffeomorphisms (gauge) invariance, not all of these perturbations
 are dynamical. Counting the degrees of freedom in a
gravitational theory implies that the metric and
dilaton fluctuation, ($h_{AB}$, $\chi$), contain 16 components, out
 of which 5 are eliminated by gauge transformation
and another 5 can be eliminated through the non-dynamical components
 of Einstein's equations. We are left with a total of
 $6$ degrees of freedom, which correspond to a five dimensional massless spin-2 field plus a scalar.

We can find the transformations of the fields defined in \eqref{split1} and \eqref{split2} by comparing them with the equations \eqref{dif h} and \eqref{dif A}. To do this, we should decompose the gauge transformations into the transverse and longitudinal directions as well, i.e.
 \be \label{difdic}
 (\delta x^\mu, \delta y) \equiv (\xi^{\mu}, \xi^5) = (\xi^{T \mu} + \nabla^\mu \xi, \xi^5)\sp \nabla_\mu\xi^{\mu}=0\,.
 \ee
According to  the diffeomorphisms in \eqref{dif A} and the decomposition \eqref{difdic}  we can write
\be\label{dela1}
\delta A_\mu =-\xi^{T'}_\m - \pa_\m \xi' -\pa_\m \xi^5 \,.
\ee
On the other hand using the relation \eqref{split1} the above equation should be equal to
\be\label{dela2}
\delta A_\mu =\pa_\m \delta W + \delta A_\m ^T  \,.
\ee
Comparing \eqref{dela1} and \eqref{dela2}, we find that a gauge transformation in the transverse and longitudinal parts could be
\be \label{Adicsdif}
\delta A^T_\mu=- \xi^{T'}_\m \sp \delta W =- \xi'-  \xi^5 \,.
\ee
For the $h_{\m\n}$ field, we obtain from \eqref{difdic} and \eqref{dif h}
 \be
 \delta h_{\m\n}
 =-2\nabla_{(\m}\, \xi_{\n)}^T -2 \nabla_\m \nabla_\n\xi -2\frac{a'}{a} \xi^5 \zeta_{\m\n}\,.\label{delh1}
 \ee
On the other hand from the decomposition of $h_{\m\n}$ in \eqref{split2} we obtain
 \be \label{delh2}
  \delta h_{\m\n}= 2\zeta_{\m\n} \delta \psi + 2\nabla_\m \nabla_\n \delta E+ 2 \nabla_{(\m} \delta V^T_{\n)}+ \delta h^{TT}_{\mu\nu}\,.
 \ee
By comparing \eqref{delh1} and \eqref{delh2} one can choose the following gauge transformation for the fields

\be \label{delth1}
\delta \psi = - \frac{a'}{a} \xi^5 \sp
\delta E = -\xi \sp
\delta V^T_\m =-\xi_\m^T \sp
\delta h^{TT}_{\m\n}= 0 \,.
\ee

%\vspace{0.3cm}
%%%%%%%%%%%%%%%%%%%%%%%%%%%%%%%%%%%%%%%%%%%%%%%%%%%%%
%%%%%%%%%%%%%%%%%%%%%%%%%%%%%%%%%%%%%%%%%%%%%%%%%%%%%
%%%%%%%%%%%%%%%%%%%%%%%%%%%%%%%%%%%%%%%%%%%%%%%%%%%%%
\subsection{The gauge-invariant  fluctuation equations} \label{simp eq}

In this section, we shall obtain the equations of motion after removing the zero modes of the fields, discussed in the previous section. To this end, we introduce some new field variables that make the equations of motion more transparent and easy to work with.
The new variables are defined as
\begin{gather}\label{chv1}
	B^T_\m\equiv A^T _\m  - {V^T_\m}'\,,
\\
	\lambda\equiv \psi-\frac{\chi}{z}\,, \label{chv2}
\\
	\gamma \equiv W-E' -\frac{a \chi}{a'z }\,, \label{chv3}
\\
	\tau \equiv 2 \psi+\phi-\frac{1}{a^3}\left(a^3\left(W-E'\right)\right)'\,,\label{chv4}
\\
	\sigma \equiv 2\frac{a'}{a}\lambda+\lambda' + \frac{\kappa}{3}\gamma-\frac{a'}{a}\left(\frac{1}{a^3}\left(a^3 \gamma\right)' \right)\,,\label{chv5}
\end{gather}
where we define $z$ in the above equations as
\be \label{defz}
z\equiv \frac{a \Phi'_0}{a'}\,.
\ee

Under the gauge transformation of the original fields in equations \eqref{difphi}, \eqref{difchi}, \eqref{Adicsdif} and \eqref{delth1} it is simple to show that all the new fields defined in \eqref{chv1}--\eqref{chv5} are gauge invariant
\be
\delta \lambda = \delta \gamma = \delta \tau = \delta \sigma  =  \delta B^T_\m = 0 \,.
\ee
\vskip 0.3cm

Starting with the equations of motion in \eqref{munu}--\eqref{dilaton}, we get the following equations for the gauge-invariant fluctuations (see the details in appendix \ref{Apgz}):
\begin{itemize}
\item Tensor mode:
\be
 h_{\mu\nu}^{TT''} + 3\frac{a'}{a} h_{\mu\nu}^{TT'} +\left( \nabla_\r \nabla^\r -\frac{2}{3}\kappa\right) h_{\mu\nu}^{TT}=0\,.\label{ATT1}\\
\ee
\item  Vector mode:
\be \label{BNes}
{B}^T_\m=0\,.
\ee
I.e. there are no propagating vector modes.
\item Scalar mode:
\be \label{Nes1}
{\tau}=0\,, \qquad {\sigma}=0 \, , \ee
\be 2\frac{a'}{a}\lambda+\lambda' + \frac{\kappa}{3}\gamma-\frac{a'}{a}\left(\frac{1}{a^3}\left(a^3 \gamma\right)' \right)=0\,, \label{Nes2}\ee
\be
\left(\nabla_\m \nabla^\m +\frac{4}{3} \kappa\right)\lambda + 4 \frac{a'}{a}\lambda'- \frac{a'}{a} \nabla_\m \nabla^\m\gamma + \left(\frac{a'}{a}\right)^2\left(\frac{z^2}{3}-4\right)\left( \frac{1}{a^3}\left(a^3 \gamma\right)'-2\lambda\right)=0\,, \label{Nes3}\ee
\be
\left(\nabla_\m \nabla^\m + 2\k\right)\lambda + \frac{1}{a^3}\left( a^3\lambda'\right)' + \frac{2z'}{3z}\left(\kappa\,\gamma+ 3 \lambda'\right)=0\,. \label{Nes4}
\ee
\end{itemize}

We should emphasis that the equations here are applicable when the dangerous zero modes are not present. They are treated in appendix \ref{SZM}.

%%%%%%%%%%%%%%%%%%%%%%%%%%%%%%%%%%%%%%%%%%%%%%%%%%%%%
%%%%%%%%%%%%%%%%%%%%%%%%%%%%%%%%%%%%%%%%%%%%%%%%%%%%%
%%%%%%%%%%%%%%%%%%%%%%%%%%%%%%%%%%%%%%%%%%%%%%%%%%%%%
\section{Separation of variables}

We are interested in solving the equations \eqref{ATT1} and \eqref{Nes2}--\eqref{Nes4} that correspond to the fields\footnote{The other scalar fields can be found after  obtaining $\lambda$ and $\gamma$. }
\be \mathcal{F}= \left\{h^{TT}_{\m\n}, \lambda, \gamma \right\} \,.
\ee
We separate variables, expressing the solution as  a function of the holographic coordinate $y$ times a function of the slice coordinate $x^\m$
\be \label{profile}
\mathcal{F} (y, x^\m)=f(y)\,  \tilde{f}(x^\m)\,.
\ee
We classify the modes according to their eigenvalues, $m^2$, of the four dimensional Laplacian operator of the slice geometry\footnote{{We are using the mostly plus metric convention in Minkowski signature and therefore $m^2$ corresponds to $-p^2$ in flat space.}}
\be \label{bskm}
\left(\nabla_\n \nabla^\n + \sigma \k\right)  \tilde{f}(x^\m)= m^2   \tilde{f}(x^\m)\,,
\ee
where  $\sigma$ is a number which depends on the mode ($\sigma=0, \frac23 $ for scalar and graviton mode respectively) and $\k$ is the four dimensional slice curvature.
Note that in the case where the slices are the flat Minkowski space-time ($\k=0$) we have
\be \label{fbox}
\nabla_\n \nabla^\n  \tilde{f} (x^\m)=m^2  \tilde{f} (x^\m)\,,
\ee
where the zero and non-zero eigenvalues of $m^2$ correspond to the massless and massive particles \cite{Kiritsis:2006ua}.

%\vspace{0.3cm}
%%%%%%%%%%%%%%%%%%%%%%%%%%%%%%%%%%%%%%%%%%%%%%%%%%%%%
%%%%%%%%%%%%%%%%%%%%%%%%%%%%%%%%%%%%%%%%%%%%%%%%%%%%%
%%%%%%%%%%%%%%%%%%%%%%%%%%%%%%%%%%%%%%%%%%%%%%%%%%%%%
\subsection{Tensor mode}\label{tenmod}

To study the 4--dimensional graviton mode, we begin with the Fierz-Pauli action
%\cite{}
\be\label{FP action}
S_{FP}=-\frac{1}{k^2} \int d^4 x \sqrt{-g}\Big( R^{(g)}-2\Lambda - \frac{M^2}{4}(h_{\m\n}{h}^{\m\n}-{h}^2) \Big)\,,
\ee
where $M$ is the mass of the graviton and $h_{\m\n}$ denotes the fluctuation around the background metric $g_{\m\n}=g_{\m\n}^{(0)}+h_{\m\n}$.
Expanding this action to quadratic order in fluctuations around the metric appropriate to the cosmological constant that solves the leading equations of motion (Minkowski, De Sitter or Anti de Sitter) we find (see details of the expansions in appendix \ref{linz})
\begin{gather}
	S_{FP}
	=-\frac{1}{k^2}\int d^4x \sqrt{-g^{(0)}}\,\bigg\{-\frac{1}{4} \nabla_\l h_{\m\n} \nabla^\l h^{\m\n}+ \frac{1}{2}\nabla ^\n h_{\m\n} \nabla_\l h^{\m\l}
	\nn\\
	-\frac{1}{2}\nabla_\m h^{\m\n} \nabla_\n h+ \frac{1}{4}\nabla_\m h \nabla^\m h	+ \frac{1}{2}  R^{(0)}_{\m\l\n\r}  h^{\m\n}  h^{\l\r} +\frac12  R^{(0)}_{\m\n}\left( {h^\m}_\l h^{\n\l}-h h^{\m\n}\right)
	\nn\\
	-\frac{1}{2}  R^{(0)}\left( \frac{1}{4}h^{\m\n}h_{\m\n}-\frac{1}{8}h^2\right)- \frac{1}{4}M^2\left(h_{\m\n}{h}^{\m\n}-{h}^2\right)+\mathcal{O}(h^3)\bigg\} \,,\label{S2}
\end{gather}
where we have
\be\label{S3}
R^{(0)}_{\mu\nu\l\r}=\frac{\Lambda}{3}\left( \zeta_{\mu\l}\zeta_{\nu\r }-\zeta_{\mu\r}\zeta_{\nu\l} \right)\sp
R^{(0)}_{\mu\nu} = \Lambda \zeta_{\mu\nu}\sp R^{(0)}=4\Lambda \,.
\ee
We impose the transverse-traceless conditions (${h}_{\m\n}= h^{TT}_{\m\n} $).
\be\label{TTC}
\nabla^\m  h^{TT}_{\m\n}=0 \sp  h^{TT\m}_{\m} =0 \,,
\ee
then the action \eqref{S2} is
\be\label{S4}
S_{FP}=\frac{1}{4 k^2} \int d^4 x \sqrt{-g^{(0)}}\left\{\nabla^\m {h}^{TT}_{\rho\sigma}\nabla_\m {h}^{TT\rho\sigma}
+\left(\frac23 \Lambda + M^2 \right) {h}_{\mu \nu}^{TT} {h}^{TT\mu \nu}\right\}\,.
\ee
Therefore, the linearized  equation of motion for ${h}_{\mu \nu}^{TT}$ is given by
\be\label{eom gra}
\left(\nabla_\r \nabla^\r -\frac23 \Lambda -M^2\right){h}_{\mu \nu}^{TT}=0 \,.
\ee
The cosmological constant is related to the slice curvature easily by comparing equations \eqref{S3} with \eqref{symspa} or by solving the background equation of motion
\be\label{S5}
R^{(0)}_{\m\n}-\frac12 \left(R^{(0)}-2\Lambda\right) g_{\m\n}^{(0)} = 0 \rightarrow R^{(0)}=R^{(\z)}=4\Lambda \rightarrow \Lambda=\kappa\,,
\ee
where we have used equation \eqref{symspa}.
Therefore, equation \eqref{eom gra} translates to
\be \label{Sp 2}
\left(\nabla_\r \nabla^\r -\frac{2}{3} \kappa- M^2\right)h^{TT}_{\m\n}(x) = 0\,.
\ee
We now return to our 5--dimensional equation for the tensor mode in \eqref{ATT1}
\be \label{eomTT}
h_{\mu\nu}^{TT''}(y,x) + 3\frac{a'}{a} h_{\mu\nu}^{TT'}(y,x) +\left( \nabla_\r \nabla^\r -\frac{2}{3}\kappa\right) h_{\mu\nu}^{TT}(y,x)=0 \,.
\ee
We separate variables in (\ref{eomTT}) as follows
\be \label{httyx}
h^{TT}_{\m\n}(y,x)=h(y)h^{TT}_{\m\n}(x)\,,
\ee
where $h^{TT}_{\m\n}(x)$ is chosen to satisfy the eigenvalue equation (\ref{Sp 2}) where $M^2$ is the eigenvalue.
In that sense, $M^2$ will be interpreted as the mass$^2$ of the 4--dimensional tensor modes.

We insert (\ref{httyx}) into equation \eqref{Sp 2} and obtain the following equation for $ h(y) $
\be \label{hy}
h''(y)+ 3 \frac{a'}{a} h'(y)+ M^2 h(y)=0\,.
\ee
This is the same equation we obtain in the flat slice case, but now, the solution for the scale factor is affected by the slice curvature and $M^2$ are the eigenvalues of the  Laplacian in curved space \eqref{Sp 2}.

If we introduce the function $B(y)$ and the wave-function $\psi_g (y) $ as
\be \label{B(y)}
\psi_g(y)= h(y)e^{-B(y)}\sp  \qquad a(y)=e^{-\frac{2}{3}B(y)}\,,
\ee
then \eqref{hy} becomes a Schrodinger-like equation for $\psi_g(y)$
\be \label{Sch-grav}
-\psi_g''(y) + V_g(y)\psi_g (y)= M^2 \psi_g (y) \,,\ee
\be \label{Vgrav}
V_g(y)= B'^2(y)-B''(y)\,.
\ee

%\vspace{0.3cm}
%%%%%%%%%%%%%%%%%%%%%%%%%%%%%%%%%%%%%%%%%%%%%%%%%%%%%
%%%%%%%%%%%%%%%%%%%%%%%%%%%%%%%%%%%%%%%%%%%%%%%%%%%%%
%%%%%%%%%%%%%%%%%%%%%%%%%%%%%%%%%%%%%%%%%%%%%%%%%%%%%
\subsection{Scalar mode} \label{smode}

For scalar fields we have equations  \eqref{Nes2}--\eqref{Nes4}.
When $\kappa=0$ then equation \eqref{Nes4} can be solved for $\lambda$ alone and then substituting into the \eqref{Nes3} we can determine $\gamma$. When $\kappa\not=0$ then we can solve equation \eqref{Nes4}  for $\g$ and substitute into the \eqref{Nes3} to obtain an equation for $\l$ only
\begin{gather}
	\left(\left(\nabla_\m \nabla^\m+\frac43\kappa\right)-\frac{\kappa z^2}{9}\right)
	\left(\left(\nabla_\n\nabla^\n+2\kappa\right)\l+\l''+\left(3\frac{a'}{a}+2\frac{z'}{z}\right)\lambda'\right)
	\nn \\
	+\frac{2\kappa a z'}{3 z a'}\left(\nabla_\m \nabla^\m+{\frac43}\kappa\right)\l+\frac{2\kappa z z'}{9}\l'=0\,.\label{eomg0}
\end{gather}
This equation can be equivalently be written as
\begin{gather}
	\left(\nabla_\m \nabla^\m+{\frac43}\kappa\right)^{\!2}\l +
	\left(\nabla_\m\nabla^\m+{\frac43}\kappa\right)\left[\l''+\left(3\frac{a'}{a}+2\frac{z'}{z}\right)\l'+\frac{\kappa}{3}\left(\frac{2 a z'}{za'}-\frac{z^2}{3}+2\right)\l\right]
	\nn\\
	-\frac{\kappa z^2}{9}\left[\l''+\frac{3a'}{a}\l'+\frac{2\kappa}{3}\l\right]=0\,.\label{eomg1a}
\end{gather}
This equation shows that we can use the separation of variables to solve the equations of motion.
We now introduce the eigenfunctions of the Laplacian as
\be\label{lap0}
\nabla_\m \nabla^\m Y_m(x)=m^2Y_m(x)\,,
\ee
and expand the two scalars in a series
\be
\lambda(y,x)={\sum_{m}}'\lambda_m(y)Y_m(x)\sp \gamma(y,x)={\sum_{m}}'\gamma_m(y)Y_m(x)\,,
\label{lap}
\ee
The prime in the sum means that we do not include  the special eigenvalues of the slice Laplacian (zero modes) that we identified in section \ref{Dec}, namely $m^2=0$ and $m^2 = -(4/3) \kappa$. For these values, the scalar-vector-tensor decomposition we used is ill-defined and we have to resort to another method (described in detailed in Appendix \ref{SZM}) to obtain decoupled equations\footnote{This eventually leads to the same  equations which one would obtain by substituting the special values of $m^2$ in the final result we will arrive at in this section, but the intermediate steps are invalid.}. Excluding the zero modes from the series and dedicating to them a separate treatment is allowed  because, at the linear level, modes corresponding to different eigenvalues do not couple to each other. 

%\FN{Is this still the case, after we cleared up how to deal with zero modes?}
We now insert (\ref{lap}) and \eqref{lap0} into (\ref{Nes2})--(\ref{Nes4}) to obtain an infinite set of radial systems for the coefficients, each for each nontrivial eigenvalue $m$:
\be\label{eoml1}
\left(m^2+\frac43\kappa\right)\left(\lambda_m-\frac{a'}{a}\gamma_m\right)+\frac{z^2 a'}{3a}\left(\lambda_m'+\frac{\kappa}{3}\gamma_m\right)=0\,,
\ee
\be
\left(m^2+2\kappa\right)\lambda_m+\lambda_m''+\left(3\frac{a'}{a}+2\frac{z'}{z}\right)\lambda_m'+\frac23\kappa\frac{z'}{z}
\gamma_m=0\,.\label{eomg1}
\ee
along with
\be
2\frac{a'}{a}\lambda_m+\lambda_m' + \frac{\kappa}{3}\gamma_m-\frac{a'}{a}\left(\frac{1}{a^3}\left(a^3 \gamma_m\right)' \right)=0\,. \label{Nes211}
\ee
We now  solve equation \eqref{eoml1} for $\g_m$ and obtain
\be\label{gamy}
\g_m=
-\frac{3 \left(z^2 a' \lambda'_m+a \left(4 \k+3 m^2\right) \lambda_m\right)}{a'\left(\k z^2-12 \k-9 m^2\right)}\,.
\ee
Substituting the above result into \eqref{eomg1} or \eqref{Nes211} gives an equation for $ \l_m $
\be \label{yprofile}
\l''_m + A \l'_m +B \l_m =0\,,
\ee
with the following coefficients
\be\label{Amassive}
A=3\frac{a'}{a}+6\frac{z'}{z}\frac{3 m^2+4\kappa}{(9 m^2 +12 \kappa -\kappa z^2)} \,,
\ee
\be \label{Bmassive}
B= m^2 +2\kappa +\frac{2\kappa a z'}{a'z} \frac{ 3 m^2 + 4\kappa}{(9 m^2 +12 \kappa -\kappa z^2)} \,.
\ee
To write the equation \eqref{yprofile} as a Schrodinger-like equation we define
\be \label{decal}
e^{h(y)}\equiv \frac{a^3 z^2}{9 m^2 + 12 \kappa - \kappa z^2} \sp \psi_s(y)\equiv e^{\frac{h(y)}{2}}\sqrt{9 m^2 + 12 \kappa}\,\l_m(y) \,.
\ee
Therefore equation \eqref{yprofile} for $\l_m(y)$ becomes the following equation for $\psi_s(y)$
\be\label{SCH}
\psi_s''(y) - V_s(y) \psi_s(y) = 0 \,,
\ee
with the effective potential
\be\label{SLpot}
V_s(y)=-\kappa\left(1+\frac{ a }{3a'}h'\right) +\frac{1}{4} {h'}^2 +\frac{1}{2}h'' -m^2\,.
\ee
As shown in Appendix \ref{SZM}, treating the zero modes correctly one arrives at the same equation, (\ref{SCH}-\ref{SLpot}),   with  $m^2$ replaced by zero or $-(4/3) \kappa$.

%\vspace{0.5cm}
%%%%%%%%%%%%%%%%%%%%%%%%%%%%%%%%%%%%%%%%%%%%%%%%%%%%%
%%%%%%%%%%%%%%%%%%%%%%%%%%%%%%%%%%%%%%%%%%%%%%%%%%%%%
%%%%%%%%%%%%%%%%%%%%%%%%%%%%%%%%%%%%%%%%%%%%%%%%%%%%%
\section{Normalizability and stability  of tensor and scalar fluctuations}
%\subsection{The normalizability of fluctuations}

If we substitute the separation of variables introduced in \eqref{profile} into the corresponding  actions for $h^{TT}_{\m\n}, \lambda $ and $ \gamma$, they separate into a $y$ and $x^\m$ dependent parts. The normalizability conditions are defined by demanding that the integrals of the actions over the $y$ coordinate be finite, which signifies that the modes propagate in the boundary directions. In the following, we shall perform these steps for the various fields.

%\vspace{0.3cm}
%%%%%%%%%%%%%%%%%%%%%%%%%%%%%%%%%%%%%%%%%%%%%%%%%%%%%
%%%%%%%%%%%%%%%%%%%%%%%%%%%%%%%%%%%%%%%%%%%%%%%%%%%%%
%%%%%%%%%%%%%%%%%%%%%%%%%%%%%%%%%%%%%%%%%%%%%%%%%%%%%
\subsection{Normalizability of tensor mode}

To obtain the normalization condition for tensor mode, we insert \eqref{httyx} into the action \eqref{acTT}
\begin{gather}
	S^{(2)TT} =  \frac{1}{2 k_5^2}\int d^4x dy\,\sqrt{-g^{(0)}}\bigg\{ a^3(y)  \Big[-\frac{1}{4}h^2(y)\nabla^\mu h^{TT}_{\rho\sigma}(x)\nabla_\mu h^{TT\rho\sigma}(x)
	\nn\\
	-\frac{1}{4} \big(h'(y)\big)^2{h^{TT}_{\rho\sigma}}(x) {h^{TT\rho\sigma}}(x)- \frac{1}{6}\kappa h^2(y)h_{\mu \nu}^{TT}(x) h^{TT\mu \nu}(x)\Big]\bigg\}\,,\label{s2tt}
\end{gather}
where by integration by parts of $(h'(y))^2$ in the second term, and using equation \eqref{hy}, we obtain
\begin{gather}
	S^{(2)TT}= - \frac{1}{8 k_5^2}\int dy \sqrt{-g^{(0)}}a^3(y)h^2(y)\int d^4x\bigg\{\nabla^\mu h^{TT}_{\rho\sigma}(x)\nabla_\mu h^{TT\rho\sigma}(x)
	\nn\\
	+\Big(\frac{2}{3}\kappa + M^2  \Big)h_{\mu \nu}^{TT}(x) h^{TT\mu \nu}(x) \bigg\}\,.\label{s2tta}
\end{gather}
The above action indicates that the tensor modes are normalizable provided that
\be \label{norgn}
\int dy a^3(y)h^2(y) < \infty\,.
\ee
We should emphasize that the action \eqref{s2tta} gives equation \eqref{Sp 2} upon variation with respect to $h_{\mu \nu}^{TT}(x)$.

In terms of the wave function $\psi_g(y) $ which is defined in \eqref{B(y)}, the normalizability condition \eqref{norgn}  reads as
\be \label{NorMg}
\int dy |\psi_g(y)|^2 < \infty.
\ee

\vspace{0.3cm}
%%%%%%%%%%%%%%%%%%%%%%%%%%%%%%%%%%%%%%%%%%%%%%%%%%%%%
%%%%%%%%%%%%%%%%%%%%%%%%%%%%%%%%%%%%%%%%%%%%%%%%%%%%%
%%%%%%%%%%%%%%%%%%%%%%%%%%%%%%%%%%%%%%%%%%%%%%%%%%%%%
\subsection{Normalizability of scalar mode}\label{Nos}

To find the action of $\l(y,x)$, we
substitute the new variables from \eqref{chv2}--\eqref{chv5} into the action \eqref{acSca}, and obtain the corresponding action for $\lambda$ and $\gamma$ scalar fields
\begin{gather}
	S^{(2)}(\lambda,\gamma) = {\frac12 k_5^2}\int d^4x dy\,\sqrt{-g^{(0)}}\frac{a^3}{9 {a'}^2}\Big(-9\left({a'}^2 z^2+\kappa a^2\right)(\pa_\m\lambda)^2
	\nn\\
	-9\kappa {a'}^2 (\pa_\m\gamma)^2 +18\kappa a a' \pa_\m\lambda\pa^\m\gamma-24\kappa^2 aa'\lambda\gamma+12\kappa\left(\kappa a^2+{a'}^2z^2\right)\lambda^2
	\nn\\
	+\kappa^2 {a'}^2 \gamma^2 \left(12-z^2\right) -3{a'}^2 z^2\lambda'\left(3\lambda'+2 \kappa\gamma \right) \Big).\label{s2sca}
\end{gather}

By substituting the relations in \eqref{lap} into the action \eqref{s2sca} and using $\g_m(y)$ in \eqref{gamy} and equation \eqref{yprofile}, we find the action of $\l$ as follow
\be
S^{(2)}_{\lambda} = \frac{1}{2\kappa_5^2}\int dy F(y) \lambda^2_m(y) \int d^4x\sqrt{-g^{(0)}} \left(\left(\pa_\m Y_m(x)\right)^2 + m^2  Y_m(x)^2 \right)\,,\label{actla1}
\ee
where
\begin{gather}
	F(y)=
	-\frac{3 a^2 \left(4 \k+3 m^2\right) z^2}{a'\left(12 \k+9 m^2-\k z^2\right)^3} \Bigg(
	\frac{27 \k a a' {z'}^2 \left(\k z^2+12 \k+9 m^2\right)}{12 \k+9 m^2-\k z^2}
\nn \\
	\quad\quad	+a a' \left(12 \k+9 m^2-\k z^2\right)^2+9 \k a z a' z''+3 k z z'\left(9 {a'}^2-2 \k a^2\right)
	\Bigg)\,.
	\label{actla2}
\end{gather}
This action works for all values of $m^2$ except at $m^2=-4\k/3$ which is identically zero.
However, as we have already supposed the scalar field does not contain the zero mode of $\Box+\frac43\k$.

Upon the definitions in \eqref{decal}, the action \eqref{actla1} changes to
\be \label{acts}
S^{(2)}_{\lambda} = \frac{1}{2k_5^2}\int dy\,\psi_s^2\, H(y)\int d^4x\sqrt{-g^{(0)}} \left(\left(\pa_\m Y_m(x)\right)^2 + m^2  Y_m(x)^2 \right)\,.
\ee
In this equation $H(y)$ is given by
\begin{gather}
	H(y)=-1 -\frac{3 \k e^{h}}{2 a^3 \left(4 \k+3 m^2\right)}\left(h''+2{h'}^2 -\left(\frac{2 \k a}{3 a'}+\frac{9 a'}{a}\right)h'\right)
	\nn \\
	-\frac{9 \k e^{h} \left(e^{h} \left(3 \left(2 \k+m^2\right) {a'}^2+\k^2 a^2\right)+2 a^3 {a'}^2+\k a^5\right)}{2 a^5 \left(4 \k+3 m^2\right) \left(a^3+\k e^{h}\right)}\,.\label{Hy}
\end{gather}
Therefore, the normalization condition for the scalar mode is
\be \label{ncsc}
\int dy\,\psi_s^2\, H(y)<\infty\,.
\ee
We should note that there is a special case where $\Phi'_0=0$. This corresponds to the exact AdS$_5 $ solution. The equations of motion for fluctuating modes and the normalization conditions for this very special case, are given in section \ref{mgads} and appendix \ref{spcase}.

\vspace{0.3cm}
%%%%%%%%%%%%%%%%%%%%%%%%%%%%%%%%%%%%%%%%%%%%%%%%%%%%%
%%%%%%%%%%%%%%%%%%%%%%%%%%%%%%%%%%%%%%%%%%%%%%%%%%%%%
%%%%%%%%%%%%%%%%%%%%%%%%%%%%%%%%%%%%%%%%%%%%%%%%%%%%%
\subsection{The stability of tensor and scalar fluctuations}\label{stability}

In this section, we study the stability of our background solutions under small field fluctuations.
We already showed that in general, the scalar and graviton modes satisfy the following Laplace equations on the 4--dimensional slices
\be\label{st1}
\nabla_\m \nabla^\m Y(x)=m^2  Y(x) \sp \left(\nabla_\r \nabla^\r -\frac{2}{3} \kappa\right)h^{TT}_{\m\n}(x) =  M^2 h^{TT}_{\m\n}(x)\,.
\ee
We can unify the above relations into a unique relation depending on the sign of the slice's curvature \cite{Ghosh:2023gvc}
\be \label{eigen Eq}
\left(\nabla_\m \nabla^\m -\frac{1}{3} s \k\right)\delta h =
\begin{cases}
-\frac{1}{\alpha^2}\left(\n^2 -\frac{9}{4}\right)\delta h \,\qquad dS  \\ \\
-k^2 \delta h \, \qquad \qquad Minkowski \,,\\ \\
\frac{1}{\alpha^2}\left(\n^2 -\frac{9}{4}\right)\delta h \,\qquad AdS
\end{cases}
\ee
where $s$ is the spin of the fluctuations ($s=0$ for scalar and $s=2$ for graviton) and $\delta h= Y , h^{TT}_{\m\n} $. Moreover, $\a$ is the curvature radius  of (A)dS slices defined in \eqref{defa}. In \eqref{eigen Eq} the parameter $\n$ is related to the mass of the fields as
\be\label{mnu}
 \n=\frac{1}{2}\sqrt{(d-1)^2+4 \mathfrak{m}^2 \alpha^2} \sp \mathfrak{m}=m,M \,.
\ee
where $M$ is the graviton masses, whereas $m$ are the scalar masses.
The stability of the solutions under the field fluctuations is mainly affected by the presence of either a ``tachyonic" or ``ghost" mode:

\begin{itemize}
\item{\bf{Tachyonic stability} }

 The criteria for the modes to be  tachyon-stable in maximally symmetric four dimensional geometries are as follows \cite{Ghosh:2023gvc}
\be \label{tac cri}
\text{The mode is tachyon-stable if}
 \begin{cases}
 |Re(\n)| \le \frac{3}{2} \qquad dS \\ \\
 k^2  \le 0 \qquad Minkowski \,, \\ \\
  Re(\n) \neq 0 \qquad AdS
 \end{cases}
\ee
Independent of the spin, the above relations translate in terms of the mass to the following equation
\be \label{tac crim}
\text{The mode is tachyon-stable if}
 \begin{cases}
 \mathfrak{m}^2\geq 0 \qquad dS \\ \\
 \mathfrak{m}^2\geq 0 \qquad Minkowski \,, \\ \\
 \mathfrak{m}^2\geq \frac34 \k \qquad AdS
 \end{cases}
\ee
where in the case of AdS space-time ($\k<0$), the above relation is the well-known BF bound.

In the dS$^{(4)}$ space-time, in order  to have a positive norm  state for graviton modes, the mass squared should satisfy
\be\label{HB}
M^2 \a^2 > 2  \quad\Rightarrow \quad M^2 > \frac23 \k\,,
\ee
which is known as  the Higuchi bound \cite{Higuchi:1986py}.
\item{\bf{Ghost stability}}

Ghost instability corresponds to a mode with a wrong sign of the kinetic term. Concerning the actions corresponding to different modes, i.e. relation \eqref{s2tta} for graviton and  \eqref{acts} for scalar we observe that the kinetic terms have the correct sign. There is no ghost instability in any one of these modes.
\end{itemize}

%\vspace{0.5cm}
%%%%%%%%%%%%%%%%%%%%%%%%%%%%%%%%%%%%%%%%%%%%%%%%%%%%%
%%%%%%%%%%%%%%%%%%%%%%%%%%%%%%%%%%%%%%%%%%%%%%%%%%%%%
%%%%%%%%%%%%%%%%%%%%%%%%%%%%%%%%%%%%%%%%%%%%%%%%%%%%%
\section{A warmup: The mass spectrum on an exact AdS space-time  }\label{mgads}

In this section, as a warm-up, we shall find the mass spectrum of the field fluctuations at a conformal fixed point, that is an extremum of the potential at which $\Phi_0= const.$
In this case,  the scalar potential effectively is\footnote{Note that we still have $\partial^2_{\Phi_0}\!V=-2m_{\Phi}^2$.},
\be \label{GeffP}
V(\Phi_0)=-\frac{d(d-1)}{\ell^2}\,,
\ee
and  the corresponding background solution is a  globally AdS$_{d+1}$ space-time with the length scale $\ell$.

In the ansatz \eqref{bac2}, the scale factor for a $d$--dimensional Minkowski and A(dS) slices is given by
\be\label{Aun}
e^{A(u)}=
\begin{cases}
 \exp\big(\!-\!\frac{u+c}{\ell}\big)\, \qquad &-\infty < u < +\infty \qquad \,\,\mathcal{M}_d \\ \cr
 \frac{\ell}{\alpha}\cosh\big(\frac{u+c}{\ell}\big)\, \qquad &-\infty < u < +\infty \qquad AdS_d \\ \cr
 \frac{\ell}{\alpha}\sinh\big(\frac{u+c}{\ell}\big)\, \qquad &-c \leq u < +\infty \quad\quad dS_d (S^d)
\end{cases} \,,
\ee
where $c$ is an integration constant and $\alpha$ is the curvature length scale of the  $d$--dimensional slices defined in \eqref{defa}.

In  conformal coordinates, defined in \eqref{metr}, the conformal factor is (the details of the calculations in this section are given in appendix \ref{spcase})
\be\label{ayn}
a(y) \equiv e^{A(y)} \equiv e^{-\frac23 B(y)}=
\begin{cases}
\ell{\big(y-y_0\big)}^{-1}\,  &y_0 < y < +\infty \quad\quad\qquad \mathcal{M}_d \\ \cr
\frac{\ell}{\alpha }\big(\sin(\frac{y-y_0}{\alpha})\big)^{-1}\,  & y_0 < y \le  y_0+\alpha \pi  \quad\quad AdS_d \\ \cr
-\frac{\ell}{\alpha}\big( \sinh(\frac{y-y_0}{\alpha})\big)^{-1}\,  & -\infty < y \leq  y_0   \quad\quad\quad dS_d (S^d)
\end{cases}\,,
\ee
where $y_0$ is a constant of integration.
\begin{itemize}
\item For flat slices, the limit $u\rightarrow - \infty$ corresponds to $y\rightarrow y_0$ and this is the boundary of AdS$_{d+1}$. On the other hand, the other boundary $u\rightarrow+\infty$ corresponds to $y\rightarrow +\infty$.

\item For AdS slices,  two boundaries at $u\rightarrow \pm \infty $ map to $y=y_0$ and $y=y_0+\pi\a$.

\item For dS slices, there is a boundary at $u\rightarrow +\infty$ or $y=y_0$ and space ends at $u=-c$ or $y\rightarrow -\infty\,$.
\end{itemize}

\vspace{0.3cm}
%%%%%%%%%%%%%%%%%%%%%%%%%%%%%%%%%%%%%%%%%%%%%%%%%%%%%
%%%%%%%%%%%%%%%%%%%%%%%%%%%%%%%%%%%%%%%%%%%%%%%%%%%%%
%%%%%%%%%%%%%%%%%%%%%%%%%%%%%%%%%%%%%%%%%%%%%%%%%%%%%
\subsection{Gravitons}\label{Gs}

In section \ref{tenmod} we found that the graviton mode obeys a Schrodinger-like equation as
\be \label{S-grn}
-\psi_g''(y) + V_g(y)\psi_g (y)= M^2 \psi_g (y) \sp
V_g(y)= B'^2-B''=\frac34 \frac{a'^2}{a^2}+\frac32 \frac{a''}{a}\,,
\ee
where $B(y)$ now is given by \eqref{ayn}. In what follows, we solve the equation above, in order  to read the mass spectrum of the graviton modes:
\begin{itemize}
\item {\bf{Minkowski slices}}
\end{itemize}
From the conformal factor in \eqref{ayn}, the potential in Schrodinger equation \eqref{S-grn} becomes (in the rest of this section  we fix $y_0=0$ for simplicity)
\be\label{Vmingln}
V_g(y)= \frac{15}{4y^2}\,,
\ee
and the solution of the wave function is
\be \label{Simingln}
\psi_g(y)=\sqrt{y}\, \Big(C_1 \, J_2(M y)+C_2 \, Y_2(M y)\Big)\,,
\ee
where $J_2$ and $Y_2 $ are the first and second kinds of the Bessel functions respectively.

To have a vev solution, we should set the source to zero by taking $C_2 =0$ at $y=0$ boundary.
 The wave function becomes
\be \label{SiNmingln}
\psi_g(y)=C_1 \,\sqrt{y}\,  J_2(M y)\,,
\ee
which is a finite function in the range $0\leq y<\infty$. Therefore, when the slices are flat, the graviton has a continuous mass spectrum and $M^2>0$ because of the tachyonic stability condition found in \eqref{tac crim}. This should follow the normalization condition \eqref{NorMg}
\be
\int dy |\psi_g (y)|^2=|C_1|^2 \int_{0}^{+\infty} dy\, y J_2(My)^2=\frac{|C_1|^2 }{M^2}\,.
\ee
We should note that for the massless graviton mode $M=0$, the wave function is
\be
\psi_g(y)=C_1 y^{\frac52}+{C_2}{y^{-\frac32}}\,,
\ee
which indicates that \eqref{S-grn} does not contain a normalizable massless mode of the graviton for flat slices.

The mass spectrum we found above, was expected since the potential in \eqref{Vmingln} diverges when $y\rightarrow 0$ and $V_g\rightarrow 0$ as $y\rightarrow +\infty$, indicating a continuous mass spectrum with $M^2>0$.
\begin{itemize}
\item {\bf{dS slices}}
\end{itemize}
From equation \eqref{ayn}, the potential \eqref{S-grn} for dS slices is
\begin{figure}[!ht]
\begin{center}
\begin{subfigure}{0.48\textwidth}
\includegraphics[width=1 \textwidth]{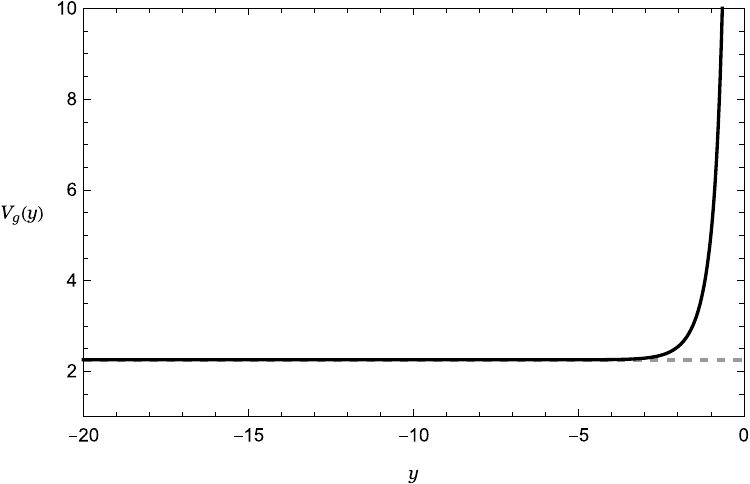}
\caption{}\label{PdSn}
\end{subfigure}\hspace{0.15cm}
\begin{subfigure}{0.50\textwidth}
\includegraphics[width=1 \textwidth]{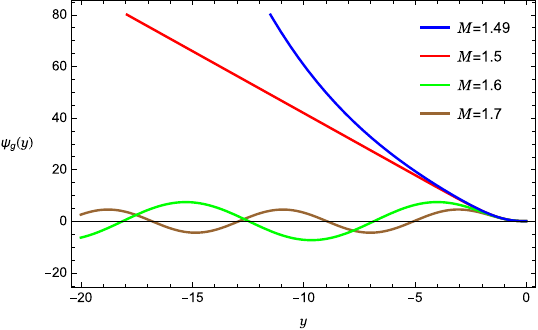}
\caption{}\label{ds numn}
\end{subfigure}
\end{center}
\caption{ \footnotesize{(a): The potential \eqref{Vdsgln} for dS slices with $\alpha=1$. $V_g(y)\rightarrow \frac 94$ when $y\rightarrow -\infty$. The mass spectrum should be continuous and a mass gap exists at $M=\frac32$. (b): Solutions of $\psi_g(y)$  for dS slices with $\alpha=1$ near the mass gap at $M=3/2$. Only above the mass gap does a continuous spectrum of the normalizable modes exist.}}
\end{figure}
\be \label{Vdsgln}
V_g(y)= \frac{3}{4 \alpha^2}\Big(3+ \frac{5}{\sinh^2 (\frac{y}{\alpha})}\Big) \,.
\ee
This potential is shown in figure \ref{PdSn}. The behavior of this potential near the UV boundary and IR end-point is
\begin{gather}
	V_g^{UV}\!(y)=\frac{15}{4 y^2}+\frac{1}{\alpha ^2}+\mathcal{O}(y^2)\sp y\rightarrow 0\,,
\\
	V_g^{IR}(y)=\frac{9}{4 \alpha ^2}+\frac{15}{\alpha ^2} e^{2y/\alpha}+ \mathcal{O}(e^{4y/\alpha})  \sp  y\rightarrow\, -\infty\,.\label{VdSAssymn}
\end{gather}
We observe that we should have a continuous mass spectrum but with a mass gap above $M^2=9/4\a^2$. We should emphasis that the mass gap is above the Higuchi bound, $M^2 \alpha^2 > 2 $.

We now solve equation \eqref{S-grn} when the mass is above or below the mass gap (the details are in appendix \ref{GAds}):
\begin{itemize}
\item When the mass is above the mass gap, near the UV boundary ($y \rightarrow 0$) the solution behaves as $\psi_g(y) \sim y^{5/2}$, if we impose the source-free boundary condition. In the IR limit ($y \rightarrow -\infty$) the wave function has a periodic behavior  $\psi_g (y) \sim \sin\big((\sqrt{ M^2- \frac{9}{4\a^2}})y\big)$. For masses above the mass gap, the wave function is finite and normalizable.

\item When the mass is below the mass gap, we know from elementary quantum mechanics that there is no normalizable solution and therefore, no propagating states.

\end{itemize}
For illustration, we have depicted the wave functions near the mass gap for four different values of masses in figure \ref{ds numn}.
For values $M^2\a^2\leq 9/4$ the wave function diverges as $y\rightarrow -\infty$ (end of space). Above the mass gap, the wave function is normalizable, it has a continues mass spectrum due to the shape of the potential and at large values of $y$, has a periodic behavior.

\begin{figure}[!t]
\centering
\includegraphics[width=0.5\textwidth]{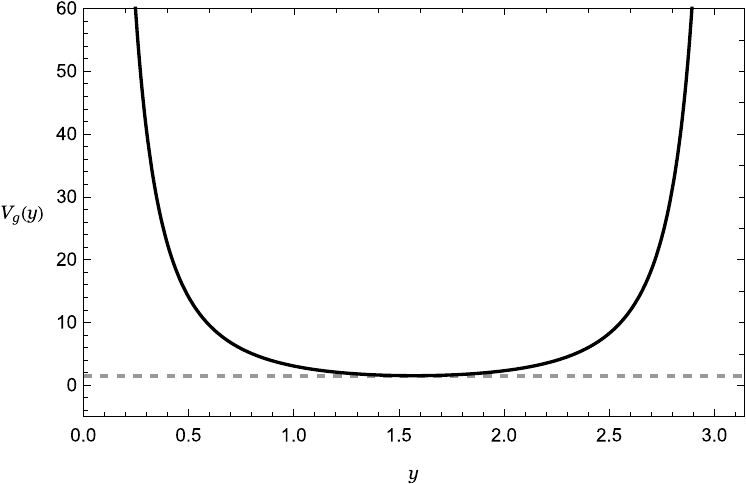}
\caption{ \footnotesize{Potential for AdS slice with $\alpha=1$, The minimum of the potential is in $(\frac{\pi}{2}, \frac32)$, and it's domain is between $(0,\pi)$ and goes to infinity at both ends. This potential gives a discrete mass spectrum with a positive mass gap.}}\label{PAdS1}
\centering
\end{figure}

\begin{itemize}
\item {\bf{AdS slices}}
\end{itemize}
The potential for AdS slices becomes (see figure \ref{PAdS1})
\be \label{Vadsgln}
V_g(y)= -\frac{3}{4 \alpha^2}\Big(3-5 \frac{1}{\sin^2 (\frac{y}{\alpha})} \Big) \,.
\ee
The behavior of $V_g(y)$ near the  UV boundaries is given by
\begin{gather}
	V_g^{UV}(y)=\frac{15}{4 y^2}-\frac{1}{\alpha ^2}+\mathcal{O}\left(y^1\right)\, \sp y\rightarrow 0 \,,
\nn\\
	V_g^{UV}(y)=\frac{15}{4 (y-\alpha \pi)^2}-\frac{1}{\alpha ^2} +\mathcal{O}\left(\left(y-\pi \alpha \right)^1\right)\, \sp  y\rightarrow \alpha \pi\,.
\end{gather}
This is a well-shaped potential with a minimum at $V(\frac{\a\pi}{2})=\frac{3}{2\a^2}$. Therefore, one expects to have a discrete mass spectrum with a mass gap that is larger than $\frac{3}{2\a^2}$.
To show this, we solve the Schrodinger equation \eqref{S-grn} which gives rise to
\be \label{Siadsgl}
\psi_g(y)=\sin\big(\frac{y}{\alpha}\big)^{1/2} \Bigg[ C_1 P_s^2\big(\cos (\frac{y}{\alpha})\big)+C_2 Q_s^2\big(\cos (\frac{y}{\alpha})\big)\Bigg]\,,
\ee
where $ P_{s}^2$ and $Q_{s}^2 $ are the associated Legendre functions with
\be \label{Legindexn}
s=\frac{1}{2}\left(\sqrt{4 \alpha ^2 M^2+9}-1\right)\,,\qquad s\geq 2\,.
\ee
 To have a regular solution at the $y=0$ boundary, we should impose the vev boundary conditions and put  $C_2=0$. The regularity of the wave function \eqref{Siadsgl} on the other boundary at $y=\alpha\pi $ implies a discrete mass spectrum through (see for example \cite{Karch:2000ct} for the same result)
\be \label{mspch1}
M^2=\frac{1}{4 \alpha^2}\left[\left(2s+1\right)^2-9 \right] \sp s\geq2 \sp s\in \mathbb{N}\,.
\ee
The minimum allowed value for the mass is above the mass gap i.e.
\be
M_{min}^2=\frac{4}{\a^2}>\frac{3}{2\a^2}\,.
\ee

\vspace{0.3cm}
%%%%%%%%%%%%%%%%%%%%%%%%%%%%%%%%%%%%%%%%%%%%%%%%%%%%%
%%%%%%%%%%%%%%%%%%%%%%%%%%%%%%%%%%%%%%%%%%%%%%%%%%%%%
%%%%%%%%%%%%%%%%%%%%%%%%%%%%%%%%%%%%%%%%%%%%%%%%%%%%%
\subsection{Scalars}\label{exsc}

Regarding the mass spectrum of the scalar fields, we note that, as we show in Appendix \ref{SAds}, when the space-time is globally AdS, the gauge invariant scalars $\gamma$ and $\lambda$ are ill-defined.
We define two new scalar fields $\chi$ and $\omega$  as in \eqref{ginv Gads}.

For the scalar field $\omega$, we find equations in \eqref{eom alpha} where after separating variables as $\omega(y,x)=\hat{\omega}(x)\omega(y)$, we obtain
\be
\omega(y)=\frac{c}{a^2} \sp \left(\nabla_\m \nabla^\m+\frac43 \k\right)\hat\omega (x)=0\,,
\label{al1}
\ee
Such a scalar field as was shown in section \ref{Nos} does not have a kinetic term and therefore does not appear in the dynamics. Moreover,
this is not an allowed mode because as we showed in section \ref{Dec}, with the decomposition of fluctuations that we have used, the scalar field does not contain the zero mode of operator $\nabla_\m \nabla^\m+\frac43 \k$.

The equation of motion for $\chi$ is given by \eqref{s5Gads}. After separating variables as $\chi(y,x)=\hat{\chi}(x)\chi(y)$ and assuming
\be \label{Nchi4d}
\nabla_\m \nabla^\m \hat{\chi}(x)= m_{\chi}^2 \, \hat{\chi}(x) \,,
\ee
the equation for $\chi(y)$ is given by \eqref{eqchiy} i.e.
\be \label{Neqchiy}
\chi''(y)+ 3 \frac{a'}{a} \chi'(y)+ \left(m_{\chi}^2-a^2(y) m^2_{\Phi}\right) \, \chi(y)=0\,.
\ee
Redefining, $\psi_s(y)=\chi(y) a(y)^\frac32$, we arrive at
\be \label{aSch-s}
-\psi_s''(y) + V_s(y)\psi_s(y)= m^2_{\chi} \psi_s(y) \,,\ee
\be \label{aV-s}
V_s(y)=\frac34 \frac{a'^2}{a^2}+\frac32 \frac{a''}{a}+ a^2(y) m_{\Phi}^2 \,.
\ee
Considering the conformal factor in \eqref{ayn} for maximally symmetric spaces, the scalar potential above is
\be\label{ps}
V_s(y)=
\begin{cases}
	\frac{15+4\ell^2 m_{\Phi}^2}{4y^2}	 & \qquad \,\,\mathcal{M}_d \\ \cr
-\frac{1}{4 \alpha^2}\Big(9- \frac{15+4\ell^2 m_{\Phi}^2}{\sin^2 (\frac{y}{\alpha})} \Big)	 & \qquad AdS_d \\ \cr
 \frac{1}{4 \alpha^2}\Big(9+ \frac{15+4\ell^2 m_{\Phi}^2}{\sinh^2 (\frac{y}{\alpha})}\Big) 	 & \quad\quad dS_d (S^d)
\end{cases} \,.
\ee
The behavior of the scalar potential and solution of Schrodinger equation is very similar to the graviton potential in subsection \eqref{Gs}. The details are as follows:

\begin{itemize}
	\item {\bf{Minkowski slices}}

The wave function is given by
\be \label{ps1}
\psi_{s}(y) = \sqrt{y}\, \Big(C_1 \, J_\n(m_\chi y)+C_2 \, Y_\n(m_\chi y)\Big)\,,
\ee
where
\be \label{ps2}
\n = \sqrt{4+\ell^2 m_{\Phi}^2}\,.
\ee

	\item {\bf{dS slices}}

In dS case the IR behavior of the potential does not change, therefore the mass gap still remains at $m^2_{\chi}=9/4\a^2$. The mass spectrum above this mass gap is continuous.	

	\item {\bf{AdS slices}}
	
In AdS space however, there is a constant shift in the minimum of the potential by $4\ell^2 m_{\Phi}^2$. The wave function becomes
\be \label{wsg}
\psi_s(y)=\sin\big(\frac{y}{\alpha}\big)^{1/2} \Bigg[ C_1\, P_s^t\big(\cos (\frac{y}{\alpha})\big)+C_2\, Q_s^t\big(\cos (\frac{y}{\alpha})\big)\Bigg]\,,
\ee
where
\be \label{ps3}
s = \frac{1}{2}\left(\sqrt{4 \alpha ^2 m_\chi^2+9}-1\right)\sp
t = \sqrt{4+\ell^2 m_{\Phi}^2}\,.
\ee
Since in associated Legendre functions $s\geq t$, we should have
\be \label{ps4}
m^2_{\chi} \geq \sqrt{\frac{\ell^2 {m^2_\Phi}+4}{\alpha ^4}}+\frac{\ell^2 {m^2_\Phi}+2}{\alpha ^2}\,.
\ee
\end{itemize}

\vspace{0.5cm}
%%%%%%%%%%%%%%%%%%%%%%%%%%%%%%%%%%%%%%%%%%%%%%%%%%%%%
%%%%%%%%%%%%%%%%%%%%%%%%%%%%%%%%%%%%%%%%%%%%%%%%%%%%%
%%%%%%%%%%%%%%%%%%%%%%%%%%%%%%%%%%%%%%%%%%%%%%%%%%%%%
\section{The mass spectrum on AdS-sliced solution }\label{mss}

In this section, we examine the mass spectrum of the modes corresponding to fluctuations around non-trivial flow  solutions with negative curvature slices.
In particular, we consider solutions that describe conformal interfaces and therefore have two boundaries.
In this section, we will use a concrete example to calculate the spectra numerically, but our comments are general and applicable to other similar cases.

\begin{figure}[!t]
\begin{center}
\includegraphics[width=0.6\textwidth]{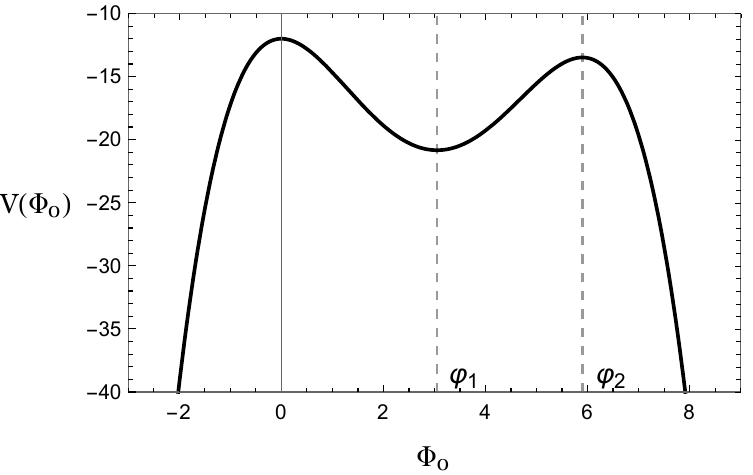}
\caption{\footnotesize{The bulk scalar potential \eqref{potqua} plotted for specific values $\ell_L=1, \ell_R= 0.94, \Delta_L=1.6$ and $\Delta_R=1.1$, The minimum is located at $\Phi_0=\varphi_1=3.05$ There are two maxima at $\Phi_0=0$ and $\Phi_0=\varphi_2=5.89 $.} }\label{Vphi}
\end{center}
\end{figure}

We shall use the following bulk scalar potential, whose holographic RG flows were studied in \cite{AdS1}
\be \label{potqua}
V(\Phi_0)=-\frac{12}{\ell_L^2}+\frac{\Delta_L(\Delta_L-4)}{\ell_L^2}\Phi_0^2-\frac{2(\varphi_1+\varphi_2)\Delta_L(\Delta_L-4)}{3\ell_L^2\varphi_1\varphi_2}\Phi_0^3+\frac{\Delta_L(\Delta_L-4)}{2\ell_L^2\varphi_1\varphi_2}\Phi_0^4,
\ee
where $\varphi_1$ and $\varphi_2$ are defined as
\be \label{potqua1}
	\varphi_1=\frac{6 \ell_R^2\sqrt{2(\ell_R^2-\ell_L^2)}\Delta_L(\Delta_L-4)}{\sqrt{\ell_R^2\Delta_L(\Delta_L-4)-\ell_L^2\Delta_R(\Delta_R-4)}\left(\ell_R^2\Delta_L(\Delta_L-4)+\ell_L^2\Delta_R(\Delta_R-4)\right)}\,,
\ee
\be
	\varphi_2=\frac{6 \sqrt{2(\ell_R^2-\ell_L^2)}}{\sqrt{\ell_R^2\Delta_L(\Delta_L-4)-\ell_L^2\Delta_R(\Delta_R-4)}}\,.\label{potqua2}
\ee
For our numerical investigations  we shall use the specific values, $\ell_L= 1,\, \ell_R= 0.94,\,  \Delta_L= 1.6$ and $\Delta_R=1.1$, (see figure \ref{Vphi})\footnote{We do not expect qualitative changes for other values of the parameters in the bulk scalar potential.}. The potential \eqref{potqua} has one minimum located at $\Phi_0=\varphi_1=3.05$ and two maxima at $\Phi_0=0$ and $\Phi_0=\varphi_2 = 5.89.$

%\vspace{0.3cm}
%%%%%%%%%%%%%%%%%%%%%%%%%%%%%%%%%%%%%%%%%%%%%%%%%%%%%
%%%%%%%%%%%%%%%%%%%%%%%%%%%%%%%%%%%%%%%%%%%%%%%%%%%%%
%%%%%%%%%%%%%%%%%%%%%%%%%%%%%%%%%%%%%%%%%%%%%%%%%%%%%
\subsection{$UV_L -UV_R $ solutions}\label{uvlrsec}

\begin{figure}[!t]
	\begin{center}
		\begin{subfigure}{0.45\textwidth}
			\includegraphics[width=1 \textwidth]{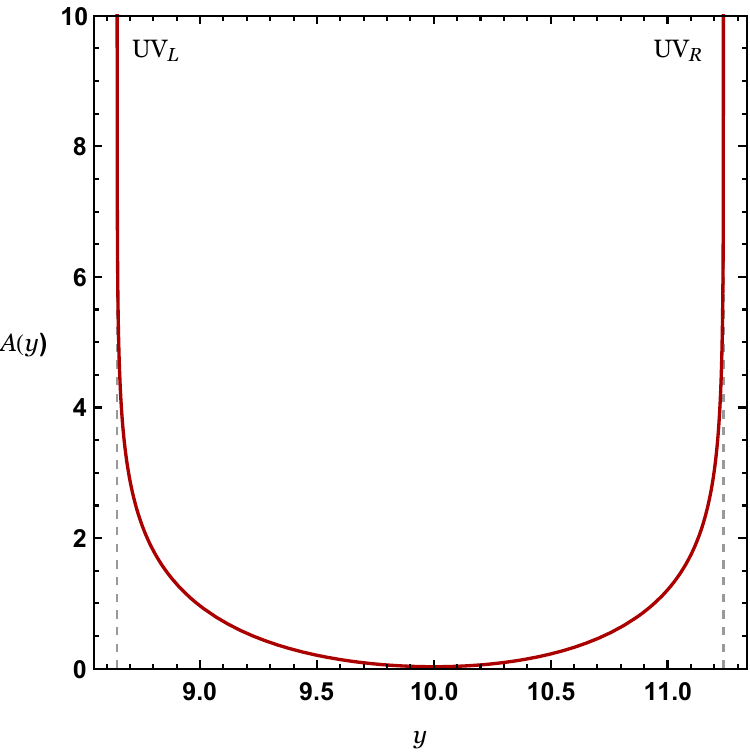}
			\caption{\footnotesize{}}\label{AdSA}
		\end{subfigure}
		\hspace{0.5cm}
		\begin{subfigure}{0.4505\textwidth}
			\includegraphics[width=1 \textwidth]{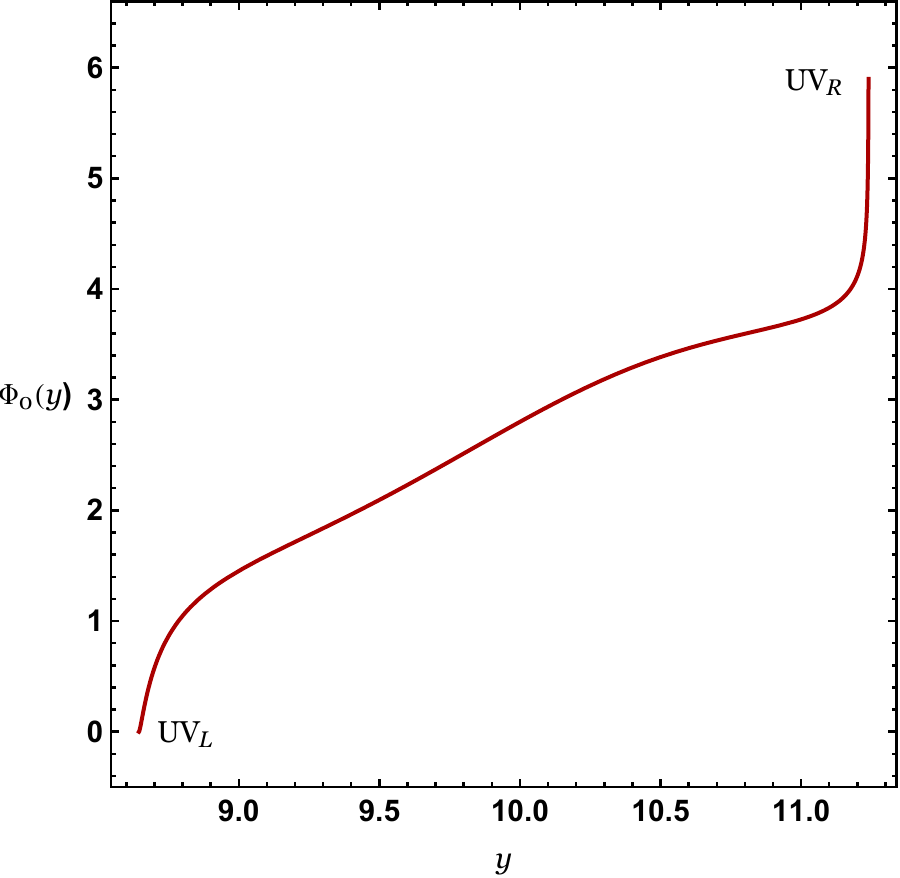}
			\caption{\footnotesize{}}\label{AdSfi}
		\end{subfigure}
		\begin{subfigure}{0.6\textwidth}
			\includegraphics[width=1 \textwidth]{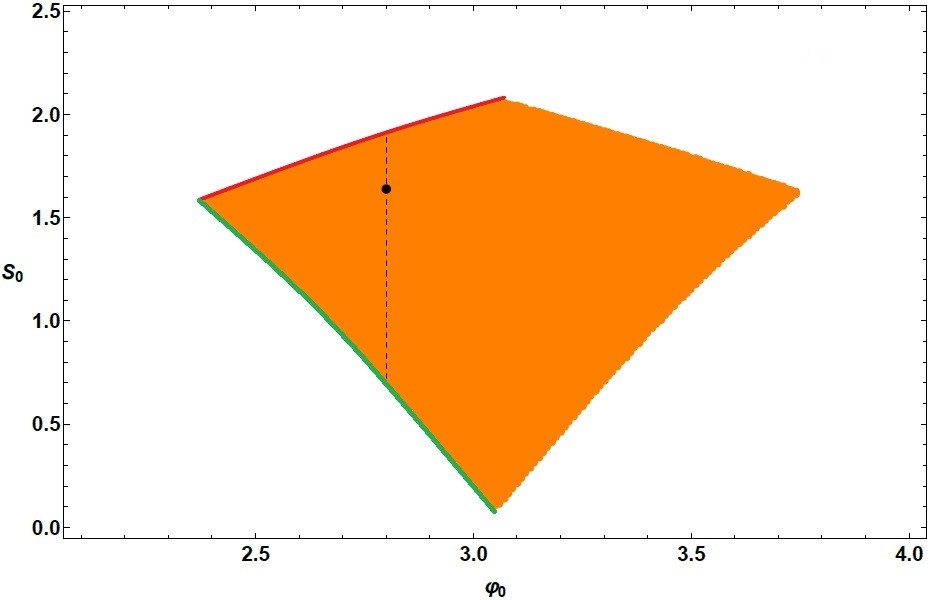}
			\caption{\footnotesize{}}\label{mapm}
		\end{subfigure}
		\caption{\footnotesize{(a): The scale factor $A(y)$ for a $UV_L-UV_R$ solution. b: $\Phi_0(y)$ is limited between two maxima of the potential in figure \ref{Vphi}. (c): The space of all solutions which are stretched between  $UV_L$ and $UV_R$ and have one $A$ bounce. Each point is one solution and the solution in figure (a) or (b) is the black dot on this map. The green boundary represents the points where solutions are fragmented into a flat and a UV-IR solution. On the red boundary, we have a vev-driven solution.} }
	\end{center}
\end{figure}

A subset of solutions for equations of motion \eqref{nsEom1}--\eqref{nsEom3}
with the negative potential in \eqref{potqua} includes those which are stretched between two UV boundaries denoted by $UV_L$ and $UV_R$.
These solutions  have at least one  A-bounce, which is defined as a point where $dA(y)/dy=0$ and signals a minimum of the scale factor. Moreover, each solution may have an $\f$-bounce where $d\Phi_0(y)/dy=0$ but $dA(y)/dy \neq 0$.

In the rest of the paper, we shall study a solution without a $\f$-bounce for simplicity.\footnote{The presence of $\f$-bounces is expected to provide features in the two-point function and shall be studied in a future publication.}
As an example of such a solution we have depicted $A(y)$ and $\Phi_0(y)$ in figures \ref{AdSA} and \ref{AdSfi}.

 The space of all such solutions is parameterized by two real initial conditions, $(\f_0, S_0)$, where $\f_0$ is the position of the $A$-bounce in field space, and $S_0$ is the value of $\dot\Phi_0$ in the same position. This space  is given by the orange region in figure \ref{mapm}. In this same map, the green boundary represents the points where $UV_L$-$UV_R$ solutions are fragmented into a flat and a UV-IR solution i.e. $\mathcal{R}_L$ asymptotes to a finite value and $\mathcal{R}_R \rightarrow 0$, \cite{AdS1}. On the red boundary however, we have a vev-driven solution where $\mathcal{R}_L\rightarrow -\infty$ and $\mathcal{R}_R$ obtains a finite value \cite{AdS1}.

For each QFT on the UV boundaries\footnote{These QFTs correspond to the dual theories on the left and right sides of the conformal interface.}, there are two parameters. $R^{UV}$ which corresponds to the curvature of the boundary metric and is defined by \eqref{yuq6},
and $\varphi_-$ which is the source of the dual operator that couples to the scalar field in the bulk. These parameters can be read from the UV expansions of the fields in the $u$ (Fefferman-Graham) coordinate, i.e. equations \eqref{yuq2} and \eqref{yuq3}. Overall, for each solution, we have three dimensionless parameters $\mathcal{R}_L$, $\mathcal{R}_R$ defined in \eqref{yuq4} and the ratio of the couplings $\xi$
\be \label{xidef}
\xi= \frac{(\varphi_{-}^{L})^{1/\Delta_L}}{(\varphi_{-}^{R})^{1/\Delta_R}}\,.
\ee
We have reviewed the numerical results for the above QFT data in appendix \ref{UVUVS}.

%\vspace{0.3cm}
%%%%%%%%%%%%%%%%%%%%%%%%%%%%%%%%%%%%%%%%%%%%%%%%%%%%%
%%%%%%%%%%%%%%%%%%%%%%%%%%%%%%%%%%%%%%%%%%%%%%%%%%%%%
%%%%%%%%%%%%%%%%%%%%%%%%%%%%%%%%%%%%%%%%%%%%%%%%%%%%%
\subsection{Graviton modes}

\begin{figure}[!ht]
\begin{center}
\begin{subfigure}{0.44\textwidth}
\includegraphics[width=0.94 \textwidth]{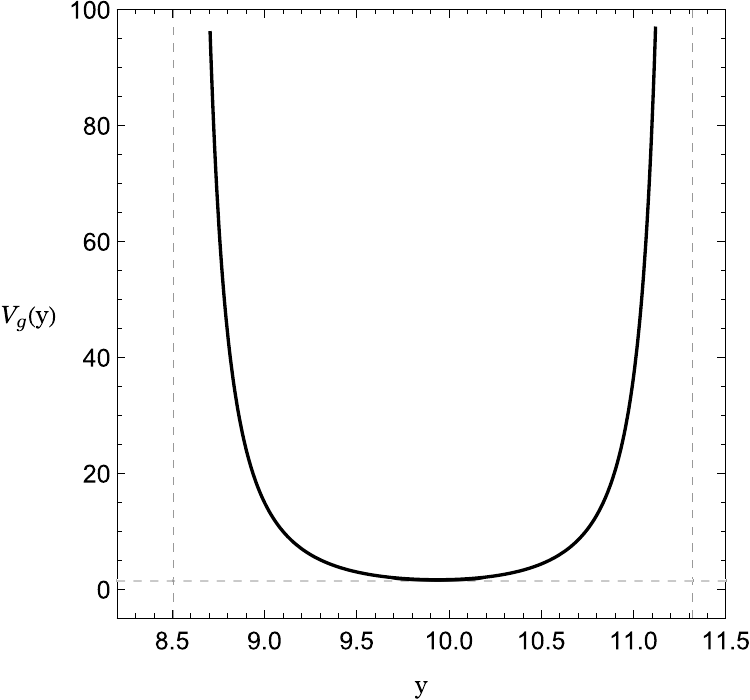}
\caption{\footnotesize{}}\label{potg}
\end{subfigure}
\begin{subfigure}{0.5\textwidth}
\includegraphics[width=1 \textwidth]{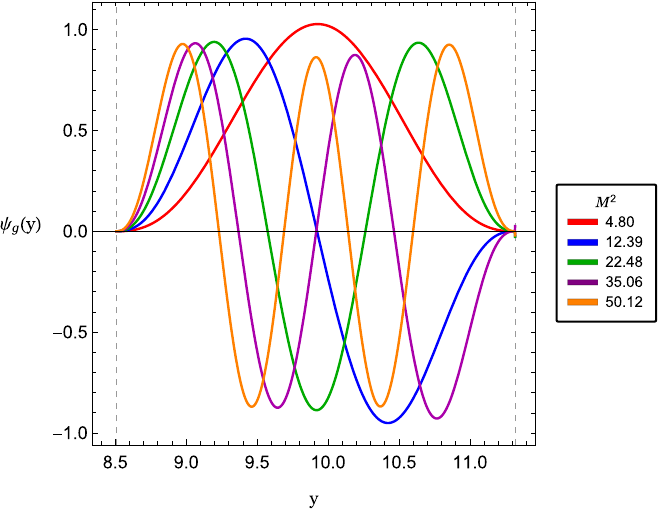}
\caption{\footnotesize{}}\label{first5}
\end{subfigure}
\end{center}
\caption{\footnotesize{(a): The graviton potential \eqref{Vgrav} for the specific solution of figure \ref{AdSA}.  (b): The first five normalizable solutions of the graviton modes. The vertical dashed lines show the locations of the left and right boundaries.}}
\end{figure}

By solving equation \eqref{Sch-grav} we obtain the mass spectrum of the normalizable solutions of the graviton modes. For example, for the specific solution in figures \ref{AdSA} and \ref{AdSfi}, the potential in the Schrodinger-like equation \eqref{Vgrav} is given in figure \ref{potg}. This is a well-shaped potential with a minimum $V_{min}=1.63$ at $y=9.94$ therefore we expect to have a discrete mass spectrum with a mass gap. To see this, we present the first five normalizable solutions $\psi_g(y)$  in figure \ref{first5}, where their normalization condition is given by \eqref{NorMg}.

One can obtain the expansion of the graviton potential \eqref{Vgrav} near the UV boundary (here we have considered that one of the boundaries is at $y=0$) by using the expansion \eqref{Ayexp}
\be\label{VexpG}
V_g(y)=\frac{15}{4 y^2} +
\begin{cases}
-\frac{(\Delta_- -2) \Delta_- ^2 }{(1+ 2 \Delta_-)}\,\varphi_-^2 \,y^{2 \Delta_- -2}+ \mathcal{O}(y^{4 \Delta_- -2}) , \qquad & 0 < \Delta_- < \frac12  \vspace{0.3cm}\\
-\frac{(\Delta_- -2) \Delta_- ^2 }{(1+ 2 \Delta_-)}\,\varphi_-^2 \,y^{2 \Delta_- -2}+ \mathcal{O}(y^0)\, \qquad & \frac12 \leq \Delta_- < 1 \vspace{0.3cm} \\
\frac{1}{12}\mathcal{R}\, \varphi_-^{2/\Delta_- }+ \mathcal{O}(y^{2 \Delta_- -2})\, \qquad & 1 < \Delta_- < 2
\end{cases}\,.
\ee

The leading term for the potential in \eqref{VexpG} always behaves as $\frac{1}{y^2}$, so for every $UV_L-UV_R$ solution, the graviton potential diverges at both boundaries. Consequently, we should expect a discrete mass spectrum for the gravitons.

Knowing the expansion of the potential \eqref{VexpG}, the wave function of the graviton modes can be computed by solving \eqref{Sch-grav}. Near $y=0$, the wave function is
\be\label{Sigrav2}
\psi_g(y)=\big(c_1 \, y^{5/2}+  \cdots\big)+\big(c_2 y^{-3/2} + \cdots\big) \,,
\ee
where $c_1$ and $c_2$ are constants of integration. To have a vev mode, we should choose $c_2=0$. The value of the $c_1$ will be fixed by the normalization condition, once we find the solution for the whole range of the $y$ coordinate.

\begin{figure}[!b]
\begin{center}
\begin{subfigure}{0.49\textwidth}
\includegraphics[width=1 \textwidth]{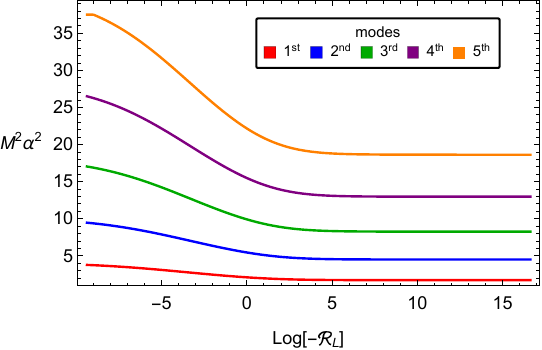}
\caption{\footnotesize{}}\label{pml}
\end{subfigure}
\begin{subfigure}{0.50\textwidth}
\includegraphics[width=1 \textwidth]{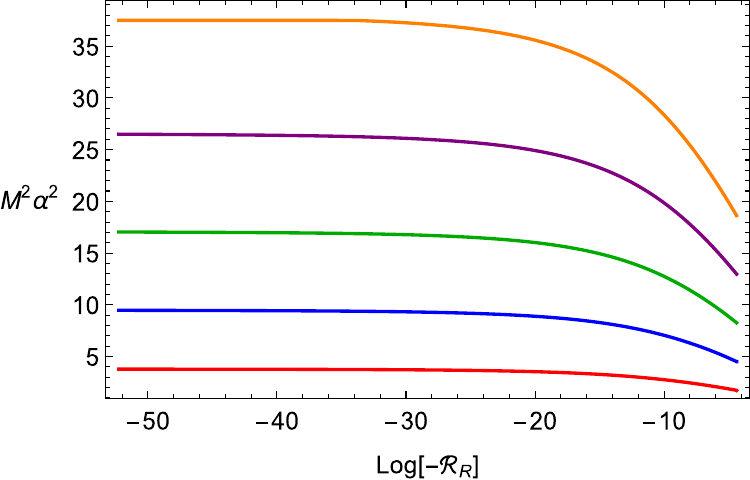}
\caption{\footnotesize{}}\label{pmr}
\end{subfigure}
\end{center}
\caption{\footnotesize{(a), (b): The mass spectrum of the first five graviton modes in terms of dimensionless curvatures for solutions on the dashed line in map \ref{mapm}.}}
\end{figure}

To find out the overall behavior of the mass spectrum for $UV_L-UV_R$ solutions, we consider a fixed value of $\Phi_0$ and all possible values of $S_0$ (that is, all solutions on the dashed line in figure \ref{mapm}). The values of $M^2$ for the first five normalizable modes are shown in figures \eqref{pml} and \eqref{pmr} as a function of the dimensionless curvatures. In both plots, the mass square for each mode monotonically decreases between two finite values by decreasing the dimensionless curvatures. It should be noted that there is a positive lower bound for all the graviton mass spectrum ($M^2 \gtrsim 3.44 $).

\begin{figure}[!t]
	\begin{center}
		\begin{subfigure}{0.6\textwidth}
			\includegraphics[width=1 \textwidth]{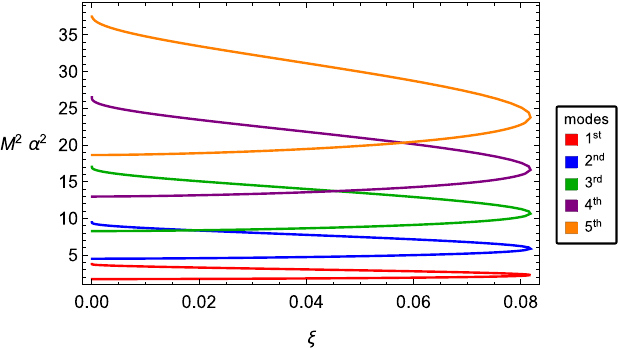}
		\end{subfigure}
	\end{center}
	\caption{\footnotesize{The physical masses as functions of the couplings  $\xi$ defined in \eqref{xidef}.}}\label{gcl}
\end{figure}

We should note that the physical mass of each mode should be measured in units of the QFT length scale.
The QFT length scale in flat space is the relevant coupling. In a curved space, it has two length scales. The extra length scale is the curvature of the manifold on which the QFT is living in the case of a single boundary solution.
 Therefore, in the single boundary case\footnote{Single boundary solutions with negative constant curvature slices exist only if $\Phi$ runs off to the boundaries of field space. Such solutions were studied in
 \cite{AdS3}.}, the physical mass of the theory defined on curvature $R^{UV}$ is given by
\be \label{physm2}
M^2_{phys}= M^2 \a^2 R^{UV}\,,
\ee
where $M$ is the mass which is appeared in \eqref{Sch-grav} and $\a$ is the fiducial length scale of the slice geometry.
If we want to compute the masses in units of the space-time curvature then the relevant ratio is
\be \label{physm22}
\frac{M^2_{phys}}{R_{UV}}= M^2 \a^2=\frac{1}{ {\cal R}} (\f_-)^\frac{-2}{ \Delta_-}~M^2_{phys}\,,
\ee
Indeed, $M\alpha$ gives masses in units of the space-time curvature.

In the case of two-boundary solutions that describe conformal interfaces, the two boundaries have different physical curvatures $R^{UV}_{L,R}$. Therefore, the only sensible way to define a dimensionless spectrum is via the dimensionless combination $M^2\a^2$.

To show the behavior of the physical mass spectrum as a function of QFT boundary data, we present figures  \ref{pml} and \ref{pmr}.
In these figures the horizontal axes have parameters that parameterize the path in solution space indicated by a dashed black line in figure \ref{mapm}. The right side of the plots corresponds to the red boundary in \ref{mapm} while the left side of the plots corresponds to the green boundary in \ref{mapm}.
 The information of the physical parameters near the green and red boundaries are summarized in the table \ref{tQFT} by using the data of the figures in Appendix \ref{UVUVS}.

The behavior of the physical masses as a function of the ratio of the couplings, $\xi$, is depicted in figure (\ref{gcl}).

\begin{table}[!t]
	\centering
	\begin{tabular}{|c|c|c|c|}
		\hline
		Parameter & green boundary  &  red boundary & fig.  \\ \hline
		$\mathcal{R}_L$ & finite & $-\infty$ & \ref{s0rl} \\ \hline
		$\varphi^L_-$ & finite & 0 & \ref{s0fl} \\ \hline
		$R^{UV}_L$ & finite & finite & \ref{s0Rluv} \\ \hline
		$\mathcal{R}_R$ & 0 & finite & \ref{s0rr} \\ \hline
		$\varphi^R_-$ & $+\infty$ & finite & \ref{s0fr} \\ \hline
				$M^2 \a^2$ & finite & finite &\ref{pml},\ref{pmr} \\ \hline
	\end{tabular}
	\caption{\footnotesize{QFT data (the dimensionless curvature, coupling and physical mass of gravitons) as one approaches the green and red boundaries of the map \ref{mapm}.}}
	\label{tQFT}
\end{table}

%\vspace{0.3cm}
%%%%%%%%%%%%%%%%%%%%%%%%%%%%%%%%%%%%%%%%%%%%%%%%%%%%%
%%%%%%%%%%%%%%%%%%%%%%%%%%%%%%%%%%%%%%%%%%%%%%%%%%%%%
%%%%%%%%%%%%%%%%%%%%%%%%%%%%%%%%%%%%%%%%%%%%%%%%%%%%%
\subsection{Scalar modes}

\begin{figure}[!hb]
\begin{center}
\begin{subfigure}{0.44\textwidth}
\includegraphics[width=1 \textwidth]{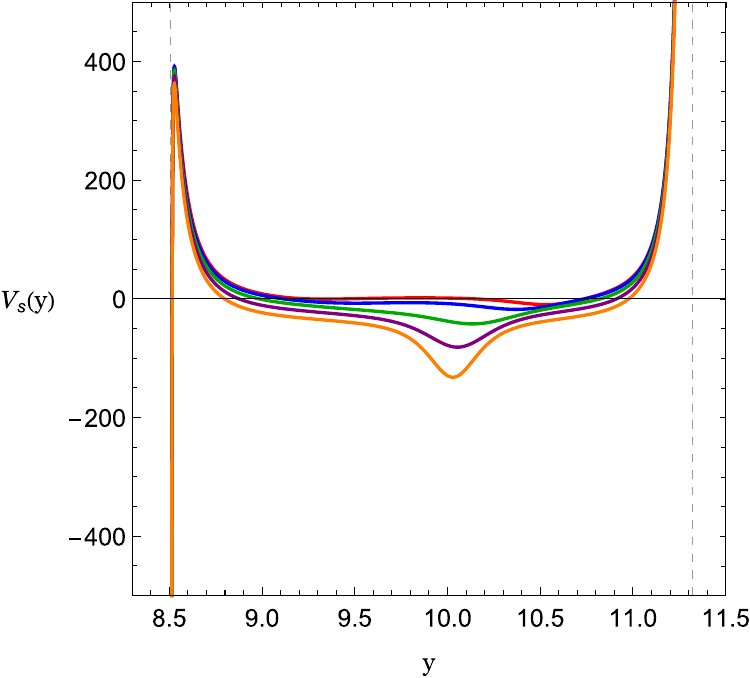}
\caption{\footnotesize{}}\label{Vsisc}
\end{subfigure}
\begin{subfigure}{0.52\textwidth}
\includegraphics[width=1 \textwidth]{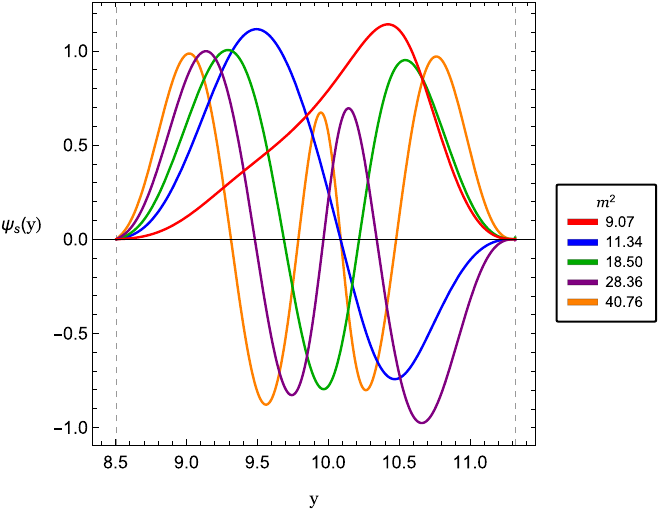}
\caption{\footnotesize{}}\label{Siscs}
\end{subfigure}
\end{center}
\caption{\footnotesize{(a) Potentials and (b) wave functions for the first five normalizable modes of the scalar field fluctuations.}}
\end{figure}

For the scalar field, we solve equation \eqref{SCH} with the potential \eqref{SLpot}. Since the potential depends on the mass, for each mode there is a different potential. The first five potential profiles, and their corresponding wave functions $\psi_s(y)$ are depicted in figures \ref{Vsisc} and \ref{Siscs} respectively. We have imposed the normalization condition \eqref{ncsc} to draw the wave functions.
%\be \label{nsc1}
%\int dy |\psi_s(y)|^2 =1 \,.
%\ee
%%%%%%%%%%%%%%%%%%%%%%%
\begin{figure}[!ht]
\begin{center}
\begin{subfigure}{0.49\textwidth}
\includegraphics[width=1 \textwidth]{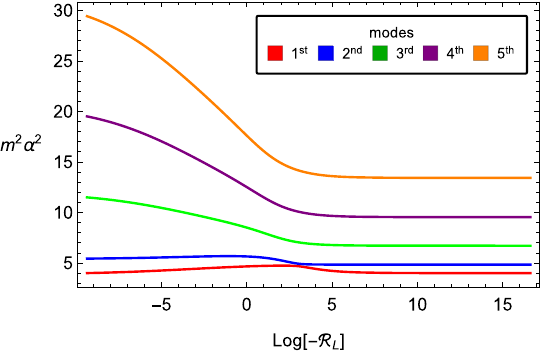}
\caption{\footnotesize{}}\label{scmL}
\end{subfigure}
\begin{subfigure}{0.5\textwidth}
\includegraphics[width=1 \textwidth]{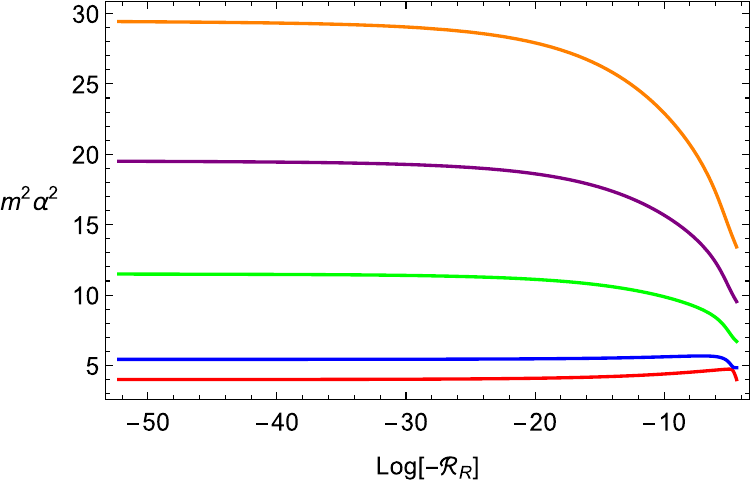}
\caption{\footnotesize{}}\label{scmR}
\end{subfigure}
\end{center}
\caption{\footnotesize{(a), (b): The mass of the first five scalar modes in terms of dimensionless curvatures.}}
\end{figure}

The expansion of the scalar potential near the UV boundaries is (see Appendix \ref{UVexp} for more details)
\be\label{VexpS}
V_s(y)=\frac{1}{4y^2}(2 \Delta_- -5) (2 \Delta_- -3) +
\begin{cases}
V_1 \,y^{2 \Delta_- -2}+ \mathcal{O}(y^{ 4\Delta_--2}) , \qquad & 0 < \Delta_- < \frac 12  \\
V_1\,y^{2 \Delta_- -2}+ \mathcal{O}(y^0)\, \qquad & \frac 12 \leq \Delta_- < 1 \\
V_2+ \mathcal{O}(y^{2 \Delta_- -2})\, \qquad & 1 < \Delta_- < \frac32 \\
V_2+ \mathcal{O}(y^{4-2\Delta_-})\, \qquad & \frac{3}{2} \leq \Delta_- < 2
\end{cases}\,,
\ee
with $V_1$ and $V_2$ are given in \eqref{V1S} and \eqref{V2S}.

%%%%%%%%%%%%%%%%%%%%%%%%%%%%%%%%%%%%%%
As figure \ref{Vsisc} shows, all the scalar potentials, asymptote to $-\infty$ on the left boundary and to $+\infty$ on the right boundary.
The reason comes back to the leading term in the expansion \eqref{VexpS}, and the fact that in our numerical calculations $\Delta_-^{L}=1.6$ and $\Delta_-^{R}=1.1$.

The expansion of the solution of the Schrodinger equation for the scalar modes, \eqref{SCH}, near the UV boundary is
\be \label{sisca}
\psi_s(y) =c_1 \big(y^{\frac52-\Delta_-}+\cdots\big) + c_2 \big(y^{-\frac32+ \Delta_-}+\cdots\big)\,,
\ee
where $c_1$ and $c_2$ are the two constants of integration.

We should note that the normalization condition for the scalar $\hat{\lambda}$ is given by \eqref{ncsc} and we have checked this condition for different scalar modes. Integration on $y$ shows a finite value for each mode.

To show the behavior of the mass spectrum for scalar modes as functions of boundary QFT data, we present figures \ref{scmL} and \ref{scmR}.

\begin{figure}[!t]
\begin{center}
\begin{subfigure}{0.46\textwidth}
\includegraphics[width=1 \textwidth]{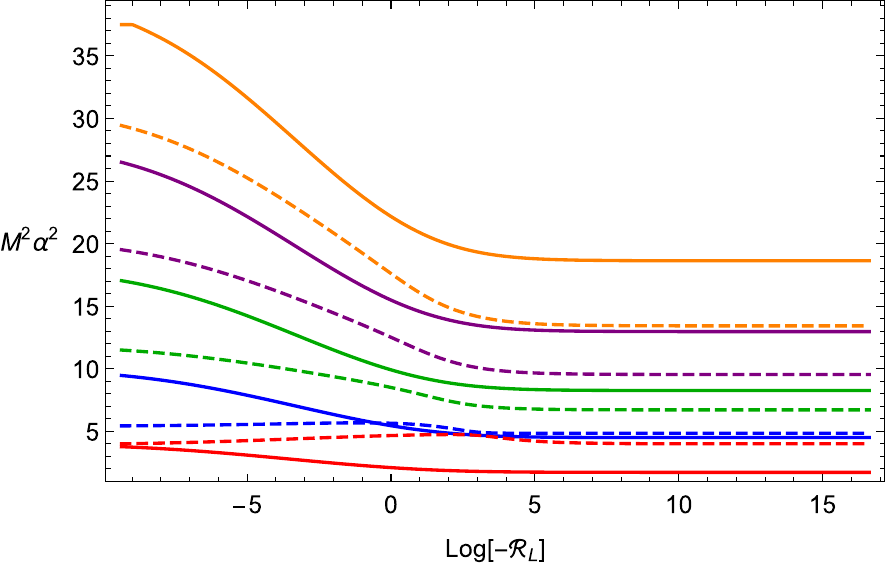}
\caption{\footnotesize{}}\label{PMLgs}
\end{subfigure}
\begin{subfigure}{0.52\textwidth}
\includegraphics[width=1 \textwidth]{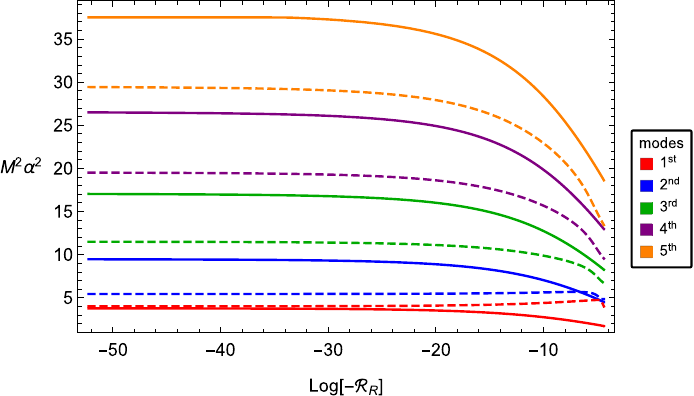}
\caption{\footnotesize{}}\label{PMRgs}
\end{subfigure}
\end{center}
\caption{\footnotesize{The comparison between the graviton mass (solid curves) and scalar mass (dashed curves).}}
\end{figure}

The numerical results show that the scalar field mass spectrum behaves similarly to the tensor mode spectrum. We also present the mass spectrum of the gravitons and scalars in the same plot in figures \ref{PMLgs} and \ref{PMRgs} for comparison.

As it can be seen in both figures, for all the values of $\mathcal{R}_L$ and  $\mathcal{R}_R$, the mass squared of the graviton for the first mode is smaller than that of the scalar field.

We should note that the minimum value for the scalar mass is at $m^2= 8.00$. For the second mode as we move to the right of the plots, for some values of dimensionless curvature, the mass square of the graviton falls so that the curve for the graviton mass squared crosses the one for the scalar and after that, it lies below the scalar mass squared. For higher modes, the graviton mass squared always lies above the scalar one.

This is an interesting effect as for Minkowski flows, it has been observed that the lightest graviton mass is always heavier that lightest scalar mass, \cite{iQCD}. Although, there is no general proof, this observation was made by calculating many different spectra.
On the other hand, we observed here that for non-zero negative curvature the lowest tensor mode can become lighter than the scalar mode.
The implications of this for theories of composite gravity  may be interesting.

\begin{figure}[!t]
\begin{center}
\includegraphics[width=0.5 \textwidth]{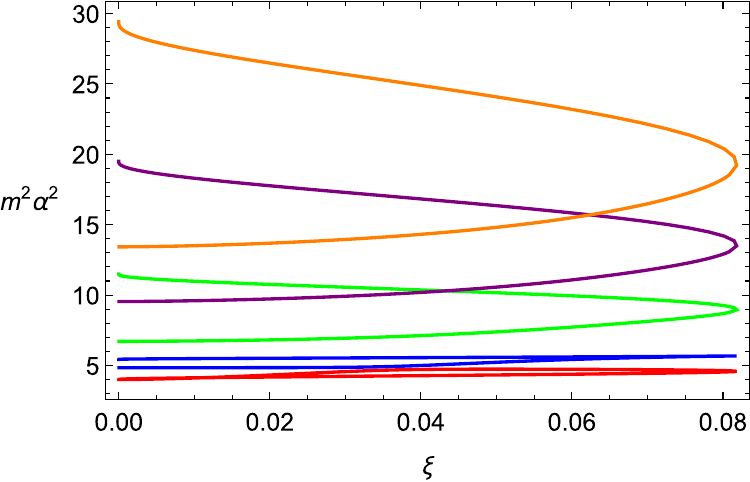}
\end{center}
\caption{\footnotesize{The behavior of the physical mass of the scalar field fluctuations as functions of the couplings  $\xi$ defined in \eqref{xidef}.}}\label{scLz}
\end{figure}

\vspace{0.5cm}
%%%%%%%%%%%%%%%%%%%%%%%%%%%%%%%%%%%%%%%%%%%%%%%%%%%%%
%%%%%%%%%%%%%%%%%%%%%%%%%%%%%%%%%%%%%%%%%%%%%%%%%%%%%
%%%%%%%%%%%%%%%%%%%%%%%%%%%%%%%%%%%%%%%%%%%%%%%%%%%%%
\section{The mass spectrum on dS-sliced solution}\label{msps}

 In this section, we shall consider the mass spectra in holographic theories on positive constant curvature manifolds. Such manifolds include the spheres in the Euclidean case and de Sitter space in Minkowski signature. We shall consider a concrete example where we do our numerical calculations, but the general case is similar.

\begin{figure}[!ht]
	\begin{center}
		\begin{subfigure}{0.45\textwidth}
			\includegraphics[width=1 \textwidth]{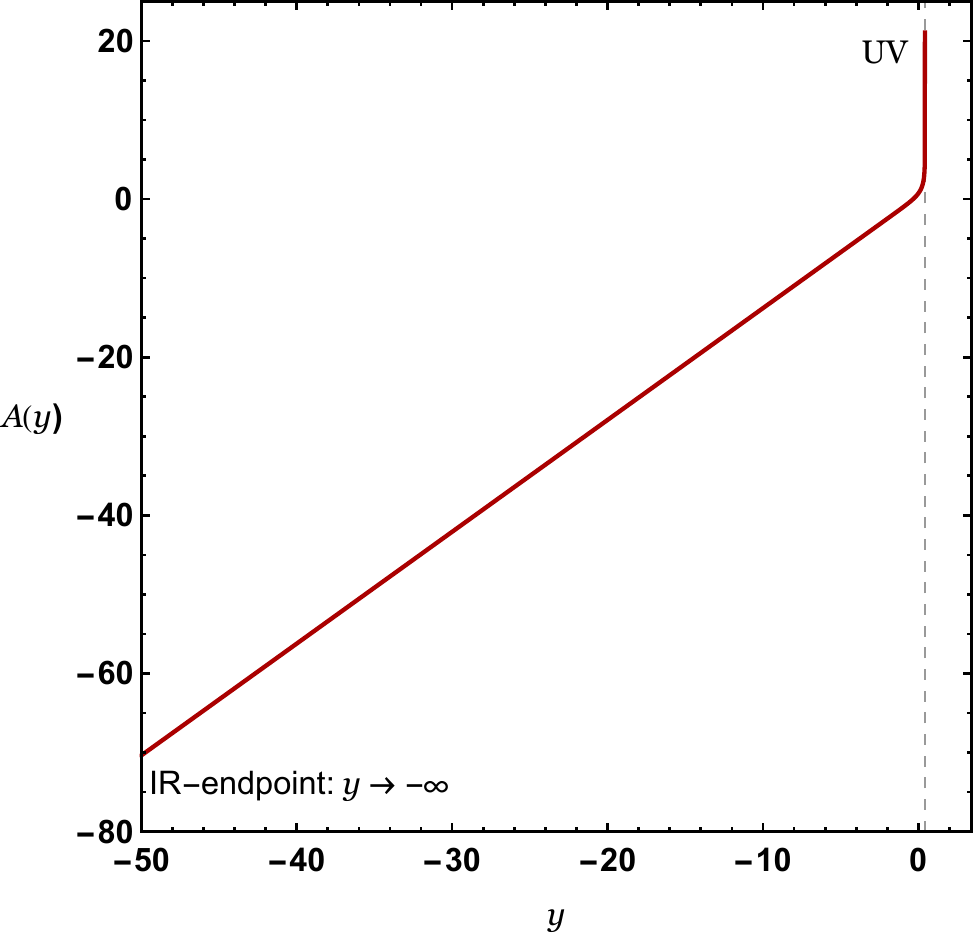}
			\caption{\footnotesize{}}\label{dSA}
		\end{subfigure}
\hspace{0.3cm}
		\begin{subfigure}{0.45\textwidth}
			\includegraphics[width=1 \textwidth]{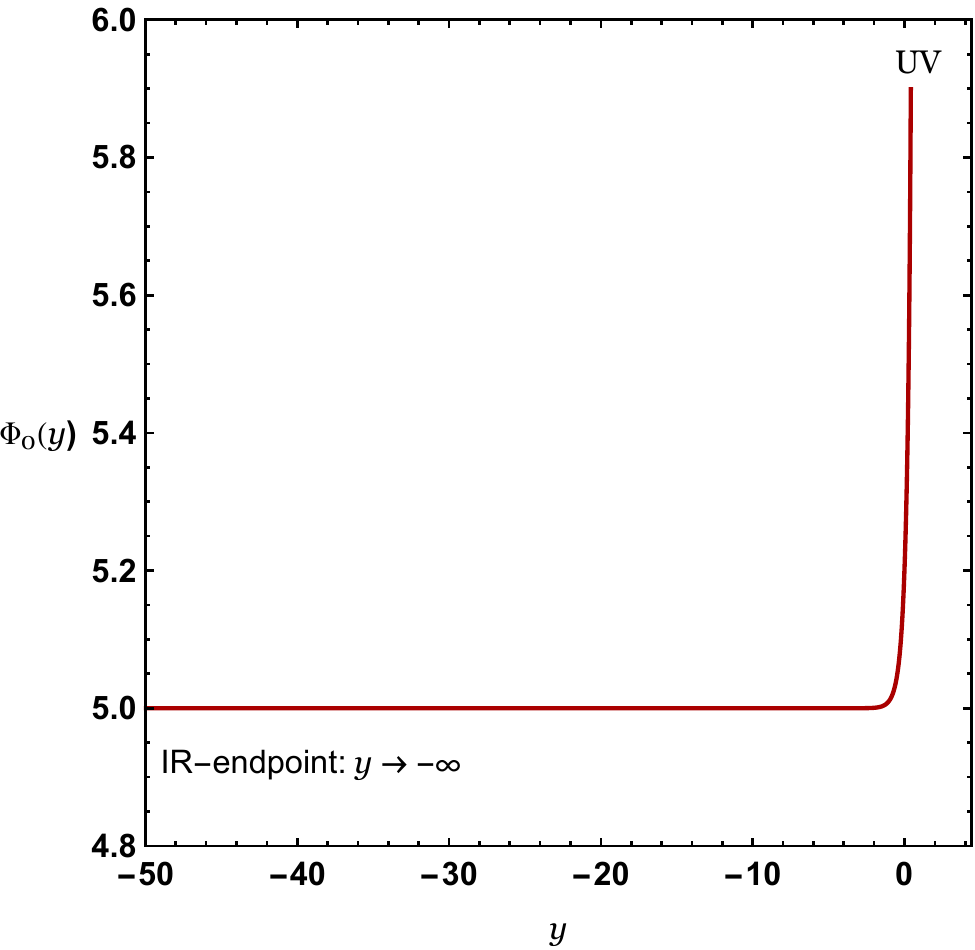}
			\caption{\footnotesize{}}\label{dSfi}
		\end{subfigure}
		\caption{\footnotesize{ $A(y)$ and $\Phi_0(y)$ for a regular UV to IR solution. The IR end-point is located at  $\Phi_0=5$ as $y\rightarrow -\infty$. The UV boundaries is located at the maximum of the potential $\Phi_0=5.89$.}}
	\end{center}
\end{figure}

 We consider the same bulk potential given in equation \eqref{potqua}. The only regular solutions are those that have a UV boundary on one side and an IR end-point on the other side.
As an example of this solution we have depicted $A(y)$ and $\Phi_0(y)$ in figures \ref{dSA} and \ref{dSfi}. In these figure the IR-endpoint is located at $\Phi_0=5$ as $y\rightarrow -\infty$. The UV boundary is at the maxima of the potential at $\Phi_0=5.89$.

\begin{figure}[!t]
\begin{center}
%\begin{subfigure}{0.46\textwidth}
\includegraphics[width=0.45 \textwidth]{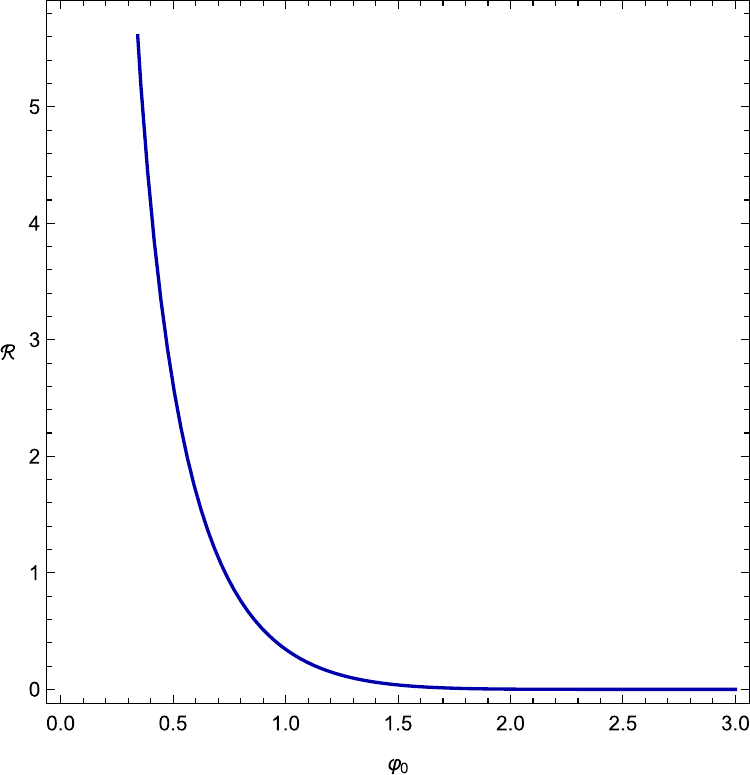}
%\end{subfigure}
\end{center}
\caption{\footnotesize{Dimensionless curvature of the UV boundary vs the position of the IR end-point $\f_0$. The data belong to the solutions with IR end-points between zero and minimum of the potential.}}\label{pos6}
\end{figure}
 Figure \ref{pos6} shows the relation between the dimensionless curvature of the UV boundary ($\mathcal{R}$) and the location of the IR end-point ($\f_0$) for solutions asymptoting to the maximum at $\Phi_0=0$. In this figure, the IR end-point is changing between the maximum of the bulk potential at $\Phi_0=0$ and its minimum at $\Phi_0=3.05$. When the IR end-point approaches the minimum of the potential we have a flat solution ($R^{UV}\rightarrow 0$).

\vspace{0.3cm}
%%%%%%%%%%%%%%%%%%%%%%%%%%%%%%%%%%%%%%%%%%%%%%%%%%%%%
%%%%%%%%%%%%%%%%%%%%%%%%%%%%%%%%%%%%%%%%%%%%%%%%%%%%%
%%%%%%%%%%%%%%%%%%%%%%%%%%%%%%%%%%%%%%%%%%%%%%%%%%%%%
\subsection{Graviton modes}

Knowing the background solution,  we can analyze the mass spectrum of the graviton and scalar field fluctuations.

For regular solutions that stretch between a UV boundary and an IR end-point, the expansion of the potential and the wave function near the UV boundary is the same as in the AdS case. For the potential we have \eqref{VexpG}
\be\label{VexpG2}
V_g(y)=\frac{15}{4 y^2} +
\begin{cases}
-\frac{(\Delta -2) \Delta_- ^2 }{(1+ 2 \Delta_-)}\,\varphi_-^2 \,y^{2 \Delta_- -2}+ \mathcal{O}(y^{4 \Delta_- -2}) , \qquad & 0 < \Delta_- < \frac12  \vspace{0.3cm}\\
-\frac{(\Delta_- -2) \Delta_- ^2 }{(1+ 2 \Delta_-)}\,\varphi_-^2 \,y^{2 \Delta_- -2}+ \mathcal{O}(y^0)\, \qquad & \frac12 \leq \Delta_- < 1 \vspace{0.3cm} \\
\frac{1}{12}\mathcal{R}\, \varphi_-^{2/\Delta_- }+ \mathcal{O}(y^{2 \Delta_- -2})\, \qquad & 1 < \Delta_- < 2
\end{cases}\,,
\ee
and the expansion of the wave function is given by \eqref{Sigrav2}
\be\label{Sigrav22}
\psi_g(y)=\big(c_1 \, y^{5/2}+  \cdots\big)+\big(c_2 y^{-3/2} + \cdots\big) \,,
\ee
where $c_1$ and $c_2$ are the two constants of integration. To have a normalizable mode, we should choose $c_2=0$.

\begin{figure}[!t]
	\begin{center}
	\includegraphics[width=0.5 \textwidth]{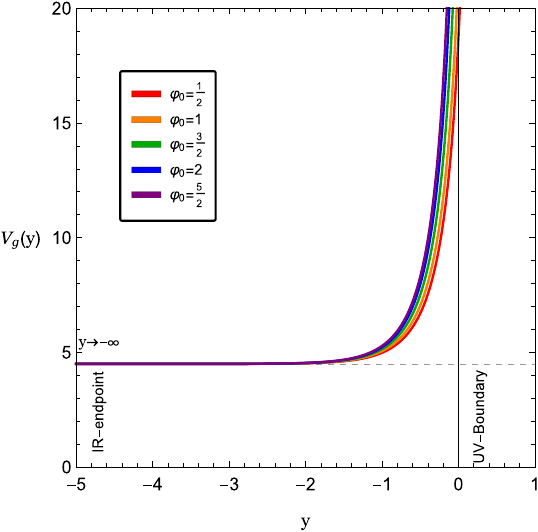}
	\end{center}
	\caption{\footnotesize{ The potential \eqref{Vgrav} of the graviton corresponding to five different values of IR end-point. The IR value of all the potentials tends to
			$\frac{9}{2}$( here we have considered $\a=1/\sqrt{2}$).}}\label{fiveVg}
\end{figure}

Figure \ref{fiveVg} shows the effective graviton potential \eqref{Vgrav} corresponding to five solutions with different IR end-points corresponding to different values of the dimensionless curvature ${\cal R}$ of the dual QFT. All these potentials have the same behavior near the IR end-point ($y\rightarrow -\infty$). To see this, using the expansion of $A(y)$ in \eqref{AyIR}, one can find the expansion of the potential in the Schrodinger-like equation i.e.
\be \label{VgIR}
V_g(y)=\frac{9}{4 \a^2} -\frac{5}{16} \frac{\bar{c}^2}{\a^2}  V_0 e^{\frac{2y}{\a}}+\mathcal{O}(e^{\frac{4y}{\a}})\sp y\to -\infty
\ee
where $\bar{c}$ and $V_0$ are the two constants appearing in the expansion of the $y$ coordinate in terms of $u$ coordinate on \eqref{yurel1} and the expansion of the potential near the IR end-point, i.e. relation \eqref{VIR}.
Since $V_0<0$, the next to leading term increases the potential $V_g$.
Therefore $y\to -\infty$ is always a local minimum of the potential $V_g$.
In our case here the potential $V_g$ turns out to be monotonic as the numerical results show.
 This suggests a mass gap $$M^2\geq \frac{9}{ 4\a^2}\,.$$
We do not know however is this is true in the general case for other bulk potentials.

The solutions of the Schr\"odinger-like equation \eqref{Sch-grav} near the IR end-point when $M^2>9/4\a^2$ are given as a linear combination of an ingoing and outgoing waves.
On the other hand they are exponentially growing or decreasing  when $M^2<9/4\a^2$ i.e.
\be \label{psgds}
\psi_g (y)=\hat{c}_1 e^{- \nu \frac{y}{\a}}+ \hat{c}_2 e^{ \nu  \frac{y}{\a}} + \cdots \sp  \n= \frac{1}{2} \sqrt{9- 4M^2 \a^2}\,.
\ee

We have the following properties for the graviton mass spectrum:

\begin{itemize}
	\item $ M^2> \frac{9}{4\a^2}$
	
	The modes are normalizable. The mass spectrum is continuous and all modes above the mass gap are in the spin-two principal series of $SO(1,4)$. We refer to appendix \ref{reps} for basic information on the different types of representations of the $SO(1,4)$ isometry group of the 4--dimensional de Sitter space.
	
	\item $\frac{2}{\a^2} <M^2\leq \frac{9}{4\a^2}$
	
    The modes are non-normalizable. The upper bound is the mass gap and the lower bound is the Higuchi bound \eqref{HB}. These modes belong the spin-two complementary series.

	\item $0 <M^2\leq \frac{2}{\a^2}$
	
	The modes in this region belongs to the Exceptional series II. For spin two fields there are just two modes in this representation with masses
	\be \label{s2exc}
	M^2\a^2 = 0,2.
	\ee
	These two modes are non-normalizable.
We conclude that the spectrum of normalizable modes is continuous with $M^2>\frac{9}{4\a^2}$.
\end{itemize}

\vspace{0.3cm}
%%%%%%%%%%%%%%%%%%%%%%%%%%%%%%%%%%%%%%%%%%%%%%%%%%%%%
%%%%%%%%%%%%%%%%%%%%%%%%%%%%%%%%%%%%%%%%%%%%%%%%%%%%%
%%%%%%%%%%%%%%%%%%%%%%%%%%%%%%%%%%%%%%%%%%%%%%%%%%%%%
\subsection{Scalar modes}

If we consider the scalar modes, the UV behavior of the potential and wave function will be the same as in the AdS slices. The expansion of the potential near the UV boundary at $y=y_0$ is given by
\be\label{VexpS2}
V_s(y)=\frac{(2 \Delta_- -5) (2 \Delta_- -3)}{4|y-y_0|^2} +
\begin{cases}
V_1 \,|y-y_0|^{2 \Delta_- -2}+ \mathcal{O}|y-y_0|^{ 4\Delta_- -2}  & 0 < \Delta_- < \frac 12  \\
V_1\,|y-y_0|^{2 \Delta_- -2}+ \mathcal{O}|y-y_0|^0\, \qquad & \frac 12 \leq \Delta_- < 1 \\
V_2+ \mathcal{O}|y-y_0|^{2 \Delta_- -2}\, \qquad & 1 < \Delta_- < \frac32 \\
V_2+ \mathcal{O}|y-y_0|^{4-2\Delta_-}\, \qquad & \frac{3}{2} \leq \Delta_- < 2
\end{cases},
\ee
with $V_1$ and $V_2$ are given in \eqref{V1S} and \eqref{V2S}.
The UV behavior of the wave function is
\be \label{sisca2}
\psi_s(y) =c_1 \big(|y-y_0|^{\frac52-\Delta_-}+\cdots\big) + c_2 \big(|y-y_0|^{-\frac32+ \Delta_-}+\cdots\big)\,,
\ee
where $c_1$ and $c_2$ are the two constants of integration.

The IR behavior of the potential is given by the following expansion
\be \label{VscIR}
V_s(y)=\frac{9}{4 \a^2}-m^2-\frac{77}{144} \frac{\bar{c}^2}{\a^2} V_0 e^{\frac{2y}{\a}}+\mathcal{O}(e^{\frac{4y}{\a}})\,.
\ee
Near the IR end-point the wave function is  similar to the wave function of the gravitons i.e.
\be \label{pssds}
\psi_s (y)=\hat{c}_1 e^{- \nu \frac{y}{\a}}+ \hat{c}_2 e^{ \nu  \frac{y}{\a}} + \cdots \sp  \n= \frac{1}{2} \sqrt{9- 4m^2 \a^2}\,,
\ee
where $\hat{c}_1$ and $\hat{c}_2$ are constants of integration.

\begin{figure}[!ht]
	\begin{center}
		\begin{subfigure}{0.47\textwidth}
			\includegraphics[width=1 \textwidth]{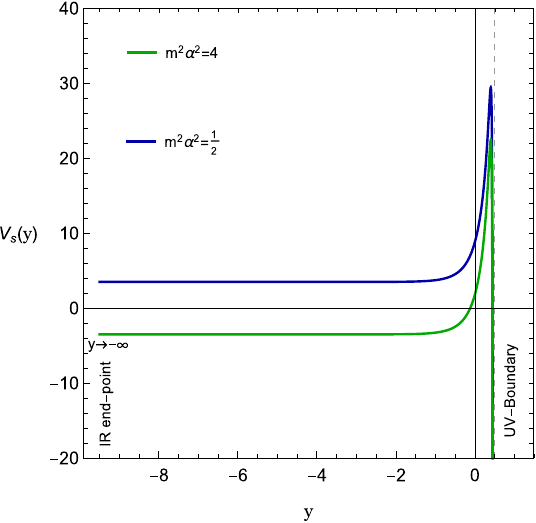}
			\caption{\footnotesize{}}\label{pos4}
		\end{subfigure}
		\begin{subfigure}{0.47\textwidth}
			\includegraphics[width=1 \textwidth]{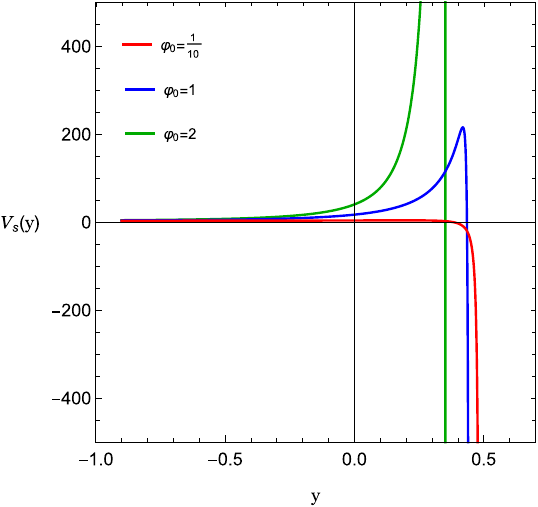}
			\caption{\footnotesize{}}\label{Psc}
		\end{subfigure}
		\begin{subfigure}{0.47\textwidth}
			\includegraphics[width=1 \textwidth]{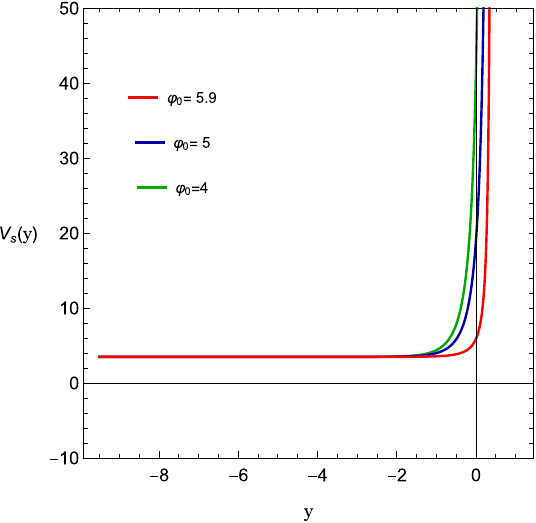}
			\caption{\footnotesize{}}\label{Psr}
		\end{subfigure}
	\end{center}
	\caption{\footnotesize{(a) The scalar potential \eqref{SLpot} for two modes with  $m^2 \a^2 =\frac12,4$ (the IR end-point is at $\f_0=\frac12$). Near the UV-boundary $V_s(y)\rightarrow -\infty$. (b) Scalar potential near the UV boundary for different values for the IR end-point $\f_0$ and for a fixed mass $m^2 \a^2=1$. In both (a) and (b) $\Delta_-=1.6$. (c) Shows three examples of scalar potential when we consider regular solutions with $\Delta_-=1.1$. The location of the UV boundary in the above figures gives the value of $y_0$ in equation \eqref{VexpS2}.
	}}
\end{figure}

To find the behavior of the scalar potential for the regular UV to IR solutions we have sketched figures \ref{pos4} and  \ref{Psc}.
Figure \ref{pos4} shows two examples of the scalar potential when $\frac32<\D_-<2$ (here we have considered  $\Delta_-=1.6$). These two potentials are sketched at a fixed value of the IR end-point\footnote{In this section $\f_0$ stands for the endpoint value of $\Phi_0$ for the flow.} $\f_0=\frac12$. Figure \ref{Psc} shows three examples of the scalar potential at a fixed value of the mass $m^2=1$ but with different values of the IR end-point. In the background potential there is a maximum at $\f=0$ and a minimum at $\f\approx 3$.  According to the figure \ref{Psc} as the IR end-point $\f_0$ moves  away from the UV boundary, the potential finds a peak very near the UV boundary. However, when the IR end-point $\f_0$ is close the UV boundary (the red curve) the peak disappears.

When  $1<\D_-<\frac32$ the scalar potential always goes to positive infinity near the UV boundary, see figure \ref{Psr} for $\Delta_-=1.1$.

 \begin{table}[!t]
	\centering
	\begin{tabular}{|c|c|c|}
		\hline
		& $1<\Delta_-<\frac32$ & $\frac32\leq\Delta_-<2$  \\ \hline
		$V_s$ in the UV limit, $y\rightarrow y_0$ eq. \eqref{VexpS2}		& $+\infty$ & $-\infty$ \\ \hline
		$V_s$ in the IR limit, $y\rightarrow -\infty$	eq. \eqref{VscIR}		& $\frac{9}{4\a^2}-m^2$ & $\frac{9}{4\a^2}-m^2$ \\ \hline
		$\psi_s$ UV boundary conditions	eq. \eqref{sisca2}		& $c_2=0$ & $c_2=0$ \\ \hline
		$\psi_s$ IR boundary conditions	$m^2<\frac{9}{4\a^2}$ eq. \eqref{pssds}		& $\hat{c}_1=0$ & $\hat{c}_1=0$ \\ \hline
		$\psi_s$ IR boundary conditions	$m^2>\frac{9}{4\a^2}$ eq. \eqref{pssds}		& -- & -- \\ \hline
	\end{tabular}
	\caption{\footnotesize{First two rows show a summary of the UV and IR behavior of the scalar potential. The last three rows show the UV and IR boundary conditions that we should impose to find a normalizable solution. The dashed line in last row are due to the fact that at IR end-point we have a linear combination of ingoing and outgoing waves and therefore we do not need to impose a boundary condition.}}\label{tab2}
\end{table}

The scalar wave function is given by solving the equation \eqref{SCH} and the solution depends on the shape of the scalar potential. Specifically we can find the expansion of the wave function near the UV boundary and IR end-point and therefore we can read the boundary conditions. These are summarized in table \ref{tab2} according to the value of $\Delta_-$. Since the scalar potential at IR reaches to the value $\frac{9}{4\a^2}-m^2$ we can split the analysis of the mass spectrum into two sets of modes:

\begin{itemize}
	\item $m^2>\frac{9}{4\a^2}$
	
	For all values of  $1<\Delta_-<2$ there is a continuous mass spectrum. The modes are normalizable provided that we consider the vev boundary condition on the UV boundary. The solution near the IR end-point behaves as a linear combination of ingoing and outgoing waves. All modes are in the scalar principal series of $SO(1,4)$ representations.

	\item $\frac{9}{4\a^2}\geq m^2$
	
	Here we have two different cases:
	
	1) When $1<\Delta_-<\frac32$ at UV boundary $V_s\rightarrow +\infty$ therefore, we can not have any normalizable mode.
	
	2) When $\frac32\leq \Delta_-<2$ at UV boundary $V_s\rightarrow -\infty$ therefore, we should have a normalizable mode if we choose the vev boundary condition on the UV boundary. However, the regularity of the solutions in the IR end-point requires to put $\hat{c}_1=0$ as a boundary condition. In this case, if we find a continuous mass spectrum for masses in  $\frac{9}{4\a^2}\geq m^2>0$ those modes would belong to the scalar complementary series of $SO(1,4)$ representations.

	For special masses with $m^2\a^2 =-j(j+3)\leq 0, \,\, j=0,1,2,\cdots.$ the modes are in the discrete series representations. Except the massless mode, the rest are tachyonic-unstable according to \eqref{tac crim}.

For our example we found no normalizable modes with $m^2<\frac{9}{ 4\a^2}$.

\end{itemize}

%\section{Conclusion and Outlook}

%%%%%%%%%%%%%%%%%%%%%%%%%%%%%%%%%%%%%%%%%%%%%%%%%%%%%
%%%%%%%%%%%%%%%%%%%%%%%%%%%%%%%%%%%%%%%%%%%%%%%%%%%%%
%%%%%%%%%%%%%%%%%%%%%%%%%%%%%%%%%%%%%%%%%%%%%%%%%%%%%

\vspace{0.5cm}
\section*{Acknowledgments}
\addcontentsline{toc}{section}{Acknowledgments}

We would like to thank Dio Anninos, Tarek Anous, Roberto Emparan, Carlos Hoyos, Takaki Ishii, Romuald
Janik, Matti Jarvinen, Javier Mas, Vasilis Niarchos, Angel Paredes, Achilleas Porfyriadis,
Edwan Preau, Alfonso Ramallo, Matthew Roberts, Jorge Russo, Christopher
Rosen, Javier Subils.

This work is partially supported by the European MSCA grant HORIZON MSCA-
2022-PF-01-01 ``BlackHoleChaos" No.101105116 and by the H.F.R.I call ``Basic
research Financing (Horizontal support of all Sciences)" under the National Recovery
and Resilience Plan ``Greece 2.0" funded by the European Union  \\
-NextGenerationEU
(H.F.R.I. Project Number: 15384.),  the In2p3 grant ``Extreme
Dynamics'' and the ANR grant ``XtremeHolo'' (ANR project n.284452)

The work of A.G. and P.M. is supported by Ferdowsi University of Mashhad under the grant 3/58422 (1401/08/07).

\appendix
\renewcommand{\theequation}{\thesection.\arabic{equation}}
\addcontentsline{toc}{section}{Appendix\label{app}}

\section*{Appendix}

%%%%%%%%%%%%%%%%%%%%%%%%%%%%%%%%%%%%%%%%%%%%%%%%%%%%%
%%%%%%%%%%%%%%%%%%%%%%%%%%%%%%%%%%%%%%%%%%%%%%%%%%%%%
%%%%%%%%%%%%%%%%%%%%%%%%%%%%%%%%%%%%%%%%%%%%%%%%%%%%%
\section{Some useful formulae for maximally symmetric slices}

In a four dimensional maximally symmetric space
\be\label{symsp}
R_{\mu\nu\rho\sigma}=\frac{ R^{(\zeta)}}{12}\left( \zeta_{\mu\rho}\zeta_{\nu\sigma }-\zeta_{\mu\sigma}\zeta_{\nu\rho} \right)\sp
R_{\mu\nu}= R^{(\zeta)}_{\mu\nu} = \kappa \zeta_{\mu\nu}\sp R^{(\zeta)}=4\kappa \,,
\ee
We obtain the following useful identities:
	
\begin{gather}
	\left[\nabla_{\m},\nabla_{\n}\right]T_{\r\s}= -{R^\l}_{\r\m\n}T_{\l\s} - {R^\l}_{\s\m\n}T_{\r\l}
\nn \\
	= -\frac{\k}{3} \left(\z_{\r\n} T_{\m\s}
	-\z_{\r\m} T_{\n\s}
	+\z_{\s\n} T_{\r\m}
	-\z_{\s\m} T_{\r\n}
	\right)\,. \label{id1}
\end{gather}

\be \label{id2}
\left[\nabla_{\m},\nabla_{\n}\right]T_{\r}= -{R^\l}_{\r\m\n}T_{\l}
= -\frac{\k}{3} \left(\z_{\r\n}T_\m - \z_{\r\m} T_\n\right)\,.
\ee

\be \label{id3}
\nabla^\n \nabla_\m \nabla^\m T_\n = \left(\nabla_\m \nabla^\m + \k \right) \nabla^\n T_\n\,,
\ee

\be \label{id4}
2 \nabla^\m \nabla_{(\m} T_{\n)}= \left( \nabla_\m \nabla^\m + \k \right)T_\n + \nabla_\n \nabla^\m T_\m\,,
\ee

\be \label{id5}
\nabla_\n  \nabla_\m \nabla^\m s=\left( \nabla_\m \nabla^\m - \k\right)\nabla_\n s\,,
\ee
where $s$ is a scalar field.

\vspace{0.5cm}
\section{Derivation of the perturbations around the background solution}\label{linz}

In this appendix we compute in detail the equations of motion and Lagrangian for the background field fluctuations, used in the main text.

The metric $g_{AB}$ in terms of the background metric $g_{AB}^{(0)}$ and its fluctuations $h_{AB}$ is given by
\be\label{Per1}
ds^2= a^2 (y) g_{AB} dx^A dx^B = a^2 (y) \left(g^{(0)}_{AB} + h_{AB}(y, x^\m) \right)dx^A dx^B\,.
\ee
The inverse and determinant of the metric can be expanded around the background as follows
\be \label{Per2}
g^{AB} = g^{(0)AB} -h^{AB}+ {h^A} _C h^{CB}+\mathcal{O}(h^3)\,,
\ee
\be
\sqrt{-g} = \sqrt{g^{(0)}}\left[1 + \frac{1}{2}h -\frac{1}{4}\left( h_{AB}h^{AB}-\frac{1}{2}h^2\right)+\mathcal{O}(h^3)\right] \sp  h\equiv {h^A} _A \,. \label{Per3}
\ee
To expand the field equations up to the second order in perturbations, we make use of the following relations for the Christoffel connection, Ricci tensor, and the Ricci scalar.

The perturbation of the Christoffel connection is
\be \label{Per4}
\delta\Gamma _{AB} ^C =\Gamma^{(1)}{}^C_{AB}  + \Gamma^{(2)}{}^C_{AB}\,,
\ee
where
\begin{gather}
	\Gamma^{(1)}{}^C_{AB} = \frac{1}{2} g^{(0)CD}\left(\nabla_A h_{DB} + \nabla_B h_{AD}- \nabla_D h_{AB}\right) \,,\label{chris1}
\\
	\Gamma^{(2)}{}^C_{AB} = -\frac{1}{2} g^{(0)CM} g^{(0)DN} h_{MN}\left( \nabla_A h_{DB}+ \nabla_B h_{AD}- \nabla_D h_{AB}\right)\,.\label{chris2}
\end{gather}
Using the perturbations of the Christoffel connection, the Ricci tensor is given by
\be \label{Per5}
 R_{AB}= R^{(0)} _{AB} + R^{(1)} _{AB} + R^{(2)} _{AB}\,,
\ee
in which $ R^{(0)} _{AB} $ is the unperturbed Ricci tensor, and
\be \label{Per6}
 R^{(1)} _{AB} = \nabla_C \Gamma^{(1)}{}^C_{AB} -\nabla_B \Gamma^{(1)}{}^C_{AC} \,,
\ee
\be \label{Per7}
 R^{(2)} _{AB} =  \nabla_C \Gamma^{(2)}{}^C_{AB} - \nabla_B \Gamma^{(2)}{}^C_{AC} +\Gamma^{(1)}{}^M_{AB} \Gamma^{(1)}{}^C_{MC} - \Gamma^{(1)}{}^M_{AC} \Gamma^{(1)}{}^C_{MB} \,.
\ee
The Ricci scalar up to the second order is
\be \label{Per8}
 R= R^{(0)} + R^{(1)} + R^{(2)}\,,
\ee
where
\be \label{Per9}
R^{(0)}= g^{(0)AB} R^{(0)} _{AB}\,,
\ee
\be \label{Per10}
R^{(1)}= g^{(0)AB} R^{(1)} _{AB} - h^{AB} R^{(0)} _{AB}\,,
\ee
\be \label{Per11}
R^{(2)}= g^{(0)AB} R^{(2)} _{AB} - h^{AB} R^{(1)} _{AB}+ {h^A} _C h^{BC} R^{(0)} _{AB}\,.
\ee

\vspace{0.3cm}
%%%%%%%%%%%%%%%%%%%%%%%%%%%%%%%%%%%%%%%%%%%%%%%%%%%%%
%%%%%%%%%%%%%%%%%%%%%%%%%%%%%%%%%%%%%%%%%%%%%%%%%%%%%
%%%%%%%%%%%%%%%%%%%%%%%%%%%%%%%%%%%%%%%%%%%%%%%%%%%%%
\subsection{Equations of motion}

To obtain the equations of motion for background field fluctuations, we separate the scale factor as
\be
\label{Gandg} G_{AB}= a^2 g_{AB}\,.
\ee
The Christoffel connection of the metric is
\be\label{GABC}
{\Gamma^{(G)A}}_{BC}={{\Gamma}^{(g)A}}_{BC}+\d^{A}_{C}\,\pa_B\!\log a +\d^{A}_{B}\,\pa_C\!\log a -g_{BC}g^{AD}\,\pa_D\!\log a \,,
\ee
where ${{\Gamma}^{(g)A}}_{BC}$ is the Christoffel connection for $g_{AB}$.
We have the following relations for Ricci tensor and Ricci scalar

\begin{gather}\label{R1}
	R^{(G)}_{AB} = R^{(g)}_{AB} - g_{AB} {\nabla^{(g)2}} \log a +3\Big[\pa_A\log a\, \pa_B\log a
\nn \\
	 -g_{AB}g^{CD} \pa_C\log a\, \pa_D\log a -\nabla^{(g)}_A\nabla^{(g)}_B\log a \Big]\,,
\end{gather}
\be
R^{(G)} =a^{-2}\Big[R^{(g)}-8{\nabla^{(g)2}} \log a -12g^{AB} \pa_A\log a\, \pa_B\log a \Big]\,.\label{R2}
\ee
We expand all terms in equations  \eqref{R1} and \eqref{R2} up to the first order in fluctuations.
Equation \eqref{R1} has the following expansion\footnote{All terms with superscript $(0)$ refers to the quantities constructed out of the background metric $g_{AB}^{(0)}$. For simplicity in notation, we have removed this superscript for all the covariant derivatives.}
\be\label{lin1}
R^{(G)}_{AB} = R^{(G,0)}_{AB} +R^{(G,1)}_{AB} \,.
\ee
Upon using the perturbed equations \eqref{Per2}--\eqref{Per11} we obtain the following relations
\begin{gather}
	R^{(G,0)}_{AB} = R^{(0)}_{AB} - g^{(0)}_{AB} g^{(0)CD} \nabla_C (a^{-1}\nabla_D a) -3\nabla_A (a^{-1}\nabla_B a)
\nn\\
	-3 g^{(0)}_{AB}a^{-2}\nabla_D a \nabla^D a + 3a^{-2}\nabla_A a \nabla_B a  \,, \label{Per12}
\end{gather}
\begin{gather}
	R^{(G,1)}_{AB} =\frac{1}{2} \nabla_C \nabla_A h^C{}_B + \frac{1}{2} \nabla_C \nabla_B h^C{}_A- \frac{1}{2} \nabla_C \nabla^C h_{AB}-\frac{1}{2} \nabla_B \nabla_A h^C{}_C
\nn\\
	+ \frac{1}{a} g^{(0)}_{AB} g^{(0)CD}\Gamma^{(1)}{}^F_{CD}\nabla_F a+\big(g^{(0)}_{AB} h^{CD} - g^{(0)CD} h_{AB} \big) \nabla_C (a^{-1}\nabla_D a)
\nn\\
	+3 \Gamma^{(1)}{}^C_{AB} (a^{-1}\nabla_C a)- 3a^{-2} h_{AB} \nabla_D a \nabla^D a + 3a^{-2}g^{(0)}_{AB} h^{CD}\nabla_C a \nabla_D a\,.
\label{Per13}
\end{gather}
Also, the expansion of the Ricci scalar \eqref{R2} is
\be\label{lin2}
R^{(G)} = R^{(G,0)} + R^{(G,1)}\,,
\ee
with
\be \label{Per16}
R^{(G,0)}= a^{-2} \Big( R^{(0)} -8 g^{(0)AB} \nabla_A(a^{-1}\nabla_B a) -12g^{(0)}_{AB}a^{-2}\nabla_A a \nabla_B a  \Big)\,,
\ee
\begin{gather} \label{Per17}
	R^{(G,1)}= a^{-2}\Big[ \nabla_A \nabla _B h^{AB} -\nabla^A \nabla _A h -h^{AB}R^{(0)}_{AB} +8 h^{AB} \nabla_A (a^{-1} \nabla_B a)
\nn\\
	+ 4 g^{(0)AB} g^{(0)CD} \big( \nabla_A h_{DB}+ \nabla_B h_{AD} -\nabla_D h_{AB} \big)(a^{-1}\nabla_C a)+12 h^{AB}a^{-2}\nabla_A a \nabla_B a \Big]\,.
\end{gather}
Finally, the expansion of the energy-momentum tensor reads as
\be\label{lin3}
T^{(G)}_{AB}= T^{(G,0)}_{AB}+ T^{(G,1)}_{AB}\,,
\ee
where by using the equation \eqref{emtensor} for the Lagrangian in \eqref{action}
\be \label{Per14}
 T^{(G,0)}_{AB} = -\frac{1}{2}g^{(0)}_{AB}\nabla_C\Phi_0 \nabla^C\Phi_0 -\frac{1}{2}g^{(0)}_{AB} a^2 V(\Phi_0) - \nabla_A \Phi_0 \nabla_B \Phi_0 \,,
\ee
\begin{gather}
	T^{(G,1)}_{AB} =-\frac{1}{2}g^{(0)}_{AB}\nabla_C\Phi_0 \nabla^C\chi + g^{(0)}_{AB} h^{CD} \nabla_C\Phi_0 \nabla_D\Phi_0 -\frac{1}{2}h_{AB} \nabla_C\Phi_0 \nabla^C\Phi_0
\nn\\
	-\frac{1}{2}g^{(0)}_{AB} a^2 \partial_{\Phi_0}V(\Phi_0) \chi-\frac{1}{2}h_{AB} V(\Phi_0)
	+ \nabla_A \Phi_0 \nabla_B \chi +\nabla_B \Phi_0 \nabla_A \chi \,. \label{Per15}
\end{gather}
The equation of motion \eqref{EEin} after the expansion will be
\be  \label{EEinL}
R^{(G,1)}_{AB}-\frac{1}{2}G^{(0)}_{AB} R^{(G,1)}-\frac{1}{2}G^{(1)}_{AB} R^{(G,0)} - T^{(G,1)}_{AB} =0 \,.
\ee
By defining the following relations
\be \label{Achange1}
h_{yy}\rightarrow 2\phi \sp  h_{\m y}\rightarrow
 A_\mu \sp  h_{\m\n} \rightarrow h_{\m\n}\,,
\ee
 the equation of motion \eqref{EEinL} give rise to the following equations, after using the relations \eqref{Per13}, \eqref{Per16}, \eqref{Per17} and \eqref{Per15}
\begin{gather}
	\hspace{-2.5cm}(\mu \nu) \hspace{0.5cm} h_{\mu\nu}'' + 3\frac{a'}{ a} h_{\mu\nu}' + \nabla_\r \nabla^\r h_{\mu\nu} \!-\!2 \nabla^{\rho} \nabla_{(\mu}h_{\nu)\rho} +\nabla_\mu\nabla_\nu h + 2\nabla_\nu\nabla_\mu \phi
\nn\\
	-2 a^{-3}\left(a^3 \nabla_{(\mu} A_{\nu)}\right)'\, + \zeta_{\mu\nu}\Big[-h'' - 3\frac{a'}{a}h' - \nabla_\r \nabla^\r h +
	\nabla^\rho \nabla^\sigma h_{\rho\sigma}
\nn\\
	 -2 \nabla_\r \nabla^\r \phi + 6\frac{a'}{a}\phi' + 6\Big(\frac{a''}{a}+2 \big(\frac{a'}{a}\big)^2\Big)\phi  -2 a^{-3}\left(a^3 \Phi_0' \chi\right)'
 \nn\\
	+ 2 a^{-3}\left(a^3 \nabla_{\rho} A^{\rho}\right)'\Big]- \kappa \zeta_{\mu\nu}( 2\phi + h \big) + 2 \kappa h_{\mu\nu} =0\,,\label{Amunu1}
\end{gather}
\be\label{Amuy1}
(\mu y)  \hspace{0.5cm} \nabla_\n \nabla^\n A_\mu - \nabla^{\nu} h_{\nu\mu}'  +\nabla_{\mu}\Big[ - 6\frac{a'}{a}\phi + 2\Phi_0' \chi + h' - \nabla^{\nu} A_\nu\Big] + \kappa A_{\mu} =0 \,,
\ee
\begin{gather}
	\hspace{-2cm}(y y) \hspace{0.5cm}   -\nabla_\m \nabla^\m h + \nabla^{\mu}\nabla^\nu h_{\mu\nu} - 3\frac{a'}{a}h' + 6\Big(\frac{a''}{a}+2 \big(\frac{a'}{a}\big)^2\Big)\phi + 6 \frac{a'}{a} \nabla^\nu A_\nu
\nn\\
	+ 4 \Phi_0' \chi'- 2 a^{-3}\left(a^3 \Phi_0' \chi\right)'  +\kappa(2 \phi -h)= 0 \,.\label{Ayy1}
\end{gather}
In deriving the above equations, we have considered that $a=a(y)$. A prime denotes the derivative with respect to $y$.

By linearizing the scalar equation of motion \eqref{EPhi} we find
\be\label{ADil1}
\chi''  +  \nabla_\m \nabla^\m \chi+ 3 \frac{a'}{a}\chi' - \frac{1}{2}a^2 \partial^2_{\Phi_0}V \chi
  -\frac{2}{a^3} \left(a^3 \Phi_0' \phi\right)' +\Phi_0'\phi' + \frac{1}{2}\Phi_0'h'   -
 \Phi_0' \nabla^\mu A_\mu=0\,.
\ee

\vspace{0.3cm}
%%%%%%%%%%%%%%%%%%%%%%%%%%%%%%%%%%%%%%%%%%%%%%%%%%%%%
%%%%%%%%%%%%%%%%%%%%%%%%%%%%%%%%%%%%%%%%%%%%%%%%%%%%%
%%%%%%%%%%%%%%%%%%%%%%%%%%%%%%%%%%%%%%%%%%%%%%%%%%%%%
\subsection{Action}\label{ApB}

As mentioned earlier, another way to obtain the linearized field equations, is to expand the action \eqref{action} up to the second order in perturbations, and then use the quadratic terms to find the field equations for fluctuations. To obtain this we use $G_{AB}=a^2 g_{AB}$ in
\be \label{acta}
S=\frac{1}{2\kappa_5^2}\int d^5x\, \sqrt{-G}\left(R^{(G)}-\partial_A\Phi \partial^A\Phi-V(\Phi)\right)\,,
\ee
to find
%\be \label{qual1}
%\sqrt{-G} R^{(G)} = a^3 \sqrt{-g}\Big[R^{(g)} -8g^{AB}\nabla_A \nabla_B  \log a -12 g^{AB} (\nabla_A \log a)( \nabla_B \log a )\Big]\,.
%\ee
% Then the action \eqref{action} takes the form
\begin{gather}
	S=\frac{1}{2\k_5^2}\int d^4 x dy\, a^3\sqrt{-g}\bigg[ R^{(g)} + 12 a^{-2} g^{AB}\partial_A a \partial_B a  -8 a^{-3} \nabla_A\left(a^3 g^{AB} \nabla_B \log a \right)
\nn\\
	- g^{AB}\partial_A \Phi \partial_B \Phi - a^2 V(\Phi)\bigg]\,.\label{aact}
\end{gather}
In the above, and all the rest of this section we treat the scale factor $a$ as a general scalar field and therefore this defines the action of the appropriate covariant derivatives here and below.
Of course for our purposes, in the end $a(y)$ will a function of the Fefferman-Graham coordinate $y$.

If we expand the above action up to the second order of perturbations, we obtain
\be
S_{pert} = S^{(1)}+S^{(2)}\,,
\ee
where after some integration by parts, we find
\begin{gather}
	S^{(1)}=\frac{1}{2\kappa_5^2}\int d^5x\, \sqrt{-g^{(0)}}\bigg\{ a^3 \Big[ \frac12 h R^{(\z)} - h^{AB} R^{(0)}_{AB}+  \nabla_{A} \nabla_{B} h^{AB}-  \nabla_{A}\nabla^{A} h
\nn\\
	-2  \nabla_{A} \Phi_0 \nabla^{A}\chi + h^{AB}  \nabla_{A} \Phi_0  \nabla_{B}\Phi_0 -\frac12 h  \nabla_{A}\Phi_0  \nabla^{A} \Phi_0 	-a^2 \chi \p_{\Phi_0}V - \frac12 a^2 h V(\Phi_0)   \Big]
\nn\\
 - 6 h^{AB} \nabla_{A}a^2 \nabla_{B}a
	+ 3 h  \nabla_{A}a^2  \nabla^{A} a + 8  \nabla_{A}(a^2 h^{AB} \nabla_{B}a)  \bigg\}\,. \label{Lac1}
\end{gather}
This action can be written as a total derivative term
\begin{gather}
	S^{(1)}_{TD}=\frac{1}{2\kappa_5^2}\int d^5x\, \sqrt{-g^{(0)}} \nabla_{A} \Big( a^3 \nabla_{B} h^{AB} -   a^3 \nabla^{A} h
\nn\\
	+3 a^2h \nabla^{A}a+ 5a^2 h^{AB}  \nabla_{B} a -2  a^3 \chi  \nabla^{A} \Phi_0  \Big)\,.\label{Lac2}
\end{gather}
and
\begin{gather}
	S^{(2)}=\frac{1}{2\kappa_5^2}\int d^5x\, \sqrt{-g^{(0)}}\bigg\{ a^3 \bigg[-\frac{1}{4} \nabla_C h_{AB} \nabla^C h^{AB}+ \frac{1}{2}\nabla ^B h_{AB} \nabla_C h^{AC}
\nn\\
 	-\frac{1}{2}\nabla_A h^{AB} \nabla_B h+ \frac{1}{4}\nabla_A h \nabla^A h+ \chi \nabla_A h \nabla^A \Phi_0 + 2 h^ {AB} \nabla_A \Phi_0 \nabla_B \chi - \nabla_A \chi \nabla^A \chi
 \nn\\
	 - \frac{1}{2}a^2\chi^2 \partial^2_{\Phi_0} V(\Phi_0) \bigg]-(\nabla_A a^3) \Big(\frac{1}{2} {h^A}_B\nabla^B h +  h_{BC}\nabla^C h^{AB}\Big)-\frac{1}{2}(\nabla^C \nabla_A a^3) h^{AB} h_{BC}
\nn\\
  + \frac{1}{2}  a^3 R^{(0)}_{ACBD}  h^{AB}  h^{CD}
  +\frac{1}{4} a^3 R^{(\zeta)}\Big({h^C}_y {h^y}_C -\frac{1}{2}h h_{yy}\Big) +\frac12 a^3 R^{(0)}_{AB}\Big( {h^A}_C h^{BC}-h h^{AB}\Big)
\nn\\
	 -\frac{1}{2} a^3 R^{(\zeta)}\Big( \frac{1}{4}h^{AB}h_{AB}-\frac{1}{8}h^2\Big)\bigg\} + S^{(2)}_{TD} \,,\label{squad1}
\end{gather}
where $ S^{(2)}_{TD}$ includes the total derivative terms
\be \label{totder1}
S^{(2)}_{TD}=\frac{1}{2\kappa_5^2}\int d^5x\, \sqrt{-g^{(0)}} \nabla _A Q^A \,,
\ee
with
\begin{gather}
	Q^A= - \frac{1}{4} a^2 \Bigl(3 h (- h \nabla^{A}a + 2 h^{A}{}_{B} \nabla^{B}a) + 2 h_{BC} (3 h^{BC} \nabla^{A}a + 7 h^{AC} \nabla^{B}a)
\nn\\
	+ 2 a \bigl(h^{BC} (-2 \nabla^{A}h_{BC} + \nabla_{C}h^{A}{}_{B}) + h (\nabla^{A}h + 2 \chi \nabla^{A} \Phi_0 -  \nabla_{C}h^{AC})
\nn\\
	+ h^{AB} (-2 \nabla_{B}h + 3 \nabla_{C}h_{B}{}^{C})\bigr)\Bigr)\,.\label{totder2}
\end{gather}
In deriving the actions in \eqref{Lac2}--\eqref{totder1} we have utilized the following background equations of motion
\be \label{Eom1}
{\Phi_0'}^2 =-\frac{1}{4} R^{(\zeta)} -3\left(\frac{a''}{a} -2 \frac{a'^2}{a^2}\right)=-\kappa -3\left(\frac{a''}{a} -2 \frac{a'^2}{a^2}\right)\,,
\ee
\be  \label{Eom2}
a^2 V(\Phi_0) =\frac{3}{4} R^{(\zeta)} -3 \left( \frac{a''}{a} + 2 \frac{a'^2}{a^2}\right)\,,
\ee
\be   \label{Eom3}
\partial_{\Phi_0} V(\Phi_0)=6 a^{-3} a' \Phi_0' + 2a^{-2} \Phi''_0\,.
\ee

\vspace{0.3cm}
%%%%%%%%%%%%%%%%%%%%%%%%%%%%%%%%%%%%%%%%%%%%%%%%%%%%%
%%%%%%%%%%%%%%%%%%%%%%%%%%%%%%%%%%%%%%%%%%%%%%%%%%%%%
%%%%%%%%%%%%%%%%%%%%%%%%%%%%%%%%%%%%%%%%%%%%%%%%%%%%%
\subsection{The boundary terms}

In the previous subsection we have found some total derivative terms in \eqref{Lac2} and \eqref{totder1}. In addition to these boundary terms, we should include the Gibbons-Hawking-York (GHY) term on the boundaries to see what boundary condition leads to a well-posed action. The GHY term is
\be \label{GHY}
S_{\text{GHY}}= \frac{1}{\kappa_5^2}\int d^4x\, \sqrt{-\gamma^{(G)}}\, K^{(G)} \Big|_{B}\,,
\ee
where $B$ denotes the regulated boundary and according to the relation \eqref{Gandg} for the metric, the induced metric on this regulated boundary, $\gamma^{(G)}_{AB}$ is
\be \label{ind1}
\gamma^{(G)}_{AB}= a^2 \gamma^{(g)}_{AB}\Big|_{B} \,.
\ee
 In general,  the regulated boundary has one component in the case of flat and sphere slices, one space-like and two time-like components in the case where the slice is global de Sitter space, three space-like components when the slice manifold is a non-compact negative curvature manifold with boundary, and two components when the slice manifold is a compact negative curvature manifold.

The induced metric is defined in terms of the unit normal vector $n_A$, describing the embedding of the slices in the five dimensional space-time
\be \label{ind2}
\gamma^{(g)}_{AB}= g_{AB}- n_A n_B\,.
\ee
The extrinsic curvature  is also given by
\be \label{excur1}
 K^{(G)}_{AB}
 = \gamma^{(G)C}_{A} \gamma^{(G)D}_{B} \nabla^{(G)}_C n_D
 = a (K^{(g)}_{AB}+\gamma^{(g)}_{AB} a^{-1} n^C \nabla_C a ) \,,
\ee
with
\be \label{excur2}
K^{(g)}_{AB}= \gamma^{(g)C}_{A} \gamma^{(g)D}_{B} \nabla_C n_D \,.
\ee
To first order in perturbations, the GHY term \eqref{GHY} is
\begin{gather}
	S^{(1)}_{\text{GHY}}= \frac{1}{\kappa_5^2}\sum_{B_i}\int d^4x\,\sqrt{-\gamma^{(g)}} a^3 \Big( h K^{(g)}- h^{AB}K^{(g)}_{AB} - n^A n^B h_{AB} K^{(g)} - n^A \nabla^B h_{AB}
\nn\\
	 + n^{A} \nabla_A h + 4 a^{-1} h n^A \nabla_A a - 8 a^{-1} h^{AB} n_A \nabla_B a - \mathcal{D}_{A}(\gamma^{(g)AB} h_{BC} n^C) \Big)\Big|_{B_i}\,. \label{LGH}
\end{gather}
where $B_i$ is any component of the total boundary.

In the last line $\mathcal{D}$ is the covariant derivative with respect to the  metric $\g_{AB}^{g}$ defined in equation (\ref{ind2}) on the regulated boundary.

By writing \eqref{Lac2} as an action on the regulated boundary i.e.
\begin{gather}
	S^{(1)}_{TD}=\frac{1}{2\kappa_5^2}\sum_{B_i}\int d^4x\, \sqrt{\gamma^{(g)}} n_A\Big( a^3 \nabla_{B} h^{AB} -   a^3 \nabla^{A} h
\nn\\
	+3 a^2h \nabla^{A}a+ 5a^2 h^{AB}  \nabla_{B} a -2  a^3 \chi  \nabla^{A} \Phi_0  \Big)\Big|_{B_i}\,,\label{Lac3}
\end{gather}
and adding the linear GHY action \eqref{LGH}
\begin{gather}
	S_{TD}^{(1)} + S^{(1)}_{\text{GHY}}=\frac{1}{\kappa_5^2}\sum_{B_i}\int d^4x\,\sqrt{-\gamma^{(g)}} a^3 \Big( h K^{(g)}- h^{AB}K^{(g)}_{AB} - n^A n^B h_{AB} K^{(g)}
\nn\\
	+ 7 a^{-1} h n^A \nabla_A a -3  a^{-1} h^{AB} n_A \nabla_B a-2 \chi n^A \nabla_A \Phi_0  - \mathcal{D}_{A}(\gamma^{(g)AB} h_{BC} n^C) \Big)\Big|_{B_i}\,, \label{surfac1}
\end{gather}
which can be written as
\begin{gather}
	S_{reg-bond} = \frac{1}{\kappa_5^2}\sum_{B_i}\int d^4x\,\sqrt{-\gamma^{(g)}} a^3 \Big(\big( \gamma^{(g)AB} K^{(g)}-K^{(g)AB} + 7 \ell^3 a^{-1} g^{AB} n^C \nabla_C a
\nn\\
	-3  a^{-1}  n^A \nabla^B a \big) h_{AB} -(2  n^A \nabla_A \Phi_0)\chi  - \mathcal{D}_{A}(\gamma^{(g)AB} h_{BC} n^C) \Big)\Big|_{B_i}\,. \label{surfac2}
\end{gather}
The last term in \eqref{surfac2} is a total derivative.

\begin{itemize}
	
\item standard AdS boundaries

We have two standard AdS boundaries at $y=0$ and $y=\a\pi$ with unit normal vectors $n_A=\mp (1 , 0 , \vec{0})$.

In addition to the action in \eqref{surfac2} we should consider the boundary counter-terms (see for example \cite{C}) as well
\be\label{cont}
	S_c = \frac{1}{\k_5^2}\int d^4x \sqrt{-\g^{(g)}}a^4 \left(-\frac{6}{\ell}-\frac{\ell}{2}a^{-2} R^{\g^{(g)}}+\mathcal{O}({R^{\g^{(g)}}})^2\right)\,,
\ee
where $\g^{(g)}$ is given by
\be
ds^2_{\g^{(g)}} = (\zeta_{\m\n}+h_{\m\n}) dx^\m dx^\n\,.
\ee
If we put the regulated standard AdS boundary for example at $y=\e$ with $\e\to 0$, then
\be \label{stb1}
h_{\m\n} = \langle \delta T_{\m\n}(x) \rangle y^{4}+\cdots \big|_{y=\epsilon}  \sp   \chi = \langle O(x) \rangle y^{\Delta_+}+\cdots \big|_{y=\epsilon}\,,
\ee
as they are vev perturbations.
Here one can check that since $a\sim \e^{-1}$ then on $y=\e$ regulated AdS boundary
\be \label{stb2}
\lim_{\e\rightarrow 0} S_{y=\e} = \frac{\ell^3}{\kappa_5^2}\int d^4x \sqrt{-\gamma^{(g)}} \Big( 4\langle {\delta T^{\m}}_\m (x) \rangle  + 2 \f_- \Delta_- \langle O(x) \rangle\Big)\,.
\ee
To derive the above result we have summed over \eqref{surfac2} and \eqref{cont}.

\item side boundary

Considering the slice metric in Poincare coordinates, the bulk metric is
\be \label{sid1}
ds^2 = a(y)^2 \big(dy^2+\frac{\a^2}{z^2}(dz^2+\eta_{ij}dx^i dx^j)\big)\,,
\ee

The normal vector to the regulated side boundary at $z=\epsilon$  is given by
\be \label{sid2}
n_A = (0,\frac{\a}{\epsilon},\vec{0})\,,
\ee
therefore the extrinsic curvature defined in \eqref{excur2} is given by
\be \label{sid3}
K^{(g)}_{yy}= 0 \sp K^{(g)}_{ij}=-\frac{\a}{\e^2}\eta_{ij} \quad \rightarrow \quad K^{(g)} = -\frac{3}{\a}\,.
\ee

Considering the above values, the regulated boundary action in \eqref{surfac2} is given by
\be
	S_{z=\e} = \frac{1}{\kappa_5^2}\int dy d^3x \left(-a^3(3\frac{\a^2}{\e^3}h_{yy}+\frac{2}{\e}\eta^{ij}h_{ij})-3a^2a' \frac{\a^2}{\e^2}h_{zy}\right)\,.\label{sid4}
\ee
According to the analysis in appendix \ref{ApLaplace} near the regulated boundary at $z=\e$ we have the following behaviors:

From equations \eqref{AI1} and \eqref{AI2} for scalar vev perturbation with $m^2>0$
\be \label{sid5}
\frac{1}{\e^3}h_{yy}=\frac{2}{\e^3}\phi \sim  \e^{\frac32(\sqrt{1+\frac{4m^2\a^2}{9}}-1)}\rightarrow 0 \,.
\ee
From equation \eqref{AdS hij sol} and definitions in \eqref{hk} and \eqref{nuM} for graviton vev perturbation above the BF bound $ M^2\geq -\frac{9}{4\a^2}$ we find
\be \label{sid6}
\frac{1}{\e}h_{ij} \sim  \e^{\frac32(\sqrt{1+\frac{4m^2\a^2}{9}}-1)}\rightarrow 0\,.
\ee
Moreover, for the vector field in  our theory with mass $m^2=-2\k=\frac{6}{\a^2}$ from \eqref{Az sol} and \eqref{nud} one leads to
\be \label{sid7}
\frac{1}{\e^2}h_{zy}=\frac{1}{\e^2}A_z \sim  \e^2\rightarrow 0\,.
\ee

\end{itemize}

\vspace{0.5cm}
%%%%%%%%%%%%%%%%%%%%%%%%%%%%%%%%%%%%%%%%%%%%%%%%%%%%%
%%%%%%%%%%%%%%%%%%%%%%%%%%%%%%%%%%%%%%%%%%%%%%%%%%%%%
%%%%%%%%%%%%%%%%%%%%%%%%%%%%%%%%%%%%%%%%%%%%%%%%%%%%%
\section{Simplified equations and decoupling of the modes} \label{Apgz}

In this appendix we provide details on the derivation of the tensor and scalar equations in section \ref{flu}.

To find a simpler form for equations of motion in \eqref{munu}--\eqref{dilaton}, we introduce the following new field variables
\begin{gather}\label{chv1ap}
	B^T_\m\equiv A^T _\m  - {V^T_\m}'\,,
\\
	\lambda\equiv \psi-\frac{\chi}{z}\,, \label{chv2ap}
\\
	\gamma \equiv W-E' -\frac{a \chi}{a'z }\,, \label{chv3ap}
\\
	\tau \equiv 2 \psi+\phi-\frac{1}{a^3}\big(a^3(W-E')\big)'\,.\label{chv4ap}
\\
	\sigma \equiv 2\frac{a'}{a}\lambda+\lambda' + \frac{\kappa}{3}\gamma-\frac{a'}{a}\Big(\frac{1}{a^3}(a^3 \gamma)' \Big)\,.\label{chv5ap}
\end{gather}
As we already argued in section \ref{simp eq} the new fields defined above are gauge invariant under the five dimensional diffeomorphisms i.e.
\be
\delta \lambda=\delta \g=\delta\tau=\delta \sigma =0
\sp \delta B_\m^T = 0\,.
\label{nSdif}
\ee
In section \ref{Dec} we showed that the decomposition of the fields is reliable if we drop the zero modes of the scalar and vector fields. Considering this, we shall find the equations of motion for new fields as follows:

\vspace{0.5cm}
$\bullet$ Starting from the $\m y $ component of the Einstein equations \eqref{muy} we have
\be
(\nabla_\n \nabla^\n + \k){B}^T_\m +6\nabla_\m \Big( {\sigma} -\frac{a'}{a}{\tau} \Big)=0\,. \label{muy2a}
\ee
Taking  the divergence of the above equation and using  the identity $\nabla^\m  \nabla_\n \nabla^\n {B}^T_\m =0$ we obtain
\be\label{muy3}
\nabla_\m \nabla^\m  ({\sigma}-\frac{a'}{a}{\tau})=0\,.
\ee
Since ${\sigma}$ and ${\tau}$ do not contain the zero mode of the Laplace operator, we conclude that
\be\label{S0}
\frac{a'}{a}{\tau}-{\sigma}=0\,.
\ee
The above equation in terms of the original fields becomes
\be \label{As1ap}
0={\psi}' - \frac{a'}{a}{\phi}  +\frac13\kappa({W}-{E}') + \frac13 \Phi'_0 {\chi}\,.
\ee
Substituting \eqref{S0} into \eqref{muy2a} gives
\be\label{BTZ}
(\nabla_\n \nabla^\n + \k) {B}^T_\m=0\rightarrow {B}^T_\m=0\,,
\ee
where the last equality is due to the fact that we have considered ${B}^T_\m$ to does not have the zero mode of the $\Box+ \k$ operator.
In terms of the original fields equation \eqref{BTZ} means that
\be \label{ATVap}
{A^T_\m}= {V^T_\m}' \,.
\ee
\vspace{0.3cm}
$\bullet$
If we begin with the $\m\n$ component of the equations of motion \eqref{munu}, we have
\begin{gather}
	0= - \zeta_{\m\n} \Big[2(\nabla_\m \nabla^\m +\k){\tau}-\frac{6}{a^3}\big(a^2 a' {\tau}\big)'
	+\frac{6}{a^3}\big(a^3 {\sigma}\big)' \Big]
	+2\nabla_\m \nabla_\n {\tau}
\nn\\
	-\frac{2}{a^3}\Big[a^3\nabla_{(\m}{{B}^T_{\n)}}  \Big]'+{h^{TT}_{\m\n}}''+ 3\frac{a'}{a}{h^{TT}_{\m\n}}'+ (\nabla_\r \nabla^\r  - \frac23 \k) h^{TT}_{\m\n} \,.
	\label{munu2ap}
\end{gather}
Applying equation \eqref{S0} we find
\begin{gather}
	0= -2 \zeta_{\m\n} (\nabla_\r \nabla^\r +\k){\tau}
	+2\nabla_\m \nabla_\n {\tau}
\nn\\
	-\frac{2}{a^3}\Big[a^3\nabla_{(\m}{{B}^T_{\n)}}  \Big]'
	+{h^{TT}_{\m\n}}''+ 3\frac{a'}{a}{h^{TT}_{\m\n}}'+ (\nabla_\r \nabla^\r  - \frac23 \k) h^{TT}_{\m\n} \,.
	\label{amunu2}
\end{gather}
By taking a trace from the above equation we obtain
\be \label{Trg1}
(\nabla_\m \nabla^\m +\frac{4}{3}\kappa){\tau}=0\,.
\ee
Since ${\tau}$ does not contain the zero mode of $(\nabla_\m \nabla^\m + \frac43 \k)$ operator, we conclude that
\be\label{soltau}
\tau=0\,.
\ee
The above equation in terms the of original fields is
\be \label{As2ap}
2 {\psi}+{\phi}-\frac{1}{a^3}\big(a^3({W}-{E}')\big)'=0\,.
\ee
Because of the \eqref{S0} we will have
\be \label{sigf}
\sigma=0\quad\Rightarrow\quad 2\frac{a'}{a}{\lambda}+{\lambda}' + \frac{\kappa}{3}{\gamma}-\frac{a'}{a}\Big(\frac{1}{a^3}(a^3 {\gamma})' \Big)=0\,.
\ee
Putting equations \eqref{BTZ} and \eqref{soltau} into the equation \eqref{amunu2}, we obtain a decoupled equation of motion for graviton mode
\be
{h^{TT}_{\m\n}}''+ 3\frac{a'}{a}{h^{TT}_{\m\n}}'+ (\nabla_\r \nabla^\r  - \frac23 \k) h^{TT}_{\m\n} =0 \,.
\label{ATT1ap}
\ee
\vspace{0.5cm}
$\bullet$
The $yy$ component of the Einstein equation \eqref{yy} after applying \eqref{soltau} leads to
\be
0=(\nabla_\m \nabla^\m  +\frac{4}{3} \kappa)\lambda  + 4 \frac{a'}{a}\lambda'-\frac{a'}{a} \nabla_\m \nabla^\m \gamma
+ \big(\frac{a'}{a}\big)^2 (\frac{z^2}{3}-4)\Big( \frac{1}{a^3}(a^3 \g)'-2\lambda\Big)\,, \label{yy2a}
\ee
or equivalently
\begin{gather}
	0 =(\nabla_\m \nabla^\m +\frac{4}{3} \kappa)\psi +4 \frac{a'}{a}\psi'- \frac{a'}{a} \nabla_\m \nabla^\m(W-E')- \Big(\frac{a''}{a}+2 \big(\frac{a'}{a}\big)^2\Big)\phi
	\nn\\
	- \frac{1}{3}\kappa \phi +\frac13 a^{-3}\left(a^3 \Phi_0' \chi\right)' - \frac{2}{3} \Phi_0' \chi'\label{As3ap}\,,
\end{gather}
\vspace{0.5cm}
$\bullet$
The Dilaton equation \eqref{dilaton} after substituting from equation \eqref{soltau}, becomes
\be
0 =(\nabla_\m \nabla^\m  + 2\k)\lambda+ \frac{1}{a^3}\big( a^3\lambda'\big)' + \frac{2z'}{3z}(\kappa\g+ 3 \lambda')\,.\label{dil2a}
\ee
or in terms of the original fields
\begin{gather}
	0 =\chi'' + 3 \frac{a'}{a}\chi' + \nabla_\m \nabla^\m \chi - \frac{1}{2}a^2 \pa^2_{\Phi}V
	\chi -2 a^{-3} \left(a^3 \Phi_0' \phi\right)'
	\nn\\
	+\Phi_0'\phi' + 4\Phi_0'\psi'   - \Phi_0' \nabla_\m \nabla^\m(W-E')\,.\label{As4ap}
\end{gather}

\vspace{0.5cm}
%%%%%%%%%%%%%%%%%%%%%%%%%%%%%%%%%%%%%%%%%%%%%%%%%%%%%
%%%%%%%%%%%%%%%%%%%%%%%%%%%%%%%%%%%%%%%%%%%%%%%%%%%%%
%%%%%%%%%%%%%%%%%%%%%%%%%%%%%%%%%%%%%%%%%%%%%%%%%%%%%
\section{Zero modes}\label{SZM}

In this section we will investigate the zero modes that we found in section \ref{Dec} directly, without using any decomposition. For simplicity in notation, we rename the equations of motion \eqref{munu1}--\eqref{Dil1} as follows
\begin{gather}
	{\bf{EQ1}} =  h_{\mu\nu}'' + 3\frac{a'}{ a} h_{\mu\nu}' + \nabla_\r \nabla^\r h_{\mu\nu} \!-\!2 \nabla^{\rho} \nabla_{(\mu}h_{\nu)\rho} +\nabla_\mu\nabla_\nu h + 2\nabla_\nu\nabla_\mu \phi
	\nn\\
	-2
	a^{-3}\left(a^3 \nabla_{(\mu} A_{\nu)}\right)'\, + \zeta_{\mu\nu}\Big[-h'' - 3\frac{a'}{a}h' - \nabla_\r \nabla^\r h +
	\nabla^\rho \nabla^\sigma h_{\rho\sigma}
	\nn\\
	-2 \nabla_\r\nabla^\r \phi + 6\frac{a'}{a}\phi' + 6\Big(\frac{a''}{a}+2 \big(\frac{a'}{a}\big)^2\Big)\phi  -2 a^{-3}\left(a^3 \Phi_0' \chi\right)'
	\nn\\
	+ 2 a^{-3}\left(a^3 \nabla_{\rho} A^{\rho}\right)'\Big]- \kappa \zeta_{\mu\nu}( 2\phi + h \big) + 2 \kappa h_{\mu\nu} =0\,,\label{zq1}
\end{gather}
\be\label{zq2}
{\bf{EQ2}} =  (\nabla_\n \nabla^\n +\kappa) A_\mu - \nabla^{\nu} h_{\nu\mu}'  +\nabla_{\mu}\Big[ - 6\frac{a'}{a}\phi + 2\Phi_0' \chi + h' - \nabla^{\nu} A_\nu\Big] =0 \,,
\ee
\begin{gather}
	{\bf{EQ3}} =    -\nabla_\m \nabla^\m h + \nabla^{\mu}\nabla^\nu h_{\mu\nu} - 3\frac{a'}{a}h' + 6\Big(\frac{a''}{a}+2 \big(\frac{a'}{a}\big)^2\Big)\phi + 6 \frac{a'}{a} \nabla^\nu A_\nu
	\nn\\
	+ 4 \Phi_0' \chi'- 2 a^{-3}\left(a^3 \Phi_0' \chi\right)'  +\kappa(2 \phi -h)= 0 \,,\label{zq3}
\end{gather}
\be \label{zD}
{\bf{EQ\chi}}=\chi''+ 3 \frac{a'}{a}\chi'  + \nabla_\m \nabla^\m \chi- \frac{1}{2}a^2 \partial^2_{\Phi}V \chi
-\frac{2}{a^3} \left(a^3 \Phi_0' \phi\right)' +\Phi_0'\phi' + \frac{1}{2}\Phi_0'h'   -
\Phi_0' \nabla^\mu A_\mu=0\,.
\ee
In the following steps we will show that there are three independent scalar equations that we can construct from the above equations. To do this, for simplicity in notation we introduce the following scalars fields
\be \label{zdef}
I(y,x^{\m})\equiv \nabla_\m A^\m \sp H(y,x^{\m})\equiv \nabla^\m\nabla^\n h_{\m\n} - \nabla^{\m}\nabla_{\m}h\,.
\ee
Then for example equation \eqref{zq3} becomes
\be
{\bf{EQ3}} =   H + 6 \frac{a'}{a} I  - 3\frac{a'}{a}h'-\k h + 6\Big(\frac{a''}{a}+2 \big(\frac{a'}{a}\big)^2+\frac{\k}{3}\Big)\phi
+ 4 \Phi_0' \chi'- \frac{2}{a^3}\left(a^3 \Phi_0' \chi\right)'  = 0 \,,\label{zq3a}
\ee
By a trace of equation \eqref{zq1} we get
\begin{gather}
	{\bf{EQ4}} =2H + 18 \frac{a'}{a} I  + 6 I'-2 \k h - 9 \frac{a'}{a} {h'} - 3 h'' - 8 \big(\k - 6 (\frac{a'}{a})^2-3\frac{a''}{a}\big) \phi + 24 \frac{a'}{a} \phi'
	\nn \\
	- 6 \nabla_{\alpha }\nabla^{\alpha }\phi-8a^{-3}(a^3\Phi'_0\chi)'=0\,.\label{zq4}
\end{gather}
On the other hand, if we get $\nabla^\m\nabla^\n$ from equation \eqref{zq1} then we shall find
\begin{gather}
	{\bf{EQ5}}  =  3 \frac{a'}{a} H' +H''
	-6 \k \frac{a'}{a} I - 2\k I' + 6\big( 2(\frac{a'}{a})^2+\frac{a''}{a}\big) \nabla_{\alpha }\nabla^{\alpha }\phi + 6 \frac{a'}{a} \nabla_{\alpha }\nabla^{\alpha }\phi'
	\nn \\
	-2a^{-3}\nabla_{\a}\nabla^{\a}(a^3\Phi'_0\chi)'=0\,.\label{zq5}
\end{gather}
The divergence of the equation \eqref{zq2} also gives
\be
{\bf{EQ6}}  =-H'+ 2 \k I - 6 \frac{a'}{a} \nabla_{\alpha }\nabla^{\alpha	}\phi+2\Phi'_0 \nabla_{\a}\nabla^{\a} \chi=0\,.\label{zq6}
\ee
However, since we have the following equality
\be \label{zq8}
{\bf{EQ5}} = -3\frac{a'}{a}{\bf{EQ6}}-({\bf{EQ6}})'\,,
\ee
equation {\bf{EQ5}} is not an independent equation.

To show that \eqref{zD} is not an independent scalar equation, we define a new scalar as follow
\be\label{zq10}
Z(y,x^{\m})\equiv (\nabla_\m \nabla^\m+\k)h -\nabla^\m\nabla^\n h_{\m\n}= -H+\k h\,,
\ee
If we begin with the following combination we will find
\begin{gather}
	{\bf{EQ3}}-3\frac{a'}{\k a}	{\bf{EQ6}} = 0 \rightarrow
	\nn \\
	H+3\frac{a'}{\k a}H'-\k(h+3\frac{a'}{\k a}h')+ 2\Big(\k+6\big(\frac{a'}{a}\big)^2+3\frac{a''}{a}\Big)\phi+18\frac{a'^2}{\k a^2} \nabla_{\alpha }\nabla^{\alpha	}\phi
	\nn \\
	+ 4 \Phi_0' \chi'- 2 a^{-3}\left(a^3 \Phi_0' \chi\right)'
	-6\frac{a'}{\k a}\Phi'_0 \nabla_{\a}\nabla^{\a} \chi=0\,. \label{zq9}
\end{gather}
where we have used the background equation of motion \eqref{Eom1}. In terms of the $Z$ function defined in \eqref{zq10} the above equation is
\begin{gather}
	Z + 3\frac{a'}{\k a} Z' =
	18\frac{a'^2}{\k a^2}\Big( \nabla_{\alpha }\nabla^{\alpha}+(\frac43-\frac{ a^2{\Phi'_0}^2}{9a'^2})\k\Big)\phi
	\nn \\
	+ 2 \Phi_0' \chi'-\frac{6a'}{\k a}\Phi'_0 \left(\nabla_{\a}\nabla^{\a}+(1+\frac{ a}{3a'} \frac{\Phi''_0}{\Phi'_0})\k \right)\chi\,.\label{zq11}
\end{gather}
On the other hand, we find the following combination of equations
\begin{gather}
	{\bf{EQZ}} =	({\bf{EQ3}})' - \frac{\k-12\frac{a'^2}{a^2}+3\frac{a''}{a}}{3\frac{a'}{a}} 	{\bf{EQ3}} - \frac{a'}{a}{\bf{EQ4}} + {\bf{EQ6}} = 0 \rightarrow \label{EQZ}     \\
	Z =-\Big(\frac{12 a' \Phi''_0}{a \Phi'_0}+\frac{12 a'^2}{a^2}+2 {\Phi'_0}^2\Big) \phi - 6 \frac{a'}{a} \phi'
	\nn                                                         \\
	+6\frac{a'}{a \Phi'_0}\nabla_{\a}\nabla^{\a} \chi + 6\frac{a'}{a \Phi'_0} \chi''+\Big(\frac{18 a'^2}{a^2 \Phi '_0}+2 \Phi'_0\Big)\chi'
	\nn                               \\
	\Big(\frac{6  a'}{a}\frac{\Phi_0^{(3)}}{{\Phi'_0}^2}
	+2 \Phi''_0 (\frac{3 a'^2}{a^2 {\Phi'_0}^2}+1)
	-\frac{6 (\k a^2 a'+3 a'^3)}{a^3 \Phi'_0}\label{zq12}
	\Big)\chi\,.
\end{gather}
By using equations \eqref{zq3a}, \eqref{zq4} and \eqref{zq6} one can show that the equation of motion  {\bf{EQ$\chi$}} or \eqref{zD} is not an independent equation
\be  \label{zq13}
{\bf{EQ\chi}} = \frac{a \Phi'_0}{6 a'}({\bf{EQZ}}-{\bf{EQ3}})\,,
\ee
which for driving of the above relation we have used \eqref{Eom1} and \eqref{Eom3}.

We should also notice that there is only one independent vector equation. By acting a $\nabla^\n$ on \eqref{zq1} one obtains
\begin{gather}
	{\bf{EQV}}  = -  \frac{1}{a^3}\big(a^3 (\k A_{\mu }+\nabla_{\alpha }\nabla^{\alpha}A_{\mu }-\nabla_{\mu }\nabla_{\alpha }A^{\alpha })\big)'+\frac{1}{a^3}\big(a^3 (\nabla_{\alpha }h'_{\mu}{}^{\!\alpha }- \nabla_{\mu }h')\big)'
	\nn \\
	+ 6 (2 (\frac{a'}{a})^2 + \frac{a''}{a}) \nabla_{\mu }\phi + 6 \frac{a'}{a}\nabla_{\mu}\phi'
	-2a^{-3}\nabla_{\m}(a^3\Phi'_0\chi)'=0\,,\label{zq7}
\end{gather}
and it is easy to show that
\be \label{zq8n}
{\bf{EQV}} = -3\frac{a'}{a}{\bf{EQ2}}-({\bf{EQ2}})'\,.
\ee

\vspace{0.3cm}
%%%%%%%%%%%%%%%%%%%%%%%%%%%%%%%%%%%%%%%%%%%%%%%%%%%%%
%%%%%%%%%%%%%%%%%%%%%%%%%%%%%%%%%%%%%%%%%%%%%%%%%%%%%
%%%%%%%%%%%%%%%%%%%%%%%%%%%%%%%%%%%%%%%%%%%%%%%%%%%%%
\subsection{Gauge invariant scalars and equations of motion}

So far we have found the following facts about the number of equations of motion and scalars:
\begin{itemize}
	\item In equations of motion we have five scalars: $H, I, h, \phi$ and $\chi$, where the first two are defined in \eqref{zdef}.
	\item We have five equations of motion for scalars, {\bf{EQ3}}, {\bf{EQ4}}, {\bf{EQ5}}, {\bf{EQ6}} and {\bf{EQ$\chi$}}. However, we also have two relations between these equations, \eqref{zq8} and \eqref{zq13}. Therefore, in total we have three equations of motion for scalars, {\bf{EQ3}}, {\bf{EQ4}} and {\bf{EQ6}} which are defined in \eqref{zq3a}, \eqref{zq4} and \eqref{zq6}.
	\item Using the transformations given in equations  \eqref{dif h}--\eqref{difchi}, we have the following transformations for scalars:
	\begin{gather}
		\d H = \d (\nabla^\m\nabla^\n h_{\m\n} - \nabla^{\m}\nabla_{\m}h)
		= -2\k\nabla_{\m}\xi^\m + 6\frac{a'}{a} \nabla_\m \nabla^\m \xi^5\,, \label{gtH}
		\\
		\d I = \d (\nabla^\m A_\m) = -  \nabla_{\m }{\xi^{\m }}' -  \nabla_{\m }\nabla^{\m }\xi^5 \,, \label{gtI}
		\\
		\delta h = -2\nabla^\mu \xi_\mu - 8 \frac{a'}{a} \xi^5 \,, \label{gth}
		\\
		\delta \phi = -{\xi^5}' - \frac{a'}{a}\xi^5\,, \label{gtp}
		\\
		\delta \chi =-\Phi'_0\, \xi^5 \,. \label{gtc}		
	\end{gather}

\end{itemize}
Now suppose that all the scalars are proportional to a single  mode  i.e.
\be \label{zm}
\nabla_\m \nabla^\m s(y,x) = M_0^2 s(y,x)\,,
\ee
where $s=\{H, I, h, \phi, \chi\}$.
By separating the variables as
\be \label{sep}
s(y,x) = s(y) \tilde{s}(x)\,,
\ee
and considering the gauge transformations \eqref{gtH}--\eqref{gtI}, we will find the following gauge invariant scalar fields
\be \label{gi1}
\Gamma = H'(y) -2\k I(y) -2 M_0^2 \Phi'_0 \chi(y) + 6 M_0^2 \frac{a'}{a} \phi(y)\,,
\ee
\be  \label{gi2}
\Omega = H(y) - \k h(y) +2 (4 \k+ 3M_0^2)\frac{a'}{a \Phi'_0} \chi(y)\,,
\ee
\be  \label{gi3}
\Theta = \phi(y) + \Big(\frac{\Phi_0''}{\Phi_0 '^2}-\frac{a'}{a \Phi '}\Big)\chi(y)-\frac{1}{\Phi'_0}\chi'(y)\,,
\ee
where we have imposed \eqref{zm} for scalar Laplacian.
If we find $I(y), h(y)$ and $\phi(y)$ from \eqref{gi1}--\eqref{gi3} and then insert them into the independent equations {\bf{EQ3}}, {\bf{EQ4}} and  {\bf{EQ6}} we will find
\be \label{gi4}
 \left(\frac{6 (4 \k+3 M_0^2) a'^2}{\k a^2}-2 \Phi_0 '^2\right)\Theta -\frac{3  a'}{\k a} \Gamma+\frac{3 a' }{\k a}\Omega ' +\Omega = 0\,,
\ee
\begin{gather}
	\frac{6 (4 \k+3 M_0^2) a' }{\k a}\Theta'-\frac{2 (4 \k+3 M_0^2) }{\k a^2}  \left(a^2 \left(2 \k+\Phi_0 '^2\right)-12 a'^2\right)\Theta
\nn \\
	-\frac{3 }{\k} \Gamma '-\frac{9 a'}{\k a}  \Gamma+\frac{3 }{\k}\Omega''+\frac{9 a'}{\k a} \Omega'+2 \Omega =0\,,
	\label{gi5}
\end{gather}
\be \label{gi6}
 \Gamma = 0\,.
\ee
By inserting $\Gamma$ and $\Theta$ from \eqref{gi6} and \eqref{gi4} into \eqref{gi5} one finds a differential equation just for $\Omega$. For example, for two specific scalar zero modes this equation becomes

\begin{itemize}
	\item $M_0^2=0$:
\begin{gather}
	\Omega ''-\frac{a'}{a}\frac{\Phi_0 ' \left(12 a'^2+a^2 \left(8 \k+5 \Phi_0'^2\right)\right)+24 a a' \Phi_0 ''}{a^2 \Phi_0'^3-12 a'^2 \Phi_0 '}\Omega '
	\nn \\
	-\frac{2\k}{3}\frac{ \Phi_0' \left(24 a'^2+a^2 \left(4 \k+\Phi_0 '^2\right)\right)+12 a a' \Phi_0 ''}{ a^2 \Phi_0 '^3-12 a'^2 \Phi_0 '}  \Omega= 0 \,.
	\label{gi7}
\end{gather}

\item  $M_0^2=-\frac43 \k$:
\be \label{gi8}
\Omega ''+\frac{3 a'}{a}\Omega '+\frac{2}{3} \k \Omega = 0 \,.
\ee
\end{itemize}
Both equations above are exactly the same as the equation of motion for $\hat{\lambda}_m$ in \eqref{yprofile} with values given by \eqref{Amassive} and \eqref{Bmassive} when $m=M_0$.

\vspace{0.3cm}
%%%%%%%%%%%%%%%%%%%%%%%%%%%%%%%%%%%%%%%%%%%%%%%%%%%%%
%%%%%%%%%%%%%%%%%%%%%%%%%%%%%%%%%%%%%%%%%%%%%%%%%%%%%
%%%%%%%%%%%%%%%%%%%%%%%%%%%%%%%%%%%%%%%%%%%%%%%%%%%%%
\subsection{Lichnerowicz differential operator and its eigen-values}

Before we focus on the zero modes, we show the relation between eigen-values of the Lichnerowicz differential operator when acts on scalar, vector and tensor modes.
The Lichnerowicz differential operator $\D$ on spin 2 field is defined as \cite{Lic}
\be \label{Lich}
-\Delta h_{\m\n} = \nabla_\a\nabla^\a h_{\m\n} +2 R_{\m\r\n\s} h^{\r\s} - 2 {R^\r}_{\!(\m}h_{\n)\r} =
\nabla_\a\nabla^\a h_{\m\n} -\frac83 \k h_{\m\n} +\frac23 \k h\z_{\m\n}\,,
\ee
where in the last equality we have used the maximally symmetry properties of the slice geometry \eqref{symspa}.
For a vector field $A_\m$ and a scalar field $s$ this operator is related to the Laplace operator as follows
\be \label{Lich1}
-\Delta{A_\m} = (\nabla^\a\nabla_\a-\k) A_\m\,,
\ee
\be \label{Lich0}
 -\Delta s = \nabla^\a\nabla_\a s \,.
\ee
Suppose that the eigen-values of the Lichnerowicz operator are defined by
\be \label{lms}
-\Delta s = M_0^2 s \sp -\Delta A_\m = M_1^2 A_\m \sp
-\Delta h_{\m\n} = M_2^2 h_{\m\n}\,.
\ee
Considering \eqref{Lich1} and \eqref{lms} and by starting with the identity (\ref{id1}) $(\nabla^\m A_\m \neq 0)$
\be \label{VT2}
\nabla^{\m}\nabla_{\n}\nabla^{\n}A_\m = (\nabla_{\m}\nabla^{\m}+\k) \nabla^{\n}A_\n \quad \rightarrow \quad (\nabla_{\m}\nabla^{\m}-M_1^2)  \nabla^{\n}A_\n=0\,.
\ee
Since $\nabla^{\n}A_\n$ is a scalar then from the last equality above we conclude that
\be \label{m10}
M_1^2 = M_0^2\,.
\ee
If we start with \eqref{Lich} and with the eigen-value defined in \eqref{lms} then we will have
\be \label{VT5}
(\nabla_\a\nabla^\a-M_2^2-\frac83 \k) h_{\m\n} =  -\frac23 \k h\z_{\m\n}\,.
\ee
By using the following identity
\be
\nabla^{\n}\nabla^{\m}\nabla_{\m} h_{\n\a} = -\frac23 \k \nabla_\a h + \frac53 \k \nabla^\m h_{\a\m} + \nabla^{\n}\nabla_{\n}\nabla^{\m} h_{\a\m}\,, \label{VT6}
\ee
and inserting the value of the Laplace operator from \eqref{VT5} we find
\be \label{VT7}
\Big(\nabla^{\n}\nabla_{\n}-M_2^2-\k \Big)\nabla^{\m} h_{\a\m}=0\,.
\ee
Since $\nabla^{\m} h_{\a\m}$ is a vector therefore from \eqref{Lich1} we conclude that
\be \label{m21}
M_2^2 = M_1^2\,.
\ee
Equations \eqref{m10} and \eqref{m21} show that the eigen-values of the Lichnerowicz operator for scalar, vector and tensor modes are equal.
In summary we would have the following equations on a four dimensional maximally symmetric space
\be \label{scl}
\nabla^\a \nabla_\a s = M_0^2 s\,,
\ee
\be \label{vl}
\nabla^\a \nabla_\a A_\m = (M_0^2+\k) A_\m\,,
\ee
\be \label{tl}
(\nabla_\a\nabla^\a-M_0^2-\frac83 \k) h_{\m\n} =  -\frac23 \k h \z_{\m\n}\,.
\ee

\vspace{0.3cm}
%%%%%%%%%%%%%%%%%%%%%%%%%%%%%%%%%%%%%%%%%%%%%%%%%%%%%
%%%%%%%%%%%%%%%%%%%%%%%%%%%%%%%%%%%%%%%%%%%%%%%%%%%%%
%%%%%%%%%%%%%%%%%%%%%%%%%%%%%%%%%%%%%%%%%%%%%%%%%%%%%
\subsection{Vector zero mode}

From \eqref{Divh} one observed that the decomposition \eqref{split2} is valid as long as  $\nabla^\m h_{\m\n}$ does not have the zero mode of $-\k$ of the Laplacian vector mode. Here we are interested in this zero mode without using the decomposition \eqref{Divh}. Looking at
\eqref{VT7} we conclude that we should consider
\be \label{M2}
M_2^2 = -2\k\,.
\ee
Therefore, equations \eqref{scl}--\eqref{tl} become
\be \label{vzscl}
\nabla^\a \nabla_\a s = -2\k s\,,
\ee
\be \label{vzvl}
\nabla^\a \nabla_\a A_\m = -\k A_\m\,,
\ee
\be \label{vztl}
(\nabla_\a\nabla^\a-\frac23 \k) h_{\m\n} =  -\frac23 \k h \z_{\m\n}\,.
\ee

\vspace{0.3cm}
%%%%%%%%%%%%%%%%%%%%%%%%%%%%%%%%%%%%%%%%%%%%%%%%%%%%%
%%%%%%%%%%%%%%%%%%%%%%%%%%%%%%%%%%%%%%%%%%%%%%%%%%%%%
%%%%%%%%%%%%%%%%%%%%%%%%%%%%%%%%%%%%%%%%%%%%%%%%%%%%%

\subsubsection{Separation of the variables and equations of motion}

To solve equations of motion we first use the following separation of the variables
\be \label{sp1}
h_{\m\n}(y,x) = h(y) \tilde{h}_{\m\n}(x) \sp A_\m(y,x) = A(y) \tilde{A}_\m(x)\,,
\ee
\be \label{sp2}
s(y,x) = s(y) \tilde{s}(x) \sp  s\in \{\phi, \chi, h, H, I \} \,.
\ee
Moreover, from \eqref{sp1} and \eqref{sp2} we find
\be \label{sp3}
h^{\mu}_\m (y,x) = h(y) \tilde{s}(x)\,,
\ee
and
\be \label{sp4}
I(y,x)\equiv \nabla_\m {A}^\m(y, x)= A(y) \nabla^\m \tilde{A}_\m(x) \equiv I(y) \tilde{s}(x)\,.
\ee
With the above separation, equations \eqref{vzscl}--\eqref{vztl} give
\be \label{sp7}
(\nabla_\n\nabla^\n+2\k)\tilde{s}(x) = 0\,.
\ee
\be \label{sp6}
(\nabla_\n\nabla^\n+\k)\tilde{A}_\m(x) = 0 \,,
\ee
\be \label{sp5}
(\nabla_\a\nabla^\a-\frac23 \k) \tilde{h}_{\m\n}(x) = -\frac23 \k \tilde{s}(x)\z_{\m\n}\,,
\ee
At the first step, we start from \eqref{sp4} and use \eqref{sp7} which leads to
\begin{gather}
	\nabla^\m \tilde{A}_\m(x) = \tilde{s}(x) = -\frac{1}{2\k}\nabla^\m\nabla_\m \tilde{s}(x) \to\nabla^{\m}\big(\tilde{A}_\m(x)+\frac{1}{2\k}\nabla_\m \tilde{s}(x)\big) = 0\to
\nn \\	
	 \tilde{A}_\m(x) = -\frac{1}{2\k}\nabla_\m \tilde{s}(x) + \nabla_\m f(x) + B^T_{\m}(x)\,, \label{sp8}
\end{gather}
where $f(x)$ is a scalar function and $B^T_\m$ is a transverse vector with
\be \label{sp9}
\nabla^\m\nabla_\m f(x) = 0\sp   \nabla^{\m}B^T_{\m}(x) = 0\,.
\ee
However, all scalars of the theory satisfy \eqref{sp7} therefore, the only possible value of function $f$ is
$f(x) = 0$.
We conclude that
\be \label{sp11}
\tilde{A}_\m(x) = -\frac{1}{\k}\nabla_\m \tilde{s}(x)+B^T_{\m}(x)\,.
\ee
In the next step, we start with the vector equation {\bf{EQ2}} in \eqref{zq2}, separate the variables and use \eqref{sp6} to find
\be \label{sp12}
-h'(y)\nabla^{\nu} \tilde{h}_{\nu\mu}(x)  +\Big[ - 6\frac{a'}{a}\phi(y) + 2\Phi_0' \chi(y) + h'(y) - I(y)\Big]\nabla_{\mu}\tilde{s}(x) =0 \,.
\ee
Acting a $\nabla^\m$ on the above equation and using \eqref{sp7} one finds
\be \label{sp13}
-h'(y)\nabla^{\mu}\nabla^{\nu} \tilde{h}_{\nu\mu}(x)  -2\k\Big[ - 6\frac{a'}{a}\phi(y) + 2\Phi_0' \chi(y) + h'(y) - I(y)\Big]\tilde{s}(x) =0 \,.
\ee
The above equation is a separated differential equation with two variables $y$ and $x$, therefore if we suppose
\be \label{sp14}
\nabla^{\mu}\nabla^{\nu} \tilde{h}_{\nu\mu}(x)  = -2\k c \tilde{s}(x)\,,
\ee
where $c$ is a constant, then from \eqref{sp13}
\be \label{sp15}
 - 6\frac{a'}{a}\phi(y) + 2\Phi_0' \chi(y) + (1-c) h'(y) - I(y) = 0\,.
\ee
Putting \eqref{sp15} into \eqref{sp12} gives
\be \label{sp16}
\nabla^{\nu} \tilde{h}_{\nu\mu}(x) = c \nabla_{\mu}\tilde{s}(x)\,.
\ee
If we start with the tensor equation {\bf{EQ1}} in \eqref{zq1} and impose \eqref{VT5} with $M_2^2$ given in \eqref{M2}, after simplification we will find
\begin{gather}
	 h_{\mu\nu}'' + 3\frac{a'}{ a} h_{\mu\nu}'  -2 \nabla_{(\m} \nabla^\a h_{\nu)\a}  +\nabla_\m\nabla_\n h+ 2\nabla_\nu\nabla_\mu \phi-2
	 a^{-3}\left(a^3 \nabla_{(\mu} A_{\nu)}\right)'
\nn\\
	 + \zeta_{\mu\nu}\Big[-h'' - 3\frac{a'}{a}h' -\k h - \nabla_\r \nabla^\r h +
	\nabla^\rho \nabla^\sigma h_{\rho\sigma}-2 \nabla_\r\nabla^\r \phi + 6\frac{a'}{a}\phi'
\nn\\
	 + 6\Big(\frac{a''}{a}+2 \big(\frac{a'}{a}\big)^2-\frac{\k}{3}\Big)\phi  -2 a^{-3}\left(a^3 \Phi_0' \chi\right)'
	+ 2 a^{-3}\left(a^3 \nabla_{\rho} A^{\rho}\right)'\Big] =0\,.\label{sp17}
\end{gather}
By separating variables and using relations \eqref{sp7}, \eqref{sp11} and \eqref{sp16} we find
\begin{gather}
	\Big(h''(y) + 3\frac{a'}{a} h'(y)\Big)\tilde{h}_{\mu\nu}(x)+ \Big((1-2c)h(y)+2\phi(y)
	+\frac{1}{\k} a^{-3}\left(a^3 I(y)\right)'\Big) \nabla_\m\nabla_\n\tilde{s}(x)
\nn \\
	+ \zeta_{\mu\nu}\Big[-h''(y) - 3\frac{a'}{a}h'(y) +\k(1-2c) h(y)	+ 2 a^{-3}\left(a^3 I(y)\right)' + 6\frac{a'}{a}\phi'(y)
\nn\\
	+ 6\Big(\frac{a''}{a}+2 \big(\frac{a'}{a}\big)^2+\frac{\k}{3}\Big)\phi(y)  -2 a^{-3}\left(a^3 \Phi_0' \chi(y)\right)'
\Big]\tilde{s}(x) =  2 a^{-3}\left(a^3 I(y)\right)'\nabla_{(\mu} B^T_{\nu)}(x)\,.\label{sp18}
\end{gather}
A trace of the above equation gives
\begin{gather}
	\Big(-3 h''(y)-\frac{9 a' }{a}h'(y)+2(1-2 c) \k h(y)+ \Big(\frac{96 a'^2}{a^2}-4 \left(\k+2 \Phi_0 '^2\right)\Big)\phi (y)+\frac{24 a'}{a} \phi '(y)
\nn \\
	-8a^{-3}(a^3 \Phi_0 ' \chi(y))'+6 a^{-3}(a^3 I(y))'\Big) \tilde{s}(x)=0\,.\label{tsp18}
\end{gather}

By using equations {\bf{EQ3}} in \eqref{zq3a}, \eqref{sp15} and \eqref{tsp18} we can find the values of $I(y), \Phi''_0(y)$ and $\chi'(y)$. Inserting these values into the equation \eqref{sp18} gives
\be
\big(h''(y)+3\frac{a'}{a}h'(y)\big)\Pi_{\m\n}(x)
= 2 a^{-3}\left(a^3 I(y)\right)'\nabla_{(\mu} B^T_{\nu)}(x)\,, \label{sp19}
\ee
where
\be \label{sp19P}
\Pi_{\m\n}(x) = \tilde{h}_{\m\n}(x)+\frac{1}{2\k}(4c-1)\nabla_{\m}\nabla_\n \tilde{s}(x)+(c-\frac12)\z_{\m\n}\tilde{s}(x)\,.
\ee
It is easy to show that $\Pi_{\m\n}(x)$ is a trace-less  and transverse rank two tensor
\be \label{r2tt}
\Pi^\m_{\,\,\m}(x) = 0 \sp \nabla^\m \Pi_{\m\n}(x) = 0 \,.
\ee
The right hand side of \eqref{sp19} is trace-less and also we have
\be
\nabla^\n\nabla_{(\mu} B^T_{\nu)} = \nabla^{\n}\nabla_{\n} B^T_{\m}+[\nabla^\n , \nabla_\m] B^T_{\n}  = -\k B^T_{\m} + \k B^T_{\m} = 0\,. \label{isp20}
\ee
Equation (\ref{sp19}) can be rewritten as
\be \label{eqP1}
\left(\pa_{y}^2+3{\frac{a'}{a}\pa_y}\right)\hat \Pi_{\m\n}(y,x)=0
\ee
with
\be \label{eqP2}
 \Pi_{\m\n}(y,x)\equiv h(y)\Pi_{\m\n}(x)-2\int^y dy' I(y')\nabla_{(\mu} B^T_{\nu)}(x)\,.
 \ee
$ \Pi_{\m\n}(y,x)$ is transverse and traceless and have the eigenvalue $2\k$ of the Lichnerowicz operator.

\vspace{0.3cm}
%%%%%%%%%%%%%%%%%%%%%%%%%%%%%%%%%%%%%%%%%%%%%%%%%%%%%
%%%%%%%%%%%%%%%%%%%%%%%%%%%%%%%%%%%%%%%%%%%%%%%%%%%%%
%%%%%%%%%%%%%%%%%%%%%%%%%%%%%%%%%%%%%%%%%%%%%%%%%%%%%

\subsection{Scalar zero modes}

Before finding equations of motion for the specific case of the scalar zero modes we first write the equations of motion {\bf{EQ1}}, {\bf{EQ2}} and {\bf{EQ3}} in terms of the eigen-value of Laplace operator for scalar fields. We assume all scalar modes have the same mass
\be \label{zma}
\nabla_\m \nabla^\m s(y,x) = M_0^2 s(y,x)\,,
\ee
and use the separation of variables which we introduced in \eqref{sp1}--\eqref{sp4} to simplify equations of motion.
The vector equation {\bf{EQ2}} in \eqref{zq2} becomes
\be\label{sz5}
(M_0^2 +2 \kappa) I(y) \tilde{A}_\mu(x) -h'(y) \nabla^{\nu} \tilde{h}_{\nu\mu}(x) - \Big[ \frac{6a'}{a}\phi(y) - 2\Phi_0' \chi(y) - h'(y) + I(y)\Big]\nabla_{\mu}\tilde{s}(x) =0 \,,
\ee
Divergence of the above vector equation together with \eqref{zma} give ($\nabla^\m  \tilde{A}_\mu(x) = \tilde{s}(x)$)
\be \label{sz6}
h'(y) \nabla^{\mu}\nabla^{\nu} \tilde{h}_{\mu\nu}(x) =
\Big(2\k I(y) -\frac{6a'}{a}M_0^2 \phi(y) +2\Phi_0' M_0^2 \chi(y) +M_0^2 h'(y)\Big)\tilde{s}(x)\,.
\ee
The above equation is a separated differential equation for $y$ and $x$ variables therefore, we write
\be \label{sz7}
 \nabla^{\mu}\nabla^{\nu} \tilde{h}_{\mu\nu}(x) =
-2\k c \tilde{s}(x)\,,
\ee
where $c$ is a constant. Equation \eqref{sz6} then gives
\be \label{sz8}
 2\k I(y) -\frac{6a'}{a}M_0^2 \phi(y) +2\Phi_0' M_0^2 \chi(y) +(M_0^2+2\k c) h'(y)=0\,,
\ee
If we start with the tensor equation {\bf{EQ1}} in \eqref{zq1} we will get
\begin{gather}
	{\bf{EQT}}=\big(h''(y)+3\frac{a'}{a}h'(y)+(M_0^2+2\k)h(y)\big)\tilde{h}_{\m\n}(x)
	-2h(y)\nabla_{(\m}\nabla^\a \tilde{h}_{\n)\a} (x)
\nn \\
	+\big(h(y)+2\phi(y)\big)\nabla_\m\nabla_\n\tilde{s}(x)-2 a^{-3}(a^3 I(y))'\nabla_{(\m}\tilde{A}_{\n)}(x)
\nn\\
	+ \zeta_{\mu\nu}\Big[-h''(y) - 3\frac{a'}{a}h'(y) -(\k+M_0^2+2\k c) h(y)+ 2 a^{-3}\left(a^3 I(y)\right)' + 6\frac{a'}{a}\phi'(y)
\nn \\
	+ 6\Big(\frac{a''}{a}+2 \big(\frac{a'}{a}\big)^2-\frac{\k}{3}-\frac{M_0^2}{3}\Big)\phi(y)-2 a^{-3}\left(a^3 \Phi_0' \chi(y)\right)'
	\Big]\tilde{s}(x) =  0\,.\label{sz10}
\end{gather}
A trace of the above equation leads to
\begin{gather}
	\big((1+2c)\k+M_0^2\big)\Big(h''(y)+\frac{3 a' }{a}h'(y)+\frac{2}{3} \k h(y)\Big)
\nn \\
	-3 (M_0^2+\frac43 \k)\Big(3\frac{a'}{a}\phi'(y)
	-(2\k+{\Phi'_0}^2-12\frac{a'^2}{a^2})\phi(y)-a^{-3}(a^3 \Phi'_0 \chi(y))'\Big) = 0 \,,\label{sz11}
\end{gather}
where we have used equations \eqref{Eom1}, \eqref{zma}, \eqref{sz7} and \eqref{sz8}.

If we insert the value of $I(y)$ from equation \eqref{sz8} into the equation {\bf{EQ3}} in \eqref{zq3} then
\begin{gather}
	- (M_0^2+(2 c+1) \k) \Big(h(y)+\frac{3 a' }{\k a}h'(y)\Big)
	+ \Big(\frac{6 (4 \k+3 M_0^2) a'^2}{\k a^2}-2 \Phi_0 '^2\Big)\phi(y)
\nn \\
	+2\Phi'_0 \chi'(y) -  \Big(\frac{6 (\k+M_0^2) a' \Phi'_0}{\k a}+2 \Phi''_0 \Big)\chi(y) = 0\,.\label{sz12}
\end{gather}

In the following we will show how to separate equations of motion we found in this section for two
scalar zero modes
\be \label{zzz}
M_0^2 = 0\,, -\frac43 \k\,.
\ee

\vspace{0.3cm}
%%%%%%%%%%%%%%%%%%%%%%%%%%%%%%%%%%%%%%%%%%%%%%%%%%%%%
%%%%%%%%%%%%%%%%%%%%%%%%%%%%%%%%%%%%%%%%%%%%%%%%%%%%%
%%%%%%%%%%%%%%%%%%%%%%%%%%%%%%%%%%%%%%%%%%%%%%%%%%%%%

\subsubsection{$M_0^2 = 0$ zero mode}

For this zero mode after separation of variables and from equations \eqref{scl}--\eqref{tl} we have
\be \label{sclz1}
\nabla^\a \nabla_\a \tilde{s}(x) = 0\,,
\ee
\be \label{vla1}
\nabla^\a \nabla_\a \tilde{A}_\m (x) = \k \tilde{A}_\m (x)\,,
\ee
\be \label{tlz1}
(\nabla_\a\nabla^\a-\frac83 \k) \tilde{h}_{\m\n} (x) =  -\frac23 \k \tilde{s}(x) \z_{\m\n}\,.
\ee

In first step and for $M_0^2 = 0$, we find $\Phi''_0$ and $\chi'(y)$  from equations \eqref{sz11} and \eqref{sz12}.
We also note that equation \eqref{sz8} gives
\be \label{sz19}
I(y) = -c \, h'(y)\,.
\ee
Next, from equation \eqref{sz5} we construct the following  equation by imposing a $\nabla_\n$
\begin{gather}
	 \kappa I(y) \nabla_{(\m}\tilde{A}_{\nu)}(x) -2 h'(y) \nabla_{(\m}\nabla^\a \tilde{h}_{\n)\a}(x)
\nn \\
	- 2\Big[ \frac{6a'}{a}\phi(y) - 2\Phi_0' \chi(y) - h'(y) + I(y)\Big]\nabla_{\mu}\nabla_{\nu}\tilde{s}(x) =0 \,.\label{sz13}
\end{gather}
By using the above equation and its $y$ derivative we will find the following relation after inserting the values of $I(y)$ from \eqref{sz19} and the values of $\Phi''_0$ and $\chi'(y)$ that we found in the first step
\begin{gather}
	8\k  \phi(y) \nabla_{\mu}\nabla_{\nu}\tilde{s}(x) = -2\k (1+2c)  h(y) \nabla_{\mu}\nabla_{\nu}\tilde{s}(x)
\nn \\
	-(\frac{3a'}{a}h'+h'')\Big((2c-1)\nabla_{\mu}\nabla_{\nu}\tilde{s}(x)+4\nabla_{(\m}\nabla^\a \tilde{h}_{\n)\a}(x) +8 c\k \nabla_{(\m}\tilde{A}_{\nu)}(x)\Big)\,.\label{sz14}
\end{gather}
In the final step we insert the values of $I(y), \Phi''_0$ and $\chi'(y)$ into the equation {\bf{EQT}} in \eqref{sz10} and then use the equation \eqref{sz14} to get rid of $\nabla_{(\m}\tilde{A}_{\nu)}(x)$. We find
\be \label{sz15}
 \Big(h''(y)+\frac{3a'}{a}h'(y)+2\k h(y)\Big)\tilde{\Pi}_{\m\n}(x) = 0\,,
\ee
where
\be \label{sz16}
\tilde{\Pi}_{\m\n}(x) = 4\k \tilde{h}_{\m\n}(x)-4 \nabla_{(\m}\nabla^\a \tilde{h}_{\n)\a}(x)+(1-2c) \nabla_{\mu}\nabla_{\nu}\tilde{s}(x)-(1+2c)\k \z_{\m\n}\tilde{s}(x)\,.
\ee
The rank two tensor in \eqref{sz16} is trace-less and transverse, i.e.
\be \label{sz17}
{\tilde\Pi}^{\m}_{\,\,\m} (x) = 0 \sp \nabla^\m \tilde{\Pi}_{\m\n}(x) = 0\,.
\ee

\vspace{0.3cm}
%%%%%%%%%%%%%%%%%%%%%%%%%%%%%%%%%%%%%%%%%%%%%%%%%%%%%
%%%%%%%%%%%%%%%%%%%%%%%%%%%%%%%%%%%%%%%%%%%%%%%%%%%%%
%%%%%%%%%%%%%%%%%%%%%%%%%%%%%%%%%%%%%%%%%%%%%%%%%%%%%

\subsubsection{$M_0^2 =  -\frac43 \k$ zero mode}

By separation of variables and from equations \eqref{scl}--\eqref{tl} we find
\be \label{sclz2}
\nabla^\a \nabla_\a \tilde{s}(x) =-\frac43 \k \tilde{s}(x)\,,
\ee
\be \label{vla2}
\nabla^\a \nabla_\a \tilde{A}_\m (x) =-\frac13 \k \tilde{A}_\m (x)\,,
\ee
\be \label{tlz2}
(\nabla_\a\nabla^\a + \frac83 \k) \tilde{h}_{\m\n} (x) =  -\frac23 \k \tilde{s}(x) \z_{\m\n}\,.
\ee
For this zero mode from \eqref{sz11} one directly observes  that
\be \label{mz1}
2\big(c-\frac16\big)\k\Big(h''(y)+\frac{3 a' }{a}h'(y)+\frac{2}{3} \k h(y)\Big)=0\,.
\ee
This is in fact equation \eqref{gi8} where from \eqref{gi2}
\be \label{mz2}
\Omega = H(y)-\k h(y) = -2\k (c-\frac16) h(y)\,.
\ee
For $M_0^2 = -\frac43 \k$,  and from equations \eqref{mz1} and \eqref{sz12} one may find $h''(y)$ and $\Phi''_0$. Then from equations \eqref{sz5} we construct the following  equation by imposing a $\nabla_\n$
\begin{gather}
	{\bf{EQ}}=\frac49\big(\Phi'_0\chi(y)-3\frac{a'}{a}\phi(y)\big) \Big(3\nabla_\m\nabla_\n \tilde{s}(x)+2\k \z_{\m\n}\tilde{s}(x)\Big)
\nn \\
	+\frac19 h'(y) \Big(-18 \nabla_{(\m}\nabla^\a \tilde{h}_{\n)\a}(x)+6(1+3c)\nabla_\m\nabla_\n \tilde{s}(x)+4(2-3c)\k \nabla_{(\m}A_{\n)}\Big) = 0\,.
	\label{mz3}
\end{gather}
Using equation \eqref{sz10} and the above equation we can construct the following equation
\begin{gather}
	{\bf{EQT}}+\frac{9}{2\k}\frac{a'}{a}{\bf{EQ}}+\frac{3}{2\k}{\bf{EQ}}'=0\,,
\nn \\
	\to\quad  \hat{\Pi}_{\m\n} (x) \Big(
	\frac{(1-6 c) a' }{3 \k a}h'+\frac{2(1-6c)}{9} h(y)+\frac{4 \Phi_0 ' }{3 \k a}\left(2 \chi(y) a'+a \chi'(y)\right)
\nn \\
	-\frac{2 a' }{\k a}\phi '(y)+ \big(\frac{4}{3}-\frac{8 a'^2}{\k a^2}\big)\phi(y)
	\Big)=0\,,\label{mz4}
\end{gather}
where
\be \label{mz5}
\hat{\Pi}_{\m\n} (x)= 3\nabla_\m\nabla_\n \tilde{s}(x)+\k \z_{\m\n}\tilde{s}(x)
\ee
The combination $\hat{\Pi}_{\m\n} (x)$ is transverse and trace-less.

\vspace{0.5cm}
%%%%%%%%%%%%%%%%%%%%%%%%%%%%%%%%%%%%%%%%%%%%%%%%%%%%%
%%%%%%%%%%%%%%%%%%%%%%%%%%%%%%%%%%%%%%%%%%%%%%%%%%%%%
%%%%%%%%%%%%%%%%%%%%%%%%%%%%%%%%%%%%%%%%%%%%%%%%%%%%%

\section{Derivations around AdS space} \label{spcase}

The AdS$_5 $ space-time is obtained  by a constant background scalar field or $\Phi_0'= 0$.
 In this case, the scalar potential in the action \eqref{action} reduces effectively to a constant cosmological term $V=-\frac{d(d-1)}{\ell^2} $ with $d=4$ in our case. Considering the metric \eqref{bac2}, the scale factor for $d$--dimensional
 Minkowski and A(dS) slices is given by \cite{C}
 \be\label{Au}
 e^{A(u)}=
 \begin{cases}
 	\exp(-\frac{u+c}{\ell})\, \qquad &-\infty < u < +\infty \qquad \,\,\mathcal{M}_d \\ \cr
 	\frac{\ell}{\alpha}\cosh(\frac{u+c}{\ell})\, \qquad &-\infty < u < +\infty \qquad AdS_d \\ \cr
 	\frac{\ell}{\alpha}\sinh(\frac{u+c}{\ell})\, \qquad &-c \leq u < +\infty \quad\quad dS_d (S^d)
 \end{cases} \,,
 \ee
 where $c$ is an integration constant, $\ell$ is the AdS$_{d+1}$ length scale and $\alpha$ is the curvature length scale of the  $d$--dimensional slices. Moreover, since the $d$--dimensional slices are  maximally symmetric, we obtain
 \be \label{Riccis}
 R^{(\zeta)}_{\m\n}= \kappa \zeta_{\m\n} \sp  R^{(\zeta)}= d \kappa\,,\quad \text{with} \qquad
 \kappa =
 \begin{cases}
 	0 \qquad \qquad\quad  \mathcal{M}_d\\
 	-\frac{(d-1)}{\alpha^2} \qquad AdS_{d}\\
 	\frac{(d-1)}{\alpha^2} \quad \quad\,\, dS_{d}(S^d)
 \end{cases}\,.
 \ee
 Using $e^{-A(u)} du= dy$ one goes to the conformal coordinates
 \be \label{chuy}
 u=
 \begin{cases}
 	-c+\ell \log\frac{y-y_0}{\ell}     \qquad\quad  & y_0\leq y<+\infty\qquad\qquad \mathcal{M}_d\\ \cr
 	-c+2\ell \, \text{arctanh}\left[\tan(\frac{y-y_0}{2\a}-\frac{\pi}{4})\right]
 	& y_0\leq y\leq y_0+\pi\a \quad\quad  AdS_{d}
 	\\ \cr
 	-c+2\ell \,\text{arccoth}\big[e^{-\frac{y-y_0}{\a}}\big]
 	\qquad\qquad\qquad  &-\infty < y \leq  y_0   \quad\quad\quad
 	dS_{d}(S^d)
 \end{cases}\,,
 \ee
 where $y_0$ is a constant of integration.
 The scale factors in $y$ coordinate then become
 \be\label{ay}
 a(y) \equiv e^{A(y)}=
 \begin{cases}
 	\ell{(y-y_0)}^{-1}\,  &y_0 < y < +\infty \quad\quad\qquad \mathcal{M}_d \\ \cr
 	\frac{\ell}{\alpha }\big(\sin(\frac{y-y_0}{\alpha})\big)^{-1}\,  & y_0 < y \le  y_0+\alpha \pi  \quad\quad AdS_d \\ \cr
 	-\frac{\ell}{\alpha}\big( \sinh(\frac{y-y_0}{\alpha})\big)^{-1}\,  & -\infty < y \leq  y_0   \quad\quad\quad dS_d (S^d)
 \end{cases}\,.
 \ee
 We should note that the above solutions can be found directly from equations of motion \eqref{Eom1}--\eqref{Eom3}.

 In next sections we shall investigate the scalar and graviton modes in the above background solutions.

\vspace{0.3cm}
%%%%%%%%%%%%%%%%%%%%%%%%%%%%%%%%%%%%%%%%%%%%%%%%%%%%%
%%%%%%%%%%%%%%%%%%%%%%%%%%%%%%%%%%%%%%%%%%%%%%%%%%%%%
%%%%%%%%%%%%%%%%%%%%%%%%%%%%%%%%%%%%%%%%%%%%%%%%%%%%%

\subsection{Gravitons}\label{GAds}

In section \ref{tenmod} we found that the 4--dimensional graviton mode obeys the  Schrodinger-like equation \eqref{Sch-grav} with potential given by \eqref{Vgrav}. In this appendix, we find the mass spectrum of graviton modes for dS and AdS slices.

\begin{itemize}
	
\item dS slices
	
From equation \eqref{ayn} the potential in this space-time reads as
\be \label{Vdsgl}
V_g(y)= \frac{3}{4 \alpha^2}\Big(3+ \frac{5}{\sinh^2 (\frac{y}{\alpha})}\Big) \,.
\ee
The behavior of $V(y)$ at the UV boundary and IR end-point is
\begin{gather}
		V_g^{UV}(y)=\frac{15}{4 y^2}+\frac{1}{\alpha ^2}+O\left(y\right)\sp\qquad y\rightarrow 0\,,
\nn\\
		\label{VdSAssym}
		V_g^{IR}(y)=\frac{9}{4 \alpha ^2}+\frac{15}{\alpha ^2} e^{2y/\alpha}+ O\left(e^{4y/\alpha}\right)  \sp \quad y\rightarrow -\infty\,,
\end{gather}
When the mass is above the mass gap, we write it as
\be \label{mabove}
M^2= \frac{1}{4\a^2}(9+\n^2)\,.
\ee
The expansion of the wave function, that is the solution of \eqref{S-grn} with the potential \eqref{Vdsgl}, near the UV boundary is
\be \label{UVg ds}
\psi_g(y)= c_1 \Big( y^{-\frac32}+  \frac{5+\n^2}{16\a^2} y^{\frac12} + \mathcal{O}( \log(y) y^{\frac52} ) \Big) +c_2 \Big( y^{\frac52} - \frac{5+\n^2}{48\a^2} y^{\frac92} + \mathcal{O}(y^{\frac{13}{2}}) \Big)\,\,,
\ee
To have a finite wave function near the UV, we should set $c_1=0$ (source-free boundary condition).
	
The expansion of the wave-function near the IR end-point is
\be \label{IRg ds}
\psi_g(y)= b_1\, e^{i(\frac{ \n}{2\a})y} + b_2 \,e^{-i(\frac{ \n}{2\a})y}+ \cdots \,.
\ee
That is, in the IR limit the wave function is finite and has a periodic behavior.
	
On the other hand, when the mass is below the mass gap, by setting
\be \label{mbelow}
M^2= \frac{1}{4\a^2}(9-\n^2)\,,
\ee
the wave function will become
\be \label{dspsga}
\psi_g(y)=  \frac{c_1 e^{\frac{\nu  y}{2\alpha }} \, _2F_1\big(-\frac{3}{2},\frac{\nu -3}{2};\frac{\nu +2}{2};e^{\frac{2 y}{\alpha }}\big) +c_2  e^{-\frac{\nu  y}{2 \alpha }}\, _2F_1\big(-\frac{3}{2},-\frac{\nu }{2}-\frac{3}{2};1-\frac{\nu }{2};e^{\frac{2 y}{\alpha }}\big)}{\big(e^{\frac{2 y}{\alpha }}-1\big)^{3/2}}\,,
\ee
where $c_1$ and $c_2$ are constants of integration.
Moving towards the UV boundary at $y=0$ the leading term in the expansion of \eqref{dspsga} is
\be
\psi_g(y)=
2 \sqrt{\frac{2}{\pi }} \alpha ^{\frac32} \left(c_1\frac{ \Gamma \left(1+ \frac{\nu }{2}\right)}{\Gamma \left(\frac{5}{2}+\frac{\nu}{2}\right)}
+c_2   \frac{ \Gamma \left(1-\frac{\nu }{2}\right)}{\Gamma \left(\frac{5}{2}-\frac{\nu }{2}\right)}
\right){y ^{-\frac32}}+ \mathcal{O}(y^\frac12)\,, \label{UVg ds2}
\ee
By choosing $c_1$ such that the big parenthesis vanishes we can make the wave function finite. However, the expansion of the solution \eqref{dspsga} near the IR end-point, $y \rightarrow -\infty$, is
\be \label{IRg ds2}
\psi_g(y)= c_2 e^{-\frac{\n}{2\a}y} + \mathcal{O}e^{(-\frac{\n}{2\a}+1)y}\,.
\ee
Because $c_2$ is a free constant the wave function \eqref{dspsga} blows up at the IR end-point.  Therefore, there is no normalizable vev solution below the mass gap when we have the dS slices.

\item AdS slices
	
The potential here becomes
\be \label{Vadsgl}
V(y)= -\frac{3}{4 \alpha^2}\Big(3-5 \,\csc ^2 (\frac{y}{\alpha}) \Big) \,.
\ee
The behavior of $V_g(y)$ at two UV boundaries  is
\begin{gather}
	V_g^{UV}(y)=\frac{15}{4 y^2}-\frac{1}{\alpha  ^2}+\mathcal{O}\left(y^1\right)\, \sp y\rightarrow 0\,,
\nn\\
	V_g^{UV}(y)=\frac{15}{4 (y-\alpha \pi)^2}-\frac{1}{\alpha ^2} +\mathcal{O}\left(\left(y-\pi \alpha \right)^1\right)\, \sp  y\rightarrow \alpha \pi\,.
\end{gather}
The solution of the Schrodinger equation \eqref{S-grn} is
\be \label{Siadsgln}
\psi_g(y)=\sin(\frac{y}{\alpha})^{1/2} \Big[ C_1 P_s^2(\cos (\frac{y}{\alpha}))+C_2 Q_s^2(\cos (\frac{y}{\alpha}))\Big]\,,
\ee
where $ P_{s}^2$ and $Q_{s}^2 $ are the associated Legendre functions, with
\be \label{Legindex}
s=\frac{1}{2}\left(\sqrt{4 \alpha ^2 M^2+9}-1\right) \sp s\geq 2\,.
\ee
The expansion of the \eqref{Siadsgl} near the $y=0$ boundary is given by
\begin{gather}
	\psi_g(y) =C_2 \Big( 2 (\frac{y}{\alpha})^{-3/2} +\frac12 (s^2+s-1)(\frac{y}{\alpha})^{1/2} + \mathcal{O}(\frac{y}{\alpha})^{5/2}\Big)
\nn \\
	\qquad \,\,+C_1\Big( \frac18  (s-1) s (s+1) (s+2)(\frac{y}{\alpha})^{5/2} + \mathcal{O}(\frac{y}{\alpha})^{7/2} \Big)+\cdots\,.
\end{gather}
To obtain a regular solution we should set $C_2=0$. After eliminating the second term in \eqref{Siadsgln} the expansion near the second boundary i.e. $y=\alpha\pi $ is
\be
\psi_g(y)=-C_1\frac{4\alpha^{3/2}}{\pi} (\alpha \pi -y)^{-3/2} \sin(s  \pi)+ \mathcal{O}\left((\alpha\pi-y )^{1/2}\right)\,.
\ee
To obtain a source-free solution, at the second boundary we impose the boundary condition $\psi_g(\a\pi)=0$ which dictates that $s$ should be an integer number and therefore the mass spectrum of the graviton modes is discrete
\be \label{mspch}
M^2=\frac{1}{4 \alpha^2}\big[(2s+1)^2-9 \big] \sp s\geq2 \sp s\in \mathbb{N}\,.
\ee
To check the normalizability we also obtain that
\be
\int_0^{\a\pi}|\psi_g|^2 dy =|C_1|^2 \int_0^{\a\pi}\sin(\frac{y}{\a}) \big[P_s^2(\cos (\frac{y}{\alpha}))\big]^2 dy=\a |C_1|^2 \frac{2}{2s+1}\frac{(s+2)!}{(s-2)!}\,.
\ee
Although \eqref{mspch} gives a minimum positive mass, for the massless case we can solve the Schrodinger equation independently. The solution is
\be
\psi_g (y)= -\frac{c_1}{\sin^{3/2}(\frac{y}{\alpha})}+ \frac{c_2}{3}\frac{\cos(\frac{y}{\alpha})}{\sin^{3/2}(\frac{y}{\alpha})}\Big(\cos^{2}(\frac{y}{\alpha})-3\Big)\,.
\ee
This solution is not finite either at $y=0$ or at $y=\alpha \pi$, so there is no massless normalizable graviton mode in global AdS space.
\end{itemize}

 \vspace{0.3cm}
%%%%%%%%%%%%%%%%%%%%%%%%%%%%%%%%%%%%%%%%%%%%%
%%%%%%%%%%%%%%%%%%%%%%%%%%%%%%%%%%%%%%%%%%%%%
%%%%%%%%%%%%%%%%%%%%%%%%%%%%%%%%%%%%%%%%%%%%%

\subsection{Scalars}\label{SAds}

The scalar mode equations \eqref{As1ap}, \eqref{As2ap} and \eqref{As3ap} will reduce to the following equations by setting  $\Phi_0'= 0$
\be
\psi' - \frac{a'}{a}\phi +\frac{1}{3} \kappa(W-E')=0\,,\label{s3Gads}\ee
\be
 2  \psi+ \phi -\frac{1}{a^3} \big(a^3(W-E')\big)'=0\,,\label{s2Gads}\ee
 \be
(\nabla_\m \nabla^\m   +\frac{4}{3} \kappa)\psi +4 \frac{a'}{a}\psi'- \frac{a'}{a} \nabla_\m \nabla^\m (W-E')- \Big(\frac{a''}{a}+2 \big(\frac{a'}{a}\big)^2\Big)\phi - \frac{1}{3}\kappa \phi =0\,.\label{s4Gads}
\ee
Solving $\phi$ from \eqref{s3Gads} and inserting into the \eqref{s2Gads} and \eqref{s4Gads} and after simplifying by equation \eqref{Eom1} we find
\be \label{eom alpha}
(a^2 \omega)'=0\sp (\nabla_\m \nabla^\m +\frac43 \k)\omega=0\,,
\ee
where we have defined
\be \label{ginv Gads}
\omega \equiv \psi- \frac{a'}{a}(W-E')\,.
\ee
This is not an allowed scalar mode since it satisfies equation \eqref{eom alpha}, however, we have assumed that the scalar field should not be the zero mode of the $\nabla_\m \nabla^\m +\frac43 \k$ operator.

In addition to the above mentioned equations there is another equation, \eqref{As4ap}, where for  $\Phi_0'= 0$ reduces to
\be
\chi'' + 3 \frac{a'}{a}\chi' + (\nabla_\m \nabla^\m - a^2 m_{\Phi}^2) \chi=0\,, \label{s5Gads}
\ee
where $m_{\Phi}$ is the mass of the background scalar field.
In this case, if we decompose the scalar field as
\be \label{decomchi}
\chi (y,x)=\hat{\chi}(x)\chi (y) \,,
\ee
and  assume that its four dimensional part obeys
\be \label{chi4d}
\nabla_\m \nabla^\m  \hat{\chi}(x)= m_{\chi}^2 \, \hat{\chi}(x) \,,
\ee
then  $\chi(y)$ satisfies
\be \label{eqchiy}
\chi''(y)+ 3 \frac{a'}{a} \chi'(y)+ (m_{\chi}^2- a^2 m_{\Phi}^2) \, \chi(y)=0.
\ee
The action for $\chi$ reads as
\be\label{acchi}
S_{\chi} = - \frac{1}{ 2k_5^2}\int d^4x dy\,\sqrt{-g^{(0)}}a^3(y)\Big(\pa_\m \chi \pa^\m \chi + \chi'^2 + a^2 m_{\Phi}^2 \chi^2 \Big)\,.
\ee
By separation \eqref{decomchi} the above action can be written as
\be \label{acxi}
S_{\chi}= -\frac{1}{2 k_5^2}\int dya^3(y)\chi^2(y)\int d^4x \sqrt{-g^{(0)}}\Big(\p_\mu  \hat{\chi}(x)\p^\mu  \hat{\chi}(x)+ m^2_\chi \hat{\chi}^2(x) \Big)\,,
\ee
where we have integrated by part the $\chi'^2$ term in \eqref{acchi} and have used the equation \eqref{eqchiy}.
Therefore, the normalizability condition for this field reads as
\be
\int dy\, a^{3}(y)\chi^2(y) < \infty \,.
\ee
We note that by the same steps as subsection \ref{tenmod} for graviton, that is by defining a new scalar field as
\be \label{siki}
\psi_s(y)=\chi(y)e^{-B(y)} \sp  a(y)=e^{-\frac{2}{3}B(y)}\,,
\ee
the equation \eqref{eqchiy} can be written as
\be \label{Sch-s}
-\psi_s''(y) + V_s(y)\psi_s(y)= m^2_{\chi} \psi_s(y) \,,\ee
\be \label{V-s}
V_s(y)= B'^2(y)-B''(y)+ a^2(y) m_{\Phi}^2 \,.
\ee
According to the scale factors given in \eqref{ayn}, when the slices are de Sitter or anti-de Sitter space-time, the leading term of the above potential in the expansion near the UV boundary is
\be \label{VsL}
V_s(y)=\frac{1}{4y^2}(2 \Delta_- -5) (2 \Delta_- -3)+ \mathcal{O}(0)\,,
\ee
where $\Delta_-$ is given with \eqref{Delpm}.

 The expansion of the wave function near the UV boundary is
\be \label{WsL}
\psi_s(y) =c_1 (y^{\frac52-\Delta_-}+\cdots) + c_2 (y^{-\frac32+ \Delta_-}+\cdots)\,.
\ee

\vspace{0.5cm}
%%%%%%%%%%%%%%%%%%%%%%%%%%%%%%%%%%%%%%%%%%%%%%%%%%%%%
%%%%%%%%%%%%%%%%%%%%%%%%%%%%%%%%%%%%%%%%%%%%%%%%%%%%%
%%%%%%%%%%%%%%%%%%%%%%%%%%%%%%%%%%%%%%%%%%%%%%%%%%%%%

\section{QFT data for two boundary solutions}\label{UVUVS}

As we already mentioned in section \ref{uvlrsec},  there is one QFT on each $UV$ boundary. Each QFT is defined by two parameters, $R^{UV}$ and $\varphi_-$. For every solution, we have three dimensionless parameters $\mathcal{R}_L$, $\mathcal{R}_R$ and $\xi$ totally.

The QFT data on the boundaries are given as follows:
\begin{itemize}
\item In figures \ref{s0rl} and \ref{s0fl} we present  $\mathcal{R}_L$ and $\varphi_-^L$ as we change $S_0$ (moving vertically along the dashed line in map \ref{mapm}). Moving towards the red boundary in map \ref{mapm}, $\mathcal{R}_L\rightarrow -\infty$ while $\varphi_-^L\rightarrow 0$. This corresponds to a vev-driven solution. Moving towards the green boundary, both  $\mathcal{R}_L$ and $\varphi_-^L$ find the finite values.

\begin{figure}[!ht]
\begin{subfigure}{0.49\textwidth}
\includegraphics[width=1 \textwidth]{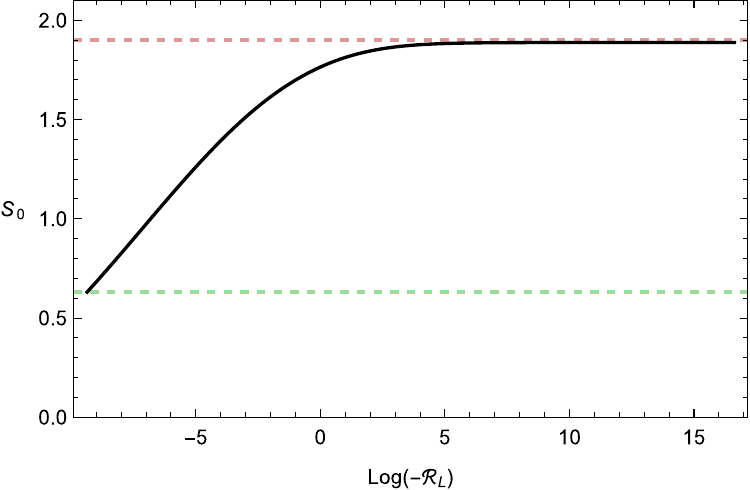}
\caption{\footnotesize{}}\label{s0rl}
\end{subfigure}
\begin{subfigure}{0.49\textwidth}
\includegraphics[width=1 \textwidth]{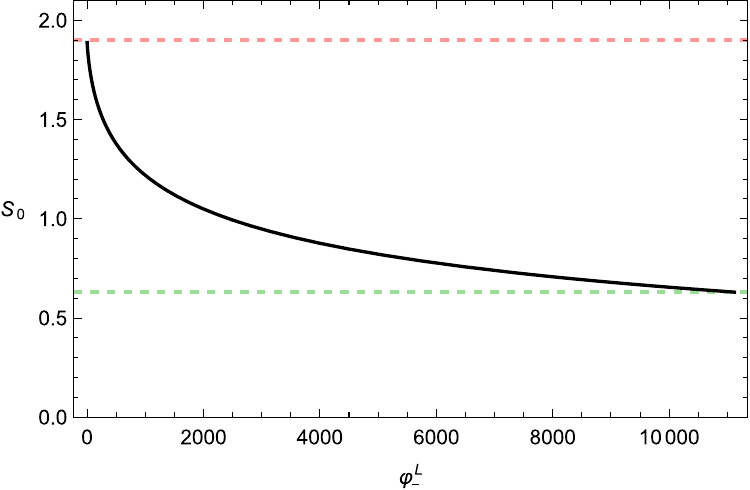}
\caption{\footnotesize{}}\label{s0fl}
\end{subfigure}
\caption{\footnotesize{(a) shows the changes in $\log(-\mathcal{R}_L)$ as we move along the dashed line in map \ref{mapm}. (b) shows the behavior of $\varphi_-^L$. The red and green dashed lines are the up-left and down-left boundaries of figure \ref{mapm} respectively. }}
\end{figure}
\begin{figure}[!ht]
\begin{subfigure}{0.49\textwidth}
\includegraphics[width=1 \textwidth]{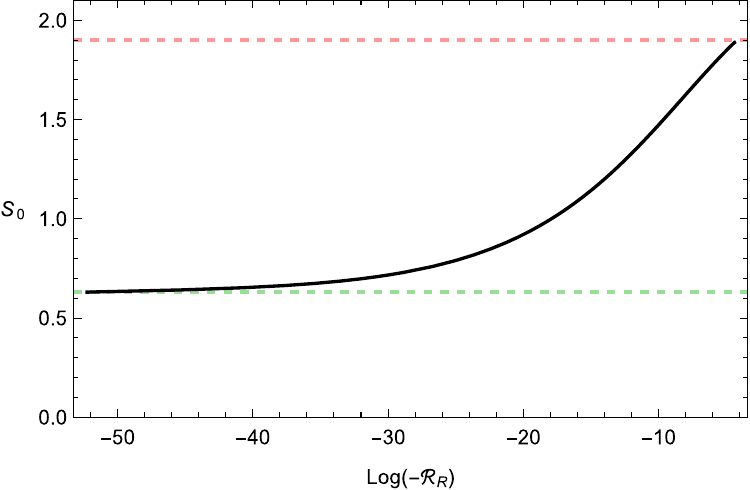}
\caption{\footnotesize{}}\label{s0rr}
\end{subfigure}
\begin{subfigure}{0.49\textwidth}
\includegraphics[width=1 \textwidth]{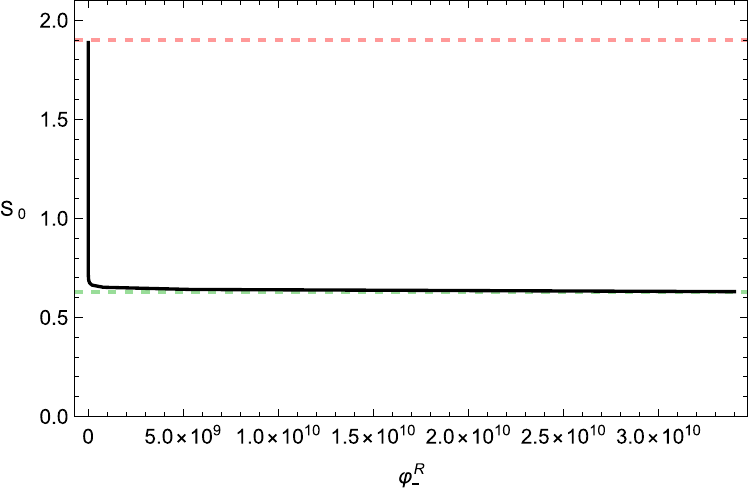}
\caption{\footnotesize{}}\label{s0fr}
\end{subfigure}
\caption{\footnotesize{(a) shows the changes in  $\log(-\mathcal{R}_R)$ as we move along the dashed line in map \ref{mapm}. (b) shows the behavior of $\varphi_-^R$.}}
\end{figure}

\begin{figure}[!ht]
\begin{subfigure}{0.49\textwidth}
\includegraphics[width=1 \textwidth]{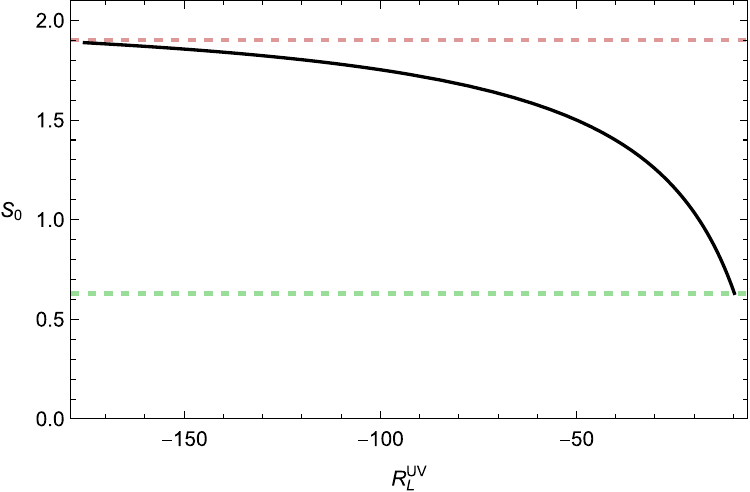}
\caption{\footnotesize{}}\label{s0Rluv}
\end{subfigure}
\begin{subfigure}{0.49\textwidth}
\includegraphics[width=1 \textwidth]{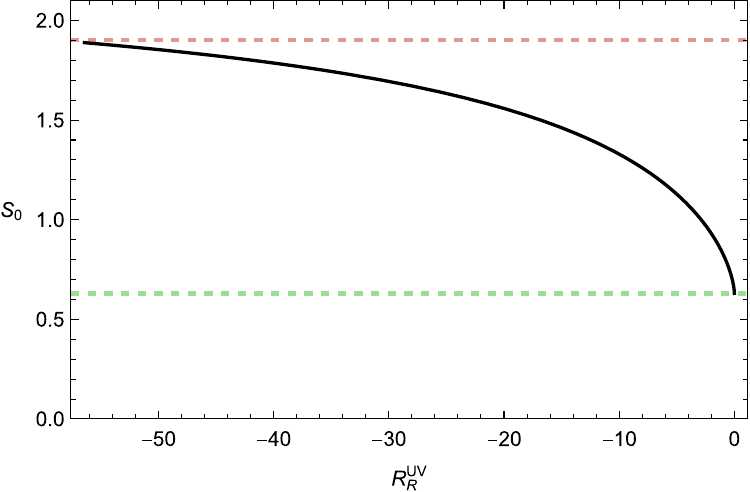}
\caption{\footnotesize{}}\label{s0RRuv}
\end{subfigure}
\caption{\footnotesize{(a) shows the changes in  $R^{UV}_L $ as we move along the dashed line in map \ref{mapm}. (b) shows the behavior of $R^{UV}_R$.}}
\end{figure}

\begin{figure}[!ht]
\centering
\begin{subfigure}{0.49\textwidth}
\includegraphics[width=1 \textwidth]{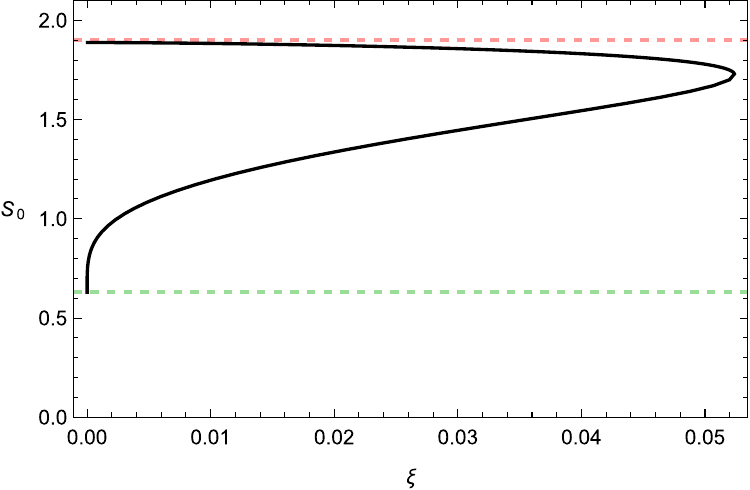}
\end{subfigure}
\caption{\footnotesize{ The ratio of the couplings in \eqref{xidef}.}}\label{cp}
\end{figure}

\item In figures \ref{s0rr} and \ref{s0fr} we present $\mathcal{R}_R$ and $\varphi_-^R$. As we increase $S_0$ and move towards the red boundary in map \ref{mapm}, $\mathcal{R}_R$ and $\varphi_-^R$ reach the constant values. However, by moving towards the green boundary,  $\mathcal{R}_R\rightarrow 0$ while $\varphi_-^R\rightarrow +\infty$.

\item Figure \ref{cp} shows the ratio of the couplings in \eqref{xidef}.
\end{itemize}

\vspace{0.5cm}
%%%%%%%%%%%%%%%%%%%%%%%%%%%%%%%%%%%%%%%%%%%%%%%%%%%%%
%%%%%%%%%%%%%%%%%%%%%%%%%%%%%%%%%%%%%%%%%%%%%%%%%%%%%
%%%%%%%%%%%%%%%%%%%%%%%%%%%%%%%%%%%%%%%%%%%%%%%%%%%%%

\section{The near UV boundary behavior of the scalar mode}\label{UVexp}

In this appendix we shall derive the near-boundary expansions of the scalar mode and its potential.

As we showed in subsection \ref{smode}, the equation for the scalar field is \eqref{SCH}
\be\label{SCHap}
\psi_s'' - V_s(y) \psi_s = 0 \,,
\ee
with the potential \eqref{SLpot}
\be  \label{SLpotap}
V_s(y)=-\kappa(1+\frac{ a }{3a'}h') +\frac{1}{4} {h'}^2 +\frac{1}{2}h'' -m^2\,.
\ee
The function $h(y)$ is defined in \eqref{decal}
\be\label{etth}
e^{h(y)}\equiv \frac{a^3 z^2}{9 m^2 + 12 \kappa - \kappa z^2}\,,
\ee
with $z(y)$ that is defined in \eqref{defz} as
\be \label{zdef2}
z\equiv \frac{a \Phi'_0}{a'}= \frac{ \Phi'_0}{A'}\,.
\ee
Knowing the expansions of  $A(y)$ and $\Phi(y)$ from \eqref{Ayexp} and \eqref{phiyexp} one can obtain the expansion of $z(y)$ in $d=4$ dimension as
\be\label{zyexp}
z(y)= -\varphi_- \Delta_- \, y^{\Delta_-} +
\begin{cases}
 \frac{ \Delta_-^2 (4\Delta_-^2 -7 \Delta_- -6) \varphi_-^3 }{12(2\Delta_- +1)(\Delta_- -1)}y^{3 \Delta_-}+ \mathcal{O}(y^{5\Delta_-})  \qquad & 0 < \Delta_- \leq \frac 12  \\ \cr
 \frac{ \Delta_-^2 (4\Delta_-^2 -7 \Delta_- -6) \varphi_-^3 }{12(2\Delta_- +1)(\Delta_- -1)}y^{3 \Delta_-}+ \mathcal{O}(y^{2+ \Delta_-})\, \qquad & \frac 12 \leq \Delta_- \leq 1  \\ \cr
-\mathcal{C} \Delta_+ \varphi_-^{(4/\Delta_-) -1} y^{\Delta_+}+ \mathcal{O}(y^{2+ \Delta_-})\, \qquad & 1 \leq \Delta_- < 2
\end{cases}\,,
\ee
where except for the leading term, the other terms in the expansion change depending on the value of $\Delta_-$.

Expansion of the function $h(y)$ near the boundary would be
\be\label{hyexp}
h(y)=\log \Big[ \frac{\ell^3 \Delta_-^2 \varphi_-^2}{9m^2 + 12 \k}\,y^{2 \Delta_- -3}\Big] +
\begin{cases}
h_1 y^{2\Delta_-} + \mathcal{O}(y^2)  \,\,\, & 0 < \Delta_- < 1 \\ \cr
\frac{2 \mathcal{C}(4-\Delta_- )}{\Delta_-}\varphi_-^{\frac{4}{\Delta_-}-2} y^{4-2\Delta_-}+ \mathcal{O}(y^2)\, \,\,\, & 1 \leq \Delta_- < 2
\end{cases}\,,
\ee
with
\be
h_1=- \frac{\Delta_- \Big(3 m^2 (9-4\Delta_-^2 + 4 \Delta_-) +2\k(\Delta_-+1)(2\Delta_-^2 -11 \Delta_-+18) \Big) \varphi_-^2}{2(9m^2 +12\k)(\Delta_- -1)(2 \Delta_- +1)}\,.
\ee
Finally, the expansion of the scalar potential near the UV boundary is
\be\label{VexpSap}
V_s(y)=\frac{1}{4y^2}(2 \Delta_- -5) (2 \Delta_- -3) +
\begin{cases}
V_1 \,y^{2 \Delta_- -2}+ \mathcal{O}(y^{ 4\Delta_--2}) , \qquad & 0 < \Delta_- < \frac 12  \\ \cr
V_1\,y^{2 \Delta_- -2}+ \mathcal{O}(y^0)\, \qquad & \frac 12 \leq \Delta_- < 1 \\ \cr
V_2+ \mathcal{O}(y^{2 \Delta_- -2})\, \qquad & 1 < \Delta_- < \frac32 \\ \cr
V_2+ \mathcal{O}(y^{4-2\Delta_-})\, \qquad & \frac{3}{2} \leq \Delta_- < 2
\end{cases}\,,
\ee
with
\be\label{V1S}
V_1=\frac{2 \Delta_- ^2 \Big(2\k (\Delta_- +1) (2 \Delta_-^2 -11\Delta_-   +18)+3 m^2(9-4\Delta_-^2 +4\Delta_-)\Big)}{ (2 \Delta_- +1) \left(9m^2+12\k \right)}\varphi_- ^2\,,
\ee
and
\begin{gather}
V_2=\frac{2\k}{3}(\Delta_- -3)-m^2- \frac{1}{32}\mathcal{R} (\Delta_-^2 + 2 \Delta_- -21 ) \varphi_-^ {2/\Delta_-} \nn\\
=-m^2- \frac{\k}{24} (3 \Delta_-^2 - 10 \Delta_- -5 ) \,.\label{V2S}
\end{gather}
The last equality comes from equations \eqref{yuq4} and \eqref{yuq1} i.e.
\be
\mathcal{R} \varphi_-^ {2/\Delta_-}=\mathcal{R} \tilde{ \varphi}_- ^{2/\Delta_-} e^{2A_-} = R^{(\zeta)}=4 \k \,.
\ee

\vspace{0.5cm}
%%%%%%%%%%%%%%%%%%%%%%%%%%%%%%%%%%%%%%%%%%%%%%%%%%%%%
%%%%%%%%%%%%%%%%%%%%%%%%%%%%%%%%%%%%%%%%%%%%%%%%%%%%%
%%%%%%%%%%%%%%%%%%%%%%%%%%%%%%%%%%%%%%%%%%%%%%%%%%%%%
\section{Laplace operator on AdS$_{d+1}$ and dS$_{d+1}$ manifold.}\label{ApLaplace}

In this appendix, we briefly review the action of the Laplace operator in AdS and dS space-times on scalar, vector, and spin 2 fields. We also examine the solutions of the zero modes which we discussed in section \ref{Dec}.

%%%%%%%%%%%%%%%%%%%%%%%%%%%%%%%%%%%%%%%%%%%%%%%%%%%%%
%%%%%%%%%%%%%%%%%%%%%%%%%%%%%%%%%%%%%%%%%%%%%%%%%%%%%
%%%%%%%%%%%%%%%%%%%%%%%%%%%%%%%%%%%%%%%%%%%%%%%%%%%%%
\subsection{AdS$_d$ slices}

Consider the AdS$_d$ space-time in Poincar\'e coordinate
\be \label{Poincare}
ds^2=\frac{\alpha^2}{z^2} \Big(dz^2+ \eta_{ij} dx^i dx^j \Big) \sp z \in [0, \infty)\,,
\ee
where $\alpha$ is the curvature length scale and $\eta_{ij} $ is the $(d-1)$--dimensional Minkowski space-time.
The non-zero components Christoffel connections associated to \eqref{Poincare} are
\be
\Gamma^z_{zz}= -\frac{1}{z}\sp \Gamma^z_{ij}=\frac{1}{z} \eta_{ij}\sp \Gamma^i_{zj}=-\frac{1}{z} {\delta^i}_{\!j}\,.
\ee

%%%%%%%%%%%%%%%%%%%%%%%%%%%%%%%%%%%%%%%%%%%%%%%%%%%%%
%%%%%%%%%%%%%%%%%%%%%%%%%%%%%%%%%%%%%%%%%%%%%%%%%%%%%
%%%%%%%%%%%%%%%%%%%%%%%%%%%%%%%%%%%%%%%%%%%%%%%%%%%%%
\subsubsection{Scalar field}\label{ApD}

The Laplacian of a scalar field in Poincare coordinates is (the indices are raised by $\eta^{ij}$)
\be \label{AdS scaLap}
\Box  \phi (z,x)=\frac{z^2}{\alpha^2}\Big(\pa_z ^2 + \pa_i \, \pa^i - \frac{d-2}{z}\pa_z \Big)\phi(z,x)\,.
\ee
For a massive scalar field, the equation of motion is
\be \label{eomsca}
(\Box -m^2)\phi(z,x) =0 \,.
\ee
By a Fourier transformation in $\mathbb{R}^{1,d-2}$,
\be\label{FT}
\phi(z,x)= \int \frac{d^{d-1}q}{(2\pi)^{d-1}}\, e^{-iq_i x^i} \phi(z,q)\,,
\ee
equation \eqref{eomsca} becomes
\be \label{ads4 scalar}
\Big(z^2(\pa_z ^2 +q^2) -z(d-2)\pa_z - \alpha^2 m^2 \Big)\phi (z,q)= 0\,,
\ee
where we have considered $q^i q_i =\eta^{ij} q_i q_j= - q^2 $.

The solution of the above equation is given in terms of the Bessel functions. In the coordinate space, the solution is
\be\label{AI1}
\phi(z,x)= \int \frac{d^{d-1}q}{(2\pi)^{d-1}}\, e^{-iq_i x^i} f(q)  z^{(d-1)/2} Z_\n (\sqrt{q^2} z)\,, \qquad
\ee
where $f(q)$ is an arbitrary function and
\be \label{AI2}
 \n=\frac{1}{2}\sqrt{(d-1)^2+4 m^2 \alpha^2}\,,
\ee
In equation \eqref{AI1}, $Z_\n$ stands for one of the two linearly independent solutions of the Bessel equation.

If we consider equation \eqref{ads4 scalar}, and perform a change as $\phi(z,q)=z^{\frac{d-2}{2}} \tilde{\phi}(z,q)$, this equation can be written as the following Schrodinger-like equation
\be \label{SCADS}
-\tilde{\phi}''(z,q)+V_{AdS}(z)\tilde{\phi}(z,q)=q^2 \tilde{\phi}(z,q) \sp V_{AdS}(z)=\frac{4m^2\a^2+d(d-2)}{4z^2}\,.
\ee
Considering $q^2=-\eta^{ij}q_i q_j>0$, the above equation shows that for each mode, the value of $q^2=q_0^2-\vec{q}^2$ is continuous for all values of the mass. This is because $q^2>0$ no matter what the sign of $V_{AdS}$ is. Therefore, the energy of each mode, i.e. $q_0$ is continuous.

Expanding solution of \eqref{ads4 scalar} near the AdS boundary at $z=0$ one finds
\be  \label{SCADS1}
\phi(z,q) = C_1 q^{-\n} z^{\frac{d-1}{2}-\n} + C_2 q^\n z^{\frac{d-1}{2}+\n}+\cdots\,.
\ee
For masses with $m^2 > 0$ we should choose $C_1=0$ (vev solutions). However for massless mode the expansion is
\be  \label{SCADS2}
\phi(z,q) = C_1 q^{-\frac{d-1}{2}} + C_2 q^{\frac{d-1}{2}} z^{d-1}+\cdots\,.
\ee

%%%%%%%%%%%%%%%%%%%%%%%%%%%%%%%%%%%%%%%%%%%%%%%%%%%%%
%%%%%%%%%%%%%%%%%%%%%%%%%%%%%%%%%%%%%%%%%%%%%%%%%%%%%
%%%%%%%%%%%%%%%%%%%%%%%%%%%%%%%%%%%%%%%%%%%%%%%%%%%%%

\subsubsection{Vector field}

The conventional Lagrangian for a massive gauge field is
\be
L=-\frac{1}{4} {F}^{\m\n}{F}_{\m\n} -\frac{1}{2} m^2 {A}_\m {A}^\m\,.
\ee
The equation of motion for the gauge field is then
\be
0=\nabla_\m {F}^{\m\n} - m^2 {A}^\n
=\nabla_\m (\nabla^\m {A}^\n -\nabla^\n {A}^\mu) -m^2 {A}^\n \,,\label{gaugeA}
\ee
in which, after taking a divergence of the first line, it implies that $\nabla_\m {A}^\m=0 $  and \eqref{gaugeA}  in a maximally symmetric space can be written as
\be \label{massiveAeq}
\nabla_\m \nabla^\m  {A}_\n -(\kappa + m^2 ){A}_\n =0 .
\ee
The term $\kappa {A}_\n $ comes from the commutation of the covariant derivatives.

The zero mode that appeared already in equation \eqref{BTZ} corresponds to a massive vector field with a mass
\be
m^2=-2\kappa\,.
\ee
To find the solution of our vector field zero mode we start with the following equation
\be \label{lapA}
(\nabla_\n \nabla^\n  + \kappa)A_\m(z,x)=0\,,
\ee
where in the  AdS$_d $ case  we have $\alpha^2 \kappa=-(d-1) $.  Following \cite{Mueck:1998iz}, we decompose the $\m$ index into $z$ and $i$ components with $i= 1,\cdots,  d-1$.
For the $A_z$ component the Laplace operator is
\be \label{AdSAz}
\nabla_\m \nabla^\m  A_z=  \frac{z^2}{\alpha^2}\Big[(\pa_z^2 + \pa_i^2)-\frac{1}{z^2}(2d-3)-\frac{1}{z}(d-4)\pa_z \Big]A_z +\frac{2 z}{\alpha^2}\pa^i A_i\,.
\ee
So the equation  \eqref{lapA} becomes
\be \label{lap Az}
\pa_z^2 A_z +  \pa_i \, \pa^i A_z- \frac{d-4}{z}\pa_z A_z + \frac{2}{z}\,  \pa_i A^i - \frac{2d-3}{z^2} A_z - \frac{d-1}{z^2}\, A_z=0\,.
\ee
The gauge condition  $\nabla_\m A^\m=0 $ now reads as
\be  \label{ads4 divA}
\pa_i A^i= -\pa _z A_z + \frac{d-2}{z} A_z\,.
\ee
By using the above gauge condition and the Fourier transform  of $A_z$, equation \eqref{lap Az}  becomes
\be \label{ads4 eom Az}
\Big[z^2\pa_z^2 + z^2 q^2 + z(2-d) \pa_z -d \Big] A_z(z,x)=0\,.
\ee
Similar to the scalar field, the solution for  $A_z$ is given by the Bessel function
\be \label{Az sol}
A_z (z,x)= \int \frac{d^{d-1}q}{(2\pi)^{d-1}}\, e^{-iq.x} \mathcal{A}_z(q) z^{(d-1)/2} Z_\n (\sqrt{q^2}z)\,,
\ee
where $\mathcal{A}_z(q)$ is an arbitrary function and
\be \label{nud}
 \n=\frac{d+1}{2}\,.
\ee
For $A_i$ components, equation \eqref{lapA} becomes
\be  \label{ads4 eom Ai}
 \Big[z^2\pa_z^2 + z^2 \pa_j \, \pa^j + (4-d)z\pa_z -2(d-1)\Big]A_i =2z\,  \pa_i A_z\,.
\ee
Now as in \cite{Mueck:1998iz} we introduce a vector field such that
\be
\tilde{A}_\m(z,x)= z A_\m(z,x)\,.
\ee
Then the equation \eqref{ads4 eom Ai} changes to
\be
 \Big[z^2(\pa_z^2 + \pa_j \, \pa^j) + z(2-d)\pa_z - d\Big]\tilde{A}_i =2z\,  \pa_i \tilde{A}_z\,.
\ee
The solution of the homogeneous part of the above equation is like the $A_z$-component, and it can be shown that the full solution  has the following form
\be \label{Ai sol}
\tilde{A}_i(z,x)= \int \frac{d^{d-1}q}{(2\pi)^{d-1}}\, e^{-iq.x} z^{(d-1)/2} \Big[  \mathcal{A}_i(q) Z_\n (\sqrt{q^2}z) - i z \mathcal{A}_z(q) \frac{q_i}{q} Z_{\n+1} (\sqrt{q^2}z)  \Big] \,.
\ee
The gauge condition \eqref{ads4 divA} in terms of $\tilde{A}_\m $ becomes
\be \label{div AT}
z(\pa_z \tilde{A}_z + \pa_i \tilde{A}_i) -(d-1)\tilde{A}_z=0\,.
\ee
Inserting \eqref{Ai sol} and \eqref{Az sol} in the above equation leads to
\be
\mathcal{A}_z = \frac{i}{2} \mathcal{A}_i \, q_i\,.
\ee

%%%%%%%%%%%%%%%%%%%%%%%%%%%%%%%%%%%%%%%%%%%%%%%%%%%%%
%%%%%%%%%%%%%%%%%%%%%%%%%%%%%%%%%%%%%%%%%%%%%%%%%%%%%
%%%%%%%%%%%%%%%%%%%%%%%%%%%%%%%%%%%%%%%%%%%%%%%%%%%%%

\subsubsection{Gravitons}

The equation of motion for a massive spin 2 field in $d=4$ is \eqref{Sp 2}. In $d-$dimensions we obtain
\be \label{lapG}
(\nabla_\r \nabla^\r  -\frac{2}{d-1} \kappa)h_{\m\n}(z,x) = M^2h_{\m\n}(z,x)\,.
\ee
By decomposing $\m$ and $\n$ indices in $z$ and $x^i$ directions we obtain three different modes $h_{zz}, h_{zi}$, and $h_{ij}$.
Moreover, we work in transverse and traceless gauges in which
\be \label{TTG}
\nabla^\m h_{\m\n}= {h^\m} _{\!\m}=0\,.
\ee
This leads to the following constraints
\begin{gather} \label{cons1}
	 z (\pa_z h_{zz}+ \pa^i h_{iz})+{h^i}_{\!i} +(3-d)h_{zz}=0\,,
 \\
	 \label{cons2}  z(\pa_z h_{iz}+ \pa^j h_{ij})+(2-d)h_{iz}=0\,,
 \\
	 \label{cons3}  h_{zz}+{h^i}_{\!i}=0\,.
\end{gather}
For the ${zz}$ component of \eqref{lapG}, the equation of motion is
\be \label{hzz eq}
 \Big[z^2\pa_z^2 + z^2 q^2 +z(2-d)\, \pa_z -M^2\a^2  \Big] h_{zz}(z,x)=0\,.
\ee
Again the solution is given in terms of the Bessel functions
\be \label{hzz sol}
 h_{zz} (z,x)= \int \frac{d^{d-1}q}{(2\pi)^{d-1}}\, e^{-iq.x} \mathcal{H}_{zz}(q) z^{(d-1)/2} Z_\n (\sqrt{q^2}z) \,,
\ee
and
\be \label{nuM}
\n= \frac{1}{2} \sqrt{(d-1)^2 +4M^2 \alpha^2} \,.
\ee
The equation of motion for $h_{iz}$ components is
\be \label{hiz0}
\Big[z^2\pa_z^2 + z^2 q^2 +z(4-d)\, \pa_z -  (d-2  + M^2 \alpha^2)\Big] h_{iz}(z,x)=2 z \pa_i h_{zz}\,.
\ee
In deriving the above equation, we have used the gauge conditions \eqref{cons1} and \eqref{cons3}.
Once again, we introduce
\be
\tilde{h}_{iz}=z h_{iz} \sp \tilde{h}_{zz}=z h_{zz} \,.
\ee
Then equation \eqref{hiz0} becomes
\be \label{hiz eq}
 \Big[z^2\pa_z^2 + z^2 q^2 + z(2-d)\, \pa_z - M^2\a^2 \Big]\,\tilde{h}_{iz}(z,x)=2z\,\pa_i \tilde{h}_{zz} \,.
\ee
The solution for the homogeneous part is the same as $h_{zz}$ and the full solution is
\be \label{hiz sol}
 \tilde{h}_{iz} (z,x)= \int \frac{d^{d-1}q}{(2\pi)^{d-1}}\, e^{-iq.x} z^{(d-1)/2}\Big[ \mathcal{H}_{iz}(q)  Z_\n (\sqrt{q^2}z)-i z \frac{q_i}{q}  \mathcal{H}_{zz}(q)  Z_{\n+1} (\sqrt{q^2}z)\Big] \,.
\ee
Using the gauge constraints \eqref{cons1} and \eqref{cons3}, there is a relation between the coefficients in \eqref{hiz sol}
\be
(3+2\n -d)\mathcal{H}_{zz}=2 i q_i \mathcal{H}_{iz}\,.
\ee
The equation of motion for $h_{ij}$ components is
\begin{gather}
	\Big[z^2\pa_z^2 + z^2 q^2 +z(6-d)\, \pa_z -(2d-6 + M^2 \alpha^2)\Big] h_{ij}(z,x)
\nn\\
	= -2 \eta_{ij}\,h_{zz} + 4 z \pa_i h_{jz}\,. \label{AdS hij eom}
\end{gather}
Again if we introduce a new tensor field as
\be \label{hk}
\tilde{h}_{\m\n}= z^2 h_{\m\n} \,,
\ee
the homogeneous part of equation for $\tilde{h}_{ij}$ becomes the same as scalar part $h_{zz}$ and we obtain
\be
\Big[z^2\pa_z^2 + z^2 q^2 +z(2-d)\, \pa_z -  M^2\a^2 \Big]\tilde{h}_{ij}(z,x)
=-2 \eta_{ij}\,\tilde{h}_{zz} + 4 z \pa_i \tilde{h}_{jz}\,.
\ee
So the solution for the homogeneous part is the same as the two previous cases and the full solution is
\begin{gather}
	 \tilde{h}_{ij} (z,x)= \int \frac{d^{d-1}q}{(2\pi)^{d-1}}\, e^{-iq.x} z^{(d-1)/2}\Big[ \mathcal{H}_{ij}(q)  Z_\n (\sqrt{q^2}z)
 \nn\\
	 -\frac{z}{q^2}\Big(\mathcal{H}_{zz}(q) \, q_i\,  q_j\, z Z_{\nu +2}(\sqrt{q^2} z)+q \big(\eta_{ij}\mathcal{H}_{zz}(q)+2 i\,\mathcal{H}_{iz}(q)\, q_j\big) Z_{\nu +1}(\sqrt{q^2} z) \Big)\Big]\,. \label{AdS hij sol}
\end{gather}

%%%%%%%%%%%%%%%%%%%%%%%%%%%%%%%%%%%%%%%%%%%%%%%%%%%%%
%%%%%%%%%%%%%%%%%%%%%%%%%%%%%%%%%%%%%%%%%%%%%%%%%%%%%
%%%%%%%%%%%%%%%%%%%%%%%%%%%%%%%%%%%%%%%%%%%%%%%%%%%%%

\subsection{dS$_d$ slices}

Now we study the dS$_d$ space-time in Poincar\'e Coordinate  with metric
\be \label{dS Poincare}
ds^2=\frac{\alpha^2}{z^2} \Big(-dz^2+ \delta_{ij} dx^i dx^j \Big)\,,
\ee
where as in AdS$_d$,  $\alpha$ is the curvature length scale, $z$ is the conformal time and $\delta_{ij} $ is the $(d-1)$--dimensional Euclidean space-time.
The non-zero components of the Christoffel connections associated to \eqref{dS Poincare} are
\be \label{dS gamas}
\Gamma^z_{zz}= -\frac{1}{z}\sp \Gamma^z_{ij}=-\frac{1}{z} \delta_{ij}\sp \Gamma^i_{zj}=-\frac{1}{z} {\delta^i}_{\!j}\,.
\ee
Here we examine the equation of motion for different fields in this space-time.

%%%%%%%%%%%%%%%%%%%%%%%%%%%%%%%%%%%%%%%%%%%%%%%%%%%%%
%%%%%%%%%%%%%%%%%%%%%%%%%%%%%%%%%%%%%%%%%%%%%%%%%%%%%
%%%%%%%%%%%%%%%%%%%%%%%%%%%%%%%%%%%%%%%%%%%%%%%%%%%%%

\subsubsection{Scalar field}\label{scds}

When the Laplace operator acts on a scalar field in this coordinate, it gives
\be\label{dS box fi}
\Box \phi(z,x) = - \frac{z^2}{\alpha^2}\Big[\pa_z^2 -\pa_i^2 -\frac{(d-2)}{z}\pa_z \Big]\phi(z,x)\,.
\ee
The equation of motion for the scalar field, i.e. equation \eqref{eomsca} after the Fourier transform \eqref{FT}, becomes
\be \label{dS eom fi}
\Big(z^2(\pa_z^2 + q_E^2 )-z(d-2)\pa_z + \alpha^2 m^2\Big)\phi(z,q)=0\,.
\ee
where we have considered the momentum in Euclidean space $q^i q_i = + q_E^2 $. The solution of the above equation is the same as in the AdS case \eqref{AI1} however with
\be \label{dS nu fi}
\n=\frac{1}{2}\sqrt{(d-1)^2-4m^2 \alpha^2}\,.
\ee
%The zero mode is also the same as in the $AdS_4$ case, equations \eqref{fi0} and \eqref{fi0AdS}.
By a change as $\phi(z,q)=z^{\frac{d-2}{2}} \tilde{\phi}(z,q)$ the equation \eqref{dS eom fi} becomes a Schrodinger equation as follow
\be \label{SCHDS}
-\tilde{\phi}''(z,q_E)+V_{dS}(z)\tilde{\phi}(z,q_E)=q_E^2 \tilde{\phi}(z,q_E) \sp V_{dS}(z)=\frac{-4m^2\a^2+d(d-2)}{4z^2}\,.
\ee
Considering $q_E^2=\delta^{ij} q_i q_j=q_0^2+{\vec{q}}^2>0$ this equation again implies that $q_E^2$ and therefore the energy spectrum of the $q_0$ is continuous for all values of the masses $m$.

Expanding solution of \eqref{dS eom fi} near $z=0$ one finds
\be  \label{SCDS1}
\phi(z,q_E) = C_1 q_E^{-\n} z^{\frac{d-1}{2}-\n} + C_2 q^\n z^{\frac{d-1}{2}+\n}+\cdots\,.
\ee
In the case of massless mode the expansion is
\be  \label{SCDS2}
\phi(z,q_E) = C_1 q_E^{-\frac{d-1}{2}} + C_2 q_E^{\frac{d-1}{2}} z^{d-1}+\cdots\,.
\ee

%%%%%%%%%%%%%%%%%%%%%%%%%%%%%%%%%%%%%%%%%%%%%%%%%%%%%
%%%%%%%%%%%%%%%%%%%%%%%%%%%%%%%%%%%%%%%%%%%%%%%%%%%%%
%%%%%%%%%%%%%%%%%%%%%%%%%%%%%%%%%%%%%%%%%%%%%%%%%%%%%

\subsubsection{Vector field}

The Laplace operator acting on  $A_z$ component of a vector field gives
\be \label{dS box Az}
\nabla_\m \nabla^\m  A_z= - \frac{z^2}{\alpha^2}\Big[(\pa_z^2 - \pa_i^2)-\frac{1}{z^2}(2d-3)-\frac{1}{z}(d-4)\pa_z \Big]A_z +\frac{2 z}{\alpha^2}\pa^i A_i\,.
\ee
There is also the transversality condition on the gauge field that reads as
\be \label{DivA}
\pa^i A_i= \pa_z A_z - \frac{1}{z}(d-2)A_z\,.
\ee
So the Laplacian of $A_z$ can be written as an operator acting on $A_z$ only
\be \label{dS box Az1}
\nabla_\m \nabla^\m  A_z=- \frac{z^2}{\alpha^2}\Big[(\pa_z^2 - \pa_i^2)-\frac{1}{z^2}-\frac{1}{z}(d-2)\pa_z \Big]A_z\,.
\ee
The equation of motion \eqref{lapA} with $\a^2 \k =d-1$, for the $A_z$ component after the Fourier transform is
\be \label{dS eom Az}
\Big[z^2(\pa_z^2 + q_E^2)+z(2-d)\pa_z -d \Big]A_z=0\,.
\ee
This equation is the same as \eqref{ads4 eom Az}, so the solution is \eqref{Az sol}.

The Laplace operator acting on $A_i$ components gives
\be \label{dS box Ai}
\nabla_\m \nabla^\m  A_i=- \frac{z^2}{\alpha^2}\Big[(\pa_z^2 - \pa_j^2)-\frac{d-1}{z^2}+\frac{(4-d)}{z}\pa_z \Big]A_i +\frac{2 z}{\alpha^2}\pa_i A_z\,.
\ee
The equation of motion for $A_i$ is
\be \label{dS eom Ai}
[z^2(\pa_z^2 + q_E^2)+ z(4-d)\pa_z -2(d-1) \Big]A_i= 2 z\pa_i A_z\,,
\ee
again this equation is the same as \eqref{ads4 eom Ai} with the solutions given by \eqref{Ai sol}.

%%%%%%%%%%%%%%%%%%%%%%%%%%%%%%%%%%%%%%%%%%%%%%%%%%%%%
%%%%%%%%%%%%%%%%%%%%%%%%%%%%%%%%%%%%%%%%%%%%%%%%%%%%%
%%%%%%%%%%%%%%%%%%%%%%%%%%%%%%%%%%%%%%%%%%%%%%%%%%%%%

\subsubsection{Gravitons}

In dS space the constraints of transverse and traceless gauge on different components of graviton is
\begin{gather} \label{cons1ds}
	 z (\pa_z h_{zz}- \pa^i h_{iz})-{h^i}_{\!i} +(3-d)h_{zz}=0\,,
 \\
	 \label{cons2ds} z(\pa_z h_{iz}- \pa^j h_{ij})+(2-d)h_{iz}=0\,,
 \\
	 \label{cons3ds} h_{zz}-{h^i}_{\!i}=0\,.
\end{gather}
The Laplace operator acting on $h_{zz}$ component in transverse and traceless gauge is
\be \label{dS box hzz}
\nabla_\m \nabla^\m  h_{zz}= - \frac{z^2}{\alpha^2}\Big[\pa_z^2 -\pa_i^2+ \frac{(2-d)}{z}\pa_z - \frac{2}{z^2}\Big]h_{zz}
\ee
So the equation of motion \eqref{lapG} for this component of the graviton is
\be \label{dS eom hzz}
\Big[z^2 \pa_z^2 + z^2 q_E^2 -z(d-2)\pa_z +  M^2\a^2 \Big]h_{zz}=0\,.
\ee
The above equation is the same as \eqref{hzz eq} except for the sign of the last term. So the solution is also the same as   \eqref{hzz sol} with
\be \label{dS nu}
 \n=\frac{1}{2}\sqrt{(d-1)^2- 4M^2 \a^2}\,.
\ee
The Laplace operator for  $h_{i z}$ component is
\be \label{dS box hiz}
\nabla_\m \nabla^\m  h_{iz}= - \frac{z^2}{\alpha^2}\Big[\pa_z^2 -\pa_i^2- \frac{d}{z^2} + \frac{(4-d)}{z}\pa_z \Big]h_{iz} + 2 \frac{z}{\a^2} \pa_i h_{zz}
\ee
The equation of motion is
\be \label{dS eom hiz}
\Big[z^2\pa_z^2 + z^2 q_E^2 +z(4-d)\, \pa_z -(d-2- M^2 \alpha^2)\Big] h_{iz}(z,x)=2 z \pa_i h_{zz}\,.
\ee
The above equation is the same as \eqref{hiz eq} except in the sign of $M^2 \a^2$. So the solution is the same as \eqref{hiz sol} but with $\n$ given in \eqref{dS nu}.

The Laplace operator for  $h_{ij}$ component is
\be \label{dS box hij}
\nabla_\m \nabla^\m  h_{ij}= - \frac{z^2}{\alpha^2}\Big[\pa_z^2 -\pa_i^2+ \frac{4-2d}{z^2} + \frac{(6-d)}{z}\pa_z \Big]h_{ij} + 4 \frac{z}{\a^2} \pa_i h_{jz}+ \frac{2}{\a^2} \delta_{ij}h_{zz}\,.
\ee
The equation of motion is
\be
\Big[z^2\pa_z^2 + z^2 q_E^2 +z(6-d)\, \pa_z -(2d-6- M^2 \alpha^2)\Big] h_{ij}(z,x)=2 \delta_{ij} h_{zz} +4 z \pa_i h_{jz}\,. \label{dS eom hij}
\ee
The above equation is the same as \eqref{AdS hij eom} except in the sign of $M^2 \a^2$ and the coefficient of $h_{zz}$ on the right-hand side. So the solution is the same as \eqref{AdS hij sol} with $-\delta_{ij}$ instead of $\eta_{ij}$ and $\n$ given in \eqref{dS nu}.

\vspace{0.5cm}
%%%%%%%%%%%%%%%%%%%%%%%%%%%%%%%%%%%%%%%%%%%%%%%%%%%%%
%%%%%%%%%%%%%%%%%%%%%%%%%%%%%%%%%%%%%%%%%%%%%%%%%%%%%
%%%%%%%%%%%%%%%%%%%%%%%%%%%%%%%%%%%%%%%%%%%%%%%%%%%%%

\section{Unitary representations of dS$_4$ and mass spectrum}\label{reps}

In this section we briefly review the unitary irreducible representations of the isometry group of the 4 dimensional de sitter space, $SO(1,4)$, for more details see \cite{Sun:2021thf}.

In a dS$_4$ space-time with the length scale $\a $, for massive/massless fields with spin $s=0$ and $s=2$, the relation between the scaling dimension $\Delta$ and the mass is
\be \label{mdelta}
\Delta= \frac32 \pm \n \rightarrow m^2 \a^2 =\Delta(3-\Delta)\,,
\ee
where we have used relations \eqref{dS nu fi} and \eqref{dS nu}.

The various distinct unitary irreducible representations classify according to the values of the above scaling dimension.
\begin{itemize}
	
	\item  Scalar principal series:
	\be \label{U2}
	\Delta= \frac32 + i \m \sp  \m\in \mathbb{R}\,,
	\ee
	where for scalar modes in 	${dS_4}$
	\be \label{U3}
	\m = \frac12\sqrt{4m^2\a^2-9} \sp m^2>\frac{4}{9\a^2}\,.
	\ee
	This representation describes a massive scalar field with a mass above $m^2 = \frac{9}{4\a^2}$ and with a continuous mass spectrum.
	
	\item Scalar complementary series:
	\be \label{U4}
	0< \Delta < 3\,.
	\ee
	It describes a massive scalar field with masses in $0< m^2 < \frac{9}{4\a^2 }$.
	
	\item Exceptional (discrete) series I:
	
	This representation describe scalar modes with angular momentum $j$ where $j=0,1,2,\cdots$. The mass spectrum is discrete
	\be \label{U5}
	m^2 \a^2= -j(j+3) \,.
	\ee
		
	\item Spin-two principal series:
	
	For spin two fields this representation describes a continuous mass spectrum similar to the scalar principal representation i.e.
	\be \label{U6}
	\Delta= \frac32 + i \m \sp  \m\in \mathbb{R}\,,
	\ee
	where
	\be \label{U7}
	\m = \frac12\sqrt{4M^2\a^2-9} \sp M^2>\frac{9}{4\a^2}\,.
	\ee

	\item Spin-two complementary series:
	
	The scaling dimension for spin-two modes in this representation falls within the following range
	\be \label{U8}
	1< \Delta < 2\,.
	\ee
	It describes a  massive spin two field with a continuous mass spectrum in between $ \frac{2}{\a^2}< M^2  < \frac{9}{4\a^2}$. The lower bound is the Higuchi bound.
	
	\item Exceptional series II:
	
	For spin two fields it contains just two discrete modes
	\be \label{U9}
	M^2 \a^2= 0, 2\,,
	\ee
	where both modes are below the Higuchi bound \eqref{HB}.
	
\end{itemize}

%%%%%%%%%%%%%%%%%%%%%%%%%%%%%%%%%%%%%%%%%%%%%%%%%%%%%
%%%%%%%%%%%%%%%%%%%%%%%%%%%%%%%%%%%%%%%%%%%%%%%%%%%%%
%%%%%%%%%%%%%%%%%%%%%%%%%%%%%%%%%%%%%%%%%%%%%%%%%%%%%

\subsection{The representation content of the scalar zero modes}

We have two zero modes for scalar fields with masses
\be\label{ma}
m^2=0 \sp m^2=-\frac{4}{3}\k =\pm \frac{4}{\a^2}\,,
\ee
where plus(minus) sign stands for four dimensional AdS(dS) slices.
\begin{itemize}
	\item $m^2=0$ on dS corresponds to the $j=0$ mode of the exceptional discrete series I, \eqref{U5}.
	
	\item $m^2=-\frac{4}{\a^2}$ on dS is a tachyon and corresponds to the $j=1$ mode of the exceptional discrete series I, \eqref{U5}.
	
	\item $m^2=0$ on
	AdS is a normalizable mode according to  \eqref{SCADS2}.
	
	\item $m^2=\frac{4}{\a^2}$ on AdS is also a normalizable mode according to \eqref{SCADS1}.
\end{itemize}

%%%%%%%%%%%%%%%%%%%%%%%%%%%%%%%%%%%%%%%%%%%%%%%%%%%%%
%%%%%%%%%%%%%%%%%%%%%%%%%%%%%%%%%%%%%%%%%%%%%%%%%%%%%
%%%%%%%%%%%%%%%%%%%%%%%%%%%%%%%%%%%%%%%%%%%%%%%%%%%%%

\end{document}